\documentclass[reqno,12pt]{amsart}

\usepackage{amssymb}
\usepackage{fullpage}
\usepackage{amsthm}
\usepackage{amsmath}
\usepackage{amsfonts}
\usepackage{latexsym}

\numberwithin{equation}{section} 
{\bf}{\it}
\newtheorem{theorem}{Theorem}[section]{\bf}{\it}
\newtheorem{lemma}[theorem]{Lemma}{\bf}{\it}
\newtheorem{corollary}[theorem]{Corollary}{\bf}{\it}
\newtheorem{proposition}[theorem]{Proposition}{\bf}{\it}
{\bf}{\it}

\def\ov{\overline}

\def\ub{\underbrace}

\def\pe{\perp}
\def\pa{\parallel}

\def\mb{\mathbf}
\def\mr{\mathrm}
\def\mk{\mathfrak}
\def\bs{\boldsymbol}

\def\Rnum{\mathbb{R}}
\def\Cnum{\mathbb{C}}

\def\i{\mr{i}}
\def\cc{{\mathcal C}}

\def\inv{{}^{-1}}

\def\p{\partial}

\def\t{{\rm t}}

\def\<{\langle}
\def\>{\rangle}

\def\ha{\tfrac{1}{2}}

\def\ee{\mr{e}}

\def\hh{\mr{h}}
\def\gg{\mr{g}}

\def\ad{{\rm ad}}
\def\Ad{{\rm Ad}}

\def\diag{{\rm diag}}
\def\tr{{\rm tr}}

\def\C{P}
\def\A{Q}
\def\S{S}
\def\CC{\bs{P}}
\def\AA{\bs{Q}}
\def\SS{\bs{S}}

\def\m{\mk{m}}
\def\h{\mk{h}}

\def\conx{\omega}
\def\e{e}

\def\tbu{\mb{u}}
\def\tbh{\mb{h}}
\def\tbw{\mb{w}}
\def\tbHH{\mb{H}}
\def\tbWW{\mb{W}}
\def\ww{\mr{w}}

\def\btbu{\ov{\mb{u}}}
\def\btbh{\ov{\mb{h}}}
\def\btbw{\ov{\mb{w}}}
\def\bww{\ov{\mr{w}}}
\def\btbHH{\ov{\mb{H}}}
\def\btbWW{\ov{\mb{W}}}

\def\u{\mr{u}}

\def\bu{\ov{\mr{u}}}

\def\Rop{{\mathcal R}}
\def\Jop{{\mathcal J}}
\def\Hop{{\mathcal H}}
\def\Eop{{\mathcal E}}
\def\Kop{{\mathcal K}}

\def\X{{\rm X}}
\def\pr{{\rm pr}}

\def\map{\gamma}

\def\bpm{\begin{pmatrix}}
\def\epm{\end{pmatrix}}

\def\eg/{e.g.}
\def\ie/{i.e.}

\def\beq{\begin{equation}}
\def\eeq{\end{equation}}

\def\secref#1{Sec.~\ref{#1}}

\def\Ref#1{Ref.~\cite{#1}}

\begin{document}
\title{Symplectically-invariant soliton equations from non-stretching geometric curve flows}

\author{
Stephen C. Anco \lowercase{\scshape{and}}
Esmaeel Asadi\\
\\\lowercase{\scshape{
Department of Mathematics, Brock University, 
St. Catharines, ON Canada}} \\
\lowercase{\scshape{
Department of Mathematics, Institute for Advanced Studies in Basic Sciences, 
Gava Zang, Zanjan, Iran}}} 
\email{sanco@brocku.ca} \email{easadi@iasbs.ac.ir}

\begin{abstract}
Bi-Hamiltonian hierarchies of symplectically-invariant soliton equations
are derived from geometric non-stretching flows of curves
in the Riemannian symmetric spaces
$Sp(n+1)/Sp(1)\times Sp(n)$ and $SU(2n)/Sp(n)$. 
The derivation uses Hasimoto variables defined by 
a moving parallel frame along the curves. 
As main results,
two new multi-component versions of the sine-Gordon (SG) equation 
and the modified Korteweg-de Vries (mKdV) equation 
exhibiting $Sp(1)\times Sp(n-1)$ invariance are obtained 
along with their bi-Hamiltonian integrability structure
consisting of a hierarchy of symmetries and conservation laws
generated by a hereditary recursion operator.
The corresponding geometric curve flows
in both $Sp(n+1)/Sp(1)\times Sp(n)$ and $SU(2n)/Sp(n)$
are shown to be described by a non-stretching wave map
and a mKdV analog of a non-stretching Schr\"odinger map.
\end{abstract}
\maketitle

\section{Introduction}

Both the modified Korteveg-de Vries (mKdV) equation
and the sine-Gordon (SG) equation
are well-known to have a geometric origin given by
certain flows of the curvature invariant of arclength-parameterized curves
in the two-dimensional geometries $\Rnum^2$ and $S^2$
\cite{GoldsteinPetrich,DoliwaSantini,LangerPerline,Gurses}. 
Similarly, the nonlinear Schrodinger (NLS) equation
has long been known to arise from a certain flow of
$U(1)$-covariants of arclength-parameterized curves
in the three-dimensional geometries $\Rnum^3$ and $SO(3)$,
where the covariants are related to
the standard curvature and torsion invariants of the curve
by the famous Hasimoto transformation 
\cite{Hasimoto,Lakshmanan,AncoMyrzakulov}. 
In all of these flows,
the equation of motion of the curve has the geometrical properties
that it preserves the arclength locally at each point on the curve
(\ie/ the motion is non-stretching)
and that it is invariant under the action of the isometry group of
the underlying Riemannian geometry.
Additionally, the differential invariant in the two-dimensional case
and the differential covariants in the three-dimensional case
have a direct geometrical meaning as the components of the Cartan connection
in a parallel frame \cite{Bishop} along the curve.

A broad generalization of such results has been obtained in recent work
\cite{AncoJGP}
using a moving parallel frame formulation for non-stretching curve flows
in Riemannian symmetric spaces $M=G/H$.
These spaces describe curved generalizations of Euclidean geometries
in which the Euclidean isometry group is replaced by a simple Lie group $G$
and the Euclidean frame rotation gauge group is replaced by
an involutive compact Lie subgroup $H$ in $G$.
In this geometric setting,
the Cartan connection components in a suitably defined parallel frame
along an arclength-parameterized curve
represent differential covariants of the curve,
which are related to standard differential invariants
by a generalized Hasimoto transformation.
For curves undergoing certain non-stretching geometric flows,
these covariants satisfy multi-component SG and mKdV equations
whose integrability structure
as given by a pair of compatible Hamiltonian operators
is encoded directly in the Cartan structure equations of the parallel frame.
In cases where $M$ additionally has a hermitian structure or a Lie group structure, 
the Hamiltonian operators also give rise to 
integrable multi-component NLS equations \cite{AncoIMA}. 
Moreover, all of these integrable multi-component equations,
along with their bi-Hamiltonian structure,
possess an explicit group invariance which arises from
the action of the equivalence group of the parallel frame.
This main result provides a geometric derivation of
many known group-invariant versions of multi-component soliton equations
as well as the possibility of deriving new versions that exhibit other
invariance groups.

For example,
there are exactly two different Riemannian symmetric spaces 
with the structure $G/SO(n)$, as given by $G=SO(n+1)$ and $G=SU(n)$
(see, \eg/ \Ref{Helgason}). 
For curves in each of these two spaces,
the components of the Cartan connection in a parallel frame yield
$n-1$ covariants that satisfy vector mKdV equations and vector SG equations
with a $SO(n-1)$ invariance group
when the curve undergoes certain non-stretching geometric flows
\cite{AncoSIGMA} (see also \Ref{SandersWang}).
This derivation geometrically accounts for
the two different rotationally-invariant vector versions of
the mKdV and SG equations obtained from symmetry-integrability classifications
\cite{SokolovWolf,AncoWolf}. 

In the present paper,
we geometrically derive 
symplectically-invariant multi-component soliton equations
from non-stretching curve flows in the Riemannian symmetric spaces
$Sp(n+1)/Sp(1)\times Sp(n)$ and $SU(2n)/Sp(n)$. 
These two geometries happen to share the same symplectic equivalence group 
$Sp(1)\times Sp(n-1)$ for parallel framings of arclength-parameterized curves. 
One main motivation for our work is the absence to-date of 
any symmetry-integrability classifications 
for multi-component versions of mKdV or SG equations with symplectic invariance.
Other work \cite{Tsuchida} with a similar motivation to ours 
has recently found multi-component symplectically-invariant 
mKdV and SG equations of derivative type 
by considering certain algebraic reductions of integrable matrix systems. 
These derivative-type soliton equations have a different form of nonlinearity
(exhibiting, in particular, a different scaling symmetry) 
than the multi-component soliton equations obtained from our results. 

For the geometry $SU(2n)/Sp(n)$, 
we obtain new symplectically-invariant mKdV and SG equations for a vector pair,
together with their symplectically-invariant
bi-Hamiltonian integrability structure.
For the geometry $Sp(n+1)/Sp(1)\times Sp(n)$,
we find symplectically-invariant mKdV and SG equations 
for a scalar pair coupled to a vector pair,
which represent the component form of new quaternionic soliton equations 
with a quaternion bi-Hamiltonian integrability structure
derived in recent work \cite{AncoAsadi} (see also \Ref{AsadiSanders})
on non-stretching curve flows in the quaternionic projective space 
$\mathbb{HP}^n  \simeq U(n+1,\mathbb{H})/U(1,\mathbb{H})\times U(n,\mathbb{H})
\simeq Sp(n+1)/Sp(1)\times Sp(n)$
(where $\mathbb H$ denotes Hamilton's quaternions). 
The symplectic invariance group of these new bi-Hamiltonian soliton equations 
arising from both geometries $SU(2n)/Sp(n)$ and $Sp(n+1)/Sp(1)\times Sp(n)$ 
is given by $Sp(1)\times Sp(n-1)$. 

There are several important ways in which our results go beyond 
previous literature on integrable systems connected with 
symmetric spaces and Lie algebras. 

In \Ref{FordyKulish,AthorneFordy}, 
multi-component NLS and mKdV equations are written down 
using a Lax pair construction based on the Lie algebra structure of 
hermitian symmetric spaces. 
This construction does not apply to the non-hermitian symmetric spaces 
$SU(2n)/Sp(n)$ and $Sp(n+1)/Sp(1)\times Sp(n)$ considered in our work
or the Riemannian symmetric spaces $SO(n+1)/SO(n)$ and $SU(n)/SO(n)$ 
in earlier work \cite{SandersWang,AncoSIGMA}. 
Thus the two different rotationally-invariant vector mKdV equations
obtained in \Ref{AncoSIGMA}
as well as the two different mKdV equations with symplectic invariance 
obtained in the present paper 
fall outside the multi-component mKdV equations constructed 
in \Ref{AthorneFordy}.

Multi-component SG equations with rotational invariance and unitary invariance
are derived in \Ref{EichenherrForger,EichenherrHonerkamp,Bakas,Wang}
using algebraic methods applied to symmetric spaces 
$SO(n+1)/SO(n)$, $SU(n)/SO(n)$, $SU(n+1)/U(n)$, $Sp(n)/U(n)$
with either a rotation gauge group $SO(n)$ or a unitary gauge group $U(n)$.
None of this work includes the symmetric spaces 
$SU(2n)/Sp(n)$ and $Sp(n+1)/Sp(1)\times Sp(n)$ 
having symplectic gauge groups. 
Hence the two symplectically-invariant SG equations obtained by us 
are different than the SG equations with rotational invariance 
and unitary invariance found in previous work. 

Likewise, the moving frame method in \Ref{AncoJGP} which we apply in this paper
differs from other geometric approaches in the literature 
\cite{ChouQu1,ChouQu2,ChouQu3,ChouQu4,TerngThorbergsson,MariBeffa1,MariBeffa2,MariBeffa3}
on curve flows. 
The approach in \Ref{ChouQu1,ChouQu2,ChouQu3,ChouQu4}
has concentrated on deriving
many known scalar soliton equations from non-stretching curve flows
in planar Klein geometries $G/H\simeq\Rnum^2$. 
This work uses a natural $G$-invariant Frenet frame 
whose Cartan connection components are differential invariants of the curve. 
In particular, for each planar geometry, 
soliton equations and their associated symmetry recursion operators
are shown to arise from the structure equations of the Frenet frame.
Such results have been extended recently to several higher dimensional 
Klein geometries. 

The work in \Ref{MariBeffa1,MariBeffa2,MariBeffa3}
gives an elegant derivation of bi-Hamiltonian structures for flows of 
differential invariants of curves in various types of homogeneous spaces $G/H$
including the Riemannian symmetric space $SO(n+1)/SO(n)$. 
A group-based moving frame is used to encode the differential invariants
in a simple way, allowing Hamiltonian structures to be induced from 
two natural Poisson brackets that exist on the dual space of 
the Lie algebra of the group of loops (\ie/ closed curves) on $G$. 
Related work appears in \Ref{TerngThorbergsson}.
In the case when $G/H$ is a Riemannian symmetric space, 
the results in \Ref{AncoJGP} can be understood as providing an explicit, simpler
formulation of these Hamiltonian structures in terms of 
geometrically natural Hasimoto variables which are represented by 
differential covariants of the underlying curves on $G$. 
In contrast with invariants, 
the Hasimoto variables are geometrically determined 
by the curve only up to a rigid action of the equivalence group of 
the moving frame along the curve, 
with the Cartan connection components of the frame obeying
algebraic properties which are a direct generalization of 
a parallel moving frame in Euclidean geometry. 
Moreover, unlike other work, 
the method of \Ref{AncoJGP} yields an explicit formulation of 
the curve flow equations on $G$ corresponding to the multi-component 
bi-Hamiltonian soliton equations satisfied by both the Hasimoto variables
and the differential invariants for the underlying curve. 

We begin in \secref{construction} by summarizing
the construction and properties of parallel moving frames
in Riemannian symmetric spaces.
In \secref{prelims} we state some essential algebraic properties of
the symplectic group $Sp(n)$
and the geometries $SU(2n)/Sp(n)$, $Sp(n+1)/Sp(1)\times Sp(n)$, 
which will be needed for this construction.
The moving parallel frame formulation for non-stretching curve flows 
in these two geometries is carried out respectively at the start of 
\secref{SU.curveflows} and \secref{Sp.curveflows}. 
In \secref{SU.ops.hier} and \secref{Sp.ops.hier}, 
we derive the $Sp(1)\times Sp(n-1)$-invariant bi-Hamiltonian operators
which are subsequently used to construct the new multi-component 
$Sp(1)\times Sp(n-1)$-invariant mKdV equations and SG equations 
obtained in \secref{SU.mkdv.eqns}--\ref{SU.sg.eqns} 
and \secref{Sp.mkdv.eqns}--\ref{Sp.sg.eqns} for each geometry. 
The corresponding geometric curve flows are worked out 
in \secref{SU.curve.flows} and \secref{Sp.curve.flows}
and shown to be a non-stretching wave map equation
and a mKdV analog of a non-stretching Schr\"odinger map equation. 
We conclude with some remarks in \secref{remarks}.

\section{Parallel moving frames and non-stretching curve flows}
\label{construction}

For a Riemannian symmetric space $M=G/H$,
defined by a simple Lie group $G$
and an involutive compact Lie subgroup $H$ in $G$,
any linear frame on $M$ provides a soldering identification between
the tangent space $T_xM$ at points $x$
and the vector space $\mk{m}=\mk{g}/\mk{h}$.
Relative to the Cartan-Killing form and Lie bracket on $\mk{g}$,
there is a decomposition $\mk{g}=\mk{h}\oplus\mk{m}$
given by a direct sum of orthogonal vector spaces
\beq
\<\mk{h},\mk{m}\> =0
\eeq
with the Lie bracket relations
\beq\label{ber.rel}
[\mk{h},\mk{h}]\subset \mk{h},
\quad
[\mk{h},\mk{m}] \subset \mk{m},
\quad [\mk{m},\mk{m}]\subset \mk{h}.
\eeq
Geometrically,
the Lie subalgebra $\mk{h}$ is identified with the generators of
isometries that leave fixed the origin $o$ in $M$,
while the vector space $\mk{m}$ is identified with the generators of
isometries that carry the origin $o$ to any point $x\neq o$ in $M$.
These isometries represent the action of the group $G$ on the space $M$,
whereby the subgroup $H$ acts as the gauge group of the frame bundle of $M$.

The Riemannian structure of the space $M=G/H$ is naturally described
\cite{KobayashiNomizu} in terms of
a $\mk{m}$-valued linear coframe $e$
and a $\mk{h}$-valued linear connection $\conx$
whose torsion and curvature
\beq\label{tor.curv}
\mk{T}:=de+\bs{[}\conx,e\bs{]},\quad
\mk{R}:=d\conx+\ha\bs{[}\conx,\conx\bs{]}
\eeq
are $2$-forms with respective values in $\mk{m}$ and $\mk{h}$,
given by the following Cartan structure equations:
\beq\label{cart.stru}
\mk{T}=0,\quad \mk{R}=-\ha\bs{[}e,e\bs{]}.
\eeq
Here $\bs{[}\cdot,\cdot\bs{]}$ denotes the Lie bracket on $\mk{g}$
composed with the wedge product on $T_x^*M$.
This structure together with the (negative-definite) Cartan-Killing form
determines a Riemannian metric and Riemannian connection
(\ie/ covariant derivative) on the space $M$ as follows:
for all $X,Y$ in $T_x M$,
\beq\label{metr.conn}
g(X,Y):=-\<e_X,e_Y\>,\quad e\rfloor\nabla_X Y:=d_X e_Y+[\conx_X,e_Y],
\eeq
where the coframe provides a soldering identification between the tangent space $T_x M$
and the vector space $\mk{m}=\mk{g}/\mk{h}$
as given by $e\rfloor X:=e_X,e\rfloor Y:=e_Y\in\mk{m}$.
The connection is metric compatible, $\nabla g=0$, and torsion-free, $T=0$,
while its curvature is covariantly constant, $\nabla  R=0$,
as given by
\beq\label{tor.curv.tensor}
e\rfloor R(X,Y)Z=[\mk{R}\rfloor(X\wedge Y),e_Z]=-[[e_X,e_Y],e_Z],
\quad
e\rfloor T(X,Y)=\mk{T}(X\wedge Y)=0,
\eeq
where $T(X,Y):=\nabla_XY-\nabla_YX-[X,Y]$ is the torsion tensor
and $R(X,Y):=[\nabla_X,\nabla_Y]-\nabla_{[X,Y]}$ is the curvature tensor.
Note the linear coframe and linear connection have gauge freedom
given by the following transformations
\beq\label{gauge.tr}
e\longrightarrow \Ad(h^{-1})e,
\quad
\conx\longrightarrow\Ad(h^{-1})\conx+h^{-1}dh
\eeq
for an arbitrary function $h:M\to H \subset G$.
These gauge transformations comprise a local ($x$-dependent) representation of
the linear transformation group $H^*=\Ad(H)$
which defines the gauge group \cite{Sharpe} of the frame bundle of $M$.
Both the metric tensor $g$ and curvature tensor $R$ on $M$ are gauge invariant.

Let $\map(x)$ be any smooth curve in $M$.
A frame consists of a set of orthonormal vectors that span the tangent space
$T_\map M$ at each point $x$ on the curve $\map$.
The Frenet equations of a frame yield a connection matrix
consisting of the set of frame components of the covariant $x$-derivative of
each frame vector along the curve \cite{Guggenheimer}.
A coframe consists of a set of orthonormal covectors that are dual to
the frame vectors relative to the Riemannian metric $g$.
Such a framing for $\map(x)$ is determined by the Lie-algebra components
of $e$ and $\conx\rfloor \map_x$
when an orthonormal basis is introduced for $\mk{m}$ and $\mk{h}$
with respect to the Cartan-Killing form,
where the Frenet equations are defined by the frame components of 
the transport equation
\beq\label{frenet.eq}
\nabla_x e=-\ad(\conx\rfloor \map_x)e
\eeq
along the curve.
In particular,
if $\{\mb{m}_l\}_{l=1,\ldots,\dim(\mk{m})}$ is any fixed orthonormal basis for $\mk{m}$,
then a frame at each point $x$ along the curve is given by
the set of vectors $X_l:=-\<e^*,\mb{m}_l\>$, $l=1,\ldots,\dim(\mk{m})$.
Here $e^*$ is a $\mk{m}$-valued linear frame 
defined as the dual to the linear coframe $e$ 
by the condition that $-\<e^*,e\>=\textrm{id}$
is the identity map on each tangent space $T_xM$
(cf \cite{AncoJGP,KobayashiNomizu}).

Now consider any smooth flow $\map(t,x)$ of a curve in $M$.
We write $X=\map_x$ for the tangent vector
and $Y=\map_t$ for the evolution vector
at each point $x$ along the curve.
Note the flow is {\it non-stretching} provided that
it preserves the $G$-invariant arclength $ds=|\map_x|dx$,
or equivalently $\nabla_t|\map_x|=0$,
in which case we have
\beq\label{non.stret}
g(\map_x,\map_x)=|\map_x|^2=1
\eeq
without loss of generality.
For flows that are transverse to the curve
(such that $X$ and $Y$ are linearly independent),
$\map(t,x)$ will describe a smooth two-dimensional surface in $M$.
The pullback of the torsion and curvature equations \eqref{cart.stru}
to this surface yields
\begin{align}
&
D_xe_t-D_te_x+[\conx_x,e_t]-[\conx_t,e_x]=0,
\label{pull.1}\\
&
D_x\conx_t-D_t\conx_x+[\conx_x,\conx_t]=-[e_x,e_t],
\label{pull.2}
\end{align}
with
\begin{align}
&
e_x:=e\rfloor X=e\rfloor \map_x,
&&
e_t:=e\rfloor Y=e\rfloor \map_t,
\label{pull.not.1}\\
&
\conx_x:=\conx\rfloor X=\conx\rfloor \map_x,
&&
\conx_t:=\conx\rfloor Y=\conx\rfloor \map_t,
\label{pull.not.2}
\end{align}
where $D_x,D_t$ denote derivative operators with respect to $x,t$.
Remarkably,
for any non-stretching curve flow,
these structure equations \eqref{pull.1}--\eqref{pull.not.2}
encode an explicit pair of bi-Hamiltonian operators
once a specific choice of frame along $\map(t,x)$ is made.

We utilize a natural choice of a moving frame defined by
the following two properties which are a direct algebraic generalization
of a parallel moving frame in Euclidean geometry \cite{AncoJGP}:
\begin{enumerate}
\item[(1)]
$e_x$ is a constant unit-norm element lying in a Cartan subspace
$\mk{a}\subset \mk{m}$ that is contained in the centralizer subspace
$\mk{m}_{\pa}$ of $e_x$, \ie/
$D_xe_x=D_te_x=0, \<e_x,e_x\>=-1$, and $\ad(\mk{m}_{\pa})e_x=0$
where $\mk{m}_{\pa}\oplus\mk{m}_{\perp}=\mk{m}$
and $\<\mk{m}_{\pa},\mk{m}_{\perp}\>=0$.
\item[(2)]
$\conx_x$ lies in the perp space $\mk{h}_{\perp}$ of the Lie subalgebra
$\mk{h}_{\pa}\subset \mk{h}$ of the linear isotropy group
$H^*_{\pa}\subset H^* = \Ad(H)$ that preserves $e_x$, \ie/
$\ad(\mk{h}_{\pa})e_x=0$ and $\<\conx_x,\mk{h}_{\pa}\>=0$
where
$\mk{h}_{\pa}\oplus\mk{h}_{\perp}=\mk{h}$
and $\<\mk{h}_{\pa},\mk{h}_{\perp}\>=0$.
\end{enumerate}

Cartan subspaces of $\mk{m}$ are defined as a maximal abelian subspace
$\mk{a}\subseteq\mk{m}$,
having the property that it is the centralizer of its elements,
$\mk{a}=\mk{m}\cap\mk{c}(\mk{a})$.
It is well-known (see, \eg/ \Ref{Helgason})
that any two Cartan subspaces are isomorphic to one another
under some linear transformation in $\Ad(H)$
and that the action of the linear transformation group $\Ad(H)$
on any Cartan subspace $\mk{a}$ generates $\mk{m}$.
The dimension of $\mk{a}$ as a vector space is equal to the rank of $\mk{m}$.

A moving frame satisfying properties $(1)$ and $(2)$ is called {\em $H$-parallel }
and its existence can be established by constructing
a suitable gauge transformation \eqref{gauge.tr} on an arbitrary frame
at each point $x$ along the curve \cite{AncoJGP}.
Specifically,
given any $\mk{m}$-valued linear coframe $\tilde{e}$
and $\mk{h}$-valued linear connection matrix $\tilde{\conx}_x$ along $\map$,
we can first find a gauge transformation such that
$h^{-1}\tilde{e}_xh=e_x$ is a constant element in any Cartan subspace
$\mk{a}\subset\mk{m}$,
as a consequence of the fact $\mk{m}={\rm Ad}(H)\mk{a}$.
The norm of $e_x$ will satisfy
$-\<e_x,e_x\>=g(\map_x,\map_x)=1$
because we have chosen an arclength parameterization \eqref{non.stret} of the curve.
We can then find a gauge transformation belonging to
the subgroup $H^*_{\pa}$ preserving $e_x$,
so that $h^{-1}D_xh+h^{-1}\tilde{\conx}_xh=\conx_x$
where $h(x)\in H^*_{\pa}$ is given by solving the linear matrix ODE
$D_xh+\tilde{\varpi}^{\pa}h=0$
in terms of the decomposition of
$\tilde{\conx}_x=\tilde{\varpi}^{\pa}+\tilde{\varpi}^{\perp}$ relative to $e_x$.
Note the solution will depend on an arbitrary initial condition
$h(x_0)\in H^*_{\pa}$, specified at some point $x=x_0$ along the curve,
which represents a rigid gauge freedom (\ie/ the equivalence group)
in the construction of the $H$-parallel moving frame.

Underpinning this construction are the Lie bracket relations on
$\mk{m}_{\pa}$, $\mk{m}_{\pe}$, $\mk{h}_{\pa}$, $\mk{h}_{\pe}$
coming from the structure of $\mk{g}$
as a symmetric Lie algebra \eqref{ber.rel}.
These relations consist of
\begin{align}
&
[\m_\pa,\m_\pa] \subseteq \h_\pa,
\quad
[\m_\pa,\h_\pa] \subseteq \m_\pa,
\quad
[\h_\pa,\h_\pa] \subseteq \h_\pa,
\label{inclusion.one}\\
&
[\h_\pa,\m_\pe] \subseteq \m_\pe,
\quad
[\h_\pa,\h_\pe] \subseteq \h_\pe,
\label{inclusion.two}\\
&
[\m_\pa,\m_\pe] \subseteq \h_\pe,
\quad
[\m_\pa,\h_\pe] \subseteq \m_\pe,
\label{inclusion.three}
\end{align}
while the remaining Lie brackets
\beq
[\m_\pe,\m_\pe],
\quad
[\h_\pe,\h_\pe],
\quad
[\m_\pe,\h_\pe]
\label{inclusion.gen}
\eeq
obey the general relations \eqref{ber.rel}.

\begin{theorem}\label{biHam.flow.eqn}
For $e_x\in\mk{a}\subset\mk{m}_{\pa}$,
let $e_t=h_{\pa}+h_{\pe}\in\mk{m}_{\pa}\oplus\mk{m}_{\pe}$,
$\conx_t=\varpi^{\pa}+\varpi^{\pe}\in\mk{h}_{\pa}\oplus\mk{h}_{\pe}$,
and $u=\conx_x \in\mk{h}_{\pe}$.
Also let $h^{\pe}=\ad(e_x)h_{\pe}\in\mk{h}_{\pe}$.
Then the Cartan structure equations \eqref{pull.1}--\eqref{pull.2}
for any $H$-parallel linear coframe $\e$ and linear connection $\conx$
pulled back to the two-dimensional surface $\map(t,x)$ in $M=G/H$
yield the flow equation \cite{AncoJGP}
\beq\label{ufloweq}
u_t = \Hop(\varpi^\pe) +h^\pe ,
\quad
\varpi^\pe = \Jop(h^\pe) ,
\eeq
where
\beq
\Hop = \Kop|_{\h_\pe} ,
\quad
\Jop= -\ad(e_{x})\inv \Kop|_{\m_\pe} \ad(e_{x})\inv
\label{HJops}
\eeq
are a bi-Hamiltonian pair of operators 
that act on $\h_\pe$-valued functions 
and are invariant under $H_\pa^*$,
as defined in terms of the linear operator
\beq\label{Kop}
\Kop := D_x +[u,\cdot\ ]_\pe -[u,D_x^{-1}[u,\cdot\ ]_\pa] .
\eeq
In particular, every linear combination of $\Hop$ and $\Jop^{-1}$ 
is a Hamiltonian operator with respect to $u$. 
\end{theorem}

We emphasize that the formulation in Theorem~\ref{biHam.flow.eqn}
applies to {\it all} non-stretching curve flows $\map(t,x)$ in $M=G/H$,
with the flow being determined by specifying $h^{\pe}$, 
or equivalently $h_{\pe}=\ad(e_x)\inv h^{\pe}$, 
freely as a function of $t$ at each point $x$ along the curve.
In particular, 
every flow equation \eqref{ufloweq} determines 
a corresponding curve flow $\map(t,x)$ through the geometrical relation
\beq\label{Yeq}
Y=-\<e^*,h_{\perp}+h_{\pa}\>=\<e^*,\mathcal{Y}(h^{\perp})\>
\eeq
in terms of the operator 
\beq
\mathcal{Y}:=D_x^{-1}[u,\ad(e_x)^{-1}\cdot\ ]_{\pa}-\ad(e_x)^{-1}\cdot
\eeq
where $e^*$ is the linear frame dual to the linear coframe $e$ along $\map$, 
with $e_x=e\rfloor X$. 
In this correspondence \eqref{Yeq}, 
$e_x$ is preserved under the action of the equivalence group $H^*_\pa$, 
while up to equivalence, 
both $e^*$ and $e$ are determined by $\conx_x$ 
from the transport equation \eqref{frenet.eq} along $\map$. 
The resulting equation of motion $\map_t=Y=\<e^*,\mathcal{Y}(h^{\perp})\>$ 
will be $G$-invariant if and only if 
$h^{\pe}$ is a $H^*_\pa$-equivariant function of 
$x$, $u$, and $x$-derivatives of $u$.
In addition, the corresponding flow on $u(t,x)$ will have a Hamiltonian structure
if and only if $\varpi^\pe = \Jop(h^\pe)$ is the variational derivative of
some $H^*_\pa$-invariant Hamiltonian function of
$x$, $u$, and $x$-derivatives of $u$.
The following general results are established in \Ref{AncoJGP}.

\begin{theorem}\label{biHam.hier}
Composition of the operators $\Hop$ and $\Jop$ yields a recursion operator
$\Rop=\Hop\Jop$ that produces a hierarchy of $H^*_\pa$-invariant flows \eqref{ufloweq}
on $u$ given in terms of
\beq\label{hperp.hier}
h^{\pe}_{(l)} = \Rop^l(u_x),
\quad
l=0,1,2,\ldots .
\eeq
Each flow in this hierarchy inherits a bi-Hamiltonian structure given by
\beq
h^{\pe}_{(l)} = \Hop(\varpi^\pe_{(l)}) = \Jop\inv(\varpi^\pe_{(l+1)}),
\quad
\varpi^\pe_{(l)} = \delta H^{(l)}/\delta u =  \Rop^{*l}(u),
\quad
l=0,1,2,\ldots
\eeq
in terms of the $H^*_\pa$-invariant Hamiltonians
\beq
H^{(l)}= \frac{-1}{1+2l}\<e_x,h_\pa^{(l)}\> ,\quad
l=0,1,2,\ldots
\eeq
where $\Rop^*=\Jop\Hop$ is the adjoint of $\Rop$.
Moreover, the kernel of the recursion operator $\Rop$ yields a further
$H^*_\pa$-invariant flow \eqref{ufloweq} on $u$ in terms of
$h^{\pe}_{(-1)}$ defined by
\beq\label{hperp.hier.-1}
\Jop(h^{\pe}_{(-1)})=0 .
\eeq
This flow has a Hamiltonian structure given by
\beq
h^{\pe}_{(-1)} = \Hop(\varpi^\pe_{(-1)}),
\quad
\varpi^\pe_{(-1)}= \delta H^{(-1)}/\delta u
\eeq
with
\beq
H^{(-1)}= \<e_x,h_\pa^{(-1)}\> .
\eeq
\end{theorem}

The bi-Hamiltonian flows \eqref{hperp.hier} and \eqref{hperp.hier.-1} 
have a geometrical formulation through the correspondence \eqref{Yeq}. 

\begin{theorem}\label{geom.curve.flows}
The hierarchy of bi-Hamiltonian flows \eqref{hperp.hier} 
corresponds to a hierarchy of non-stretching geometric curve flows in $M=G/H$ 
given by equations of motion
\beq\label{curve.flow.hier}
\map_t=Y_{(l)}(\map_x,\nabla_x\map_x,\ldots,\nabla_x^{2l}\map_x), \quad 
|\map_x|=1, 
\quad l=0,1,2,\ldots ,
\eeq
where $Y_{(l)}= \<e^*,\mathcal{Y}(h^{\perp}_{(l)})\>$. 
The additional Hamiltonian flow \eqref{hperp.hier.-1} 
corresponds to the non-stretching geometric curve flow
\beq 
\nabla_x\map_t=\nabla_x Y_{(-1)}=0, \quad 
|\map_x|=1, 
\eeq
with $Y_{(-1)}= \<e^*,\mathcal{Y}(h^{\perp}_{(-1)})\>$. 
Each equation of motion ($l=-1,0,1,2,\ldots$) is invariant with respect to 
the isometry group $G$ of $M$ and preserves the $G$-invariant arclength $x$ of
the curve $\map(t,x)$. 
\end{theorem}

Curves in $M=G/H$ can be divided into algebraic equivalence classes 
defined by the orbit of the element $e\rfloor \map_x =e_x$ 
in $\mk{a}\subset \mk{m}$ under the action of the group $H^*=Ad(H)$. 
There is a single orbit iff the rank of $\mk{m}$ is equal to $1$. 
Consequently, if $\mk{m}$ has rank greater than $1$,
then each distinct orbit in $\mk{a}\subset \mk{m}$ 
will correspond to a distinct hierarchy of bi-Hamiltonian flows \eqref{curve.flow.hier} in $M=G/H$.

\section{Algebraic preliminaries}
\label{prelims}

Recall,
the complex symplectic group $Sp(n,\Cnum)$ is the group of matrices $g$
in $GL(2n,\Cnum)$ that leaves invariant the exterior form
$z_1\wedge z_{n+1}+\cdots+z_n\wedge z_{2n}$ in terms of coordinates
$(z_1,\ldots,z_{2n})\in\Cnum^{2n}$, \ie/
\beq
g^\t Jg=J,
\eeq
where
\beq
J=\bpm 0 & I_n\\-I_n & 0\epm
\eeq
with $I_n$ denoting the identity matrix in $GL(n,\Cnum)$.
Also recall,
the complex unitary group $U(2n)$ is the group of matrices $g$
in $GL(2n,\Cnum)$ that leaves invariant the Hermitian form
$z_1\overline{z}_1+\cdots+z_{2n}\overline{z}_{2n}$, \ie/
\beq
g^\t\overline{g}=I_{2n}.
\eeq
The compact {\it symplectic group} is defined by
$Sp(n)=Sp(n,\Cnum)\cap U(2n)$.

For later convenience, we let $\mk{s}(n,\Cnum)$ denote the vector space of
symmetric matrices $\gg$ in $\mk{gl}(n,\Cnum)$), \ie/ $\gg^\t =\gg$.

A general reference for the following material is \Ref{Helgason,AncoJGP}.

\subsection{The vector space $\mk{su}(2n)/\mk{sp}(n)$}
\label{g=su}
The special unitary Lie algebra $\mk{su}(2n)$ is defined by
the matrices $\gg$ in $\mk{gl}(2n,\Cnum)$ that are skew-Hermitian and trace-free,
\ie/ $\gg^\t =-\ov{\gg}$, $\tr(\gg)=0$.
There is an involutive automorphism of $\mk{gl}(2n,\Cnum)$ given by
\beq
\sigma(\gg)=J\ov{\gg}J^{-1}
\eeq
preserving $\mk{su}(2n)\subset \mk{gl}(2n,\Cnum)$.
The matrices $\hh$ in $\mk{gl}(2n,\Cnum)$ that are skew-Hermitian, trace-free,
and invariant under $\sigma$,
\ie/ $\hh^\t =-\ov{\hh}$, $\tr(\hh)=0$, $\sigma(\hh)=\hh$,
span the compact symplectic Lie algebra $\mk{sp}(n)$.
This leads to the orthogonal decomposition of $\mk{g}=\mk{su}(2n)$
as a symmetric Lie algebra given by the eigenspaces of $\sigma$,
\beq\label{su.rep.eigen.h}
\mk{h}:= \mk{sp}(n)\subset \mk{g},
\quad
\sigma(\mk{h})=\mk{h}
\eeq
and
\beq\label{su.rep.eigen.m}
\mk{m}:=\mk{su}(2n)/\mk{sp}(n)\subset \mk{g},
\quad
\sigma(\mk{m})=-\mk{m}.
\eeq

\begin{lemma}\hfil \newline
1. The matrix representations of
the vector space $\mk{m}=\mk{su}(2n)/\mk{sp}(n)$
and the Lie subalgebra $\mk{h}= \mk{sp}(n)$ in $\mk{gl}(2n,\Cnum)$ are
respectively given by
\begin{align}
(A,B)&:=
\bpm A&B\\\ov{B}&-\ov{A}\epm
\in\mk{m},\quad
B^\t=-B,\quad A^\t=-\ov{A},\quad  \tr(A)=0,
\label{su.m}\\
(C,D)&:= \bpm C&D\\-\ov{D}&\ov{C}\epm \in\mk{h},\quad
C^\t=-\ov{C},\quad D^\t=D, 
\label{su.h}
\end{align}
where $A,B,C,D\in\mk{gl}(n,\Cnum)$.
The Lie bracket relations \eqref{ber.rel} have the matrix representation
\begin{subequations}
\begin{align}
[(A_1,B_1),(A_2,B_2)] =&
([A_1,A_2]+B_1\ov{B}_2-B_2\ov{B}_1, A_1B_2+B_2\ov{A}_1-B_1\ov{A}_2-A_2B_1)
\in\mk{h}, 
\label{su.bracket.m.m}\\
[(A_1,B_1),(C_1,D_1)] =&
([A_1,C_1]-B_1\ov{D}_1-D_1\ov{B}_1, A_1D_1+D_1\ov{A}_1+B_1\ov{C}_1-C_1B_1)
\in\mk{m},
\label{su.bracket.h.m}\\
[(C_1,D_1),(C_2,D_2)] =&
([C_1,C_2]-D_1\ov{D}_2+D_2\ov{D}_1, C_1D_2-D_2\ov{C}_1+D_1\ov{C}_2-C_2D_1)
\in\mk{h}.
\label{su.bracket.h.h}
\end{align}
\end{subequations}
2. The restriction of the Cartan-Killing form on $\mk{g}=\mk{su}(2n)$
to $\mk{m}=\mk{su}(2n)/\mk{sp}(n)$
yields a negative-definite inner product
\beq\label{su.kill.on.m}
\<(A_1,B_1),(A_2,B_2)\>=4n\big(2\tr(A_1A_2)+\tr(\ov{B}_1B_2+B_1\ov{B}_2)\big) .
\eeq
3. The (real) dimension of $\mk{m}=\mk{su}(2n)/\mk{sp}(n)$ is $(n-1)(2n+1)$
and its rank is $n-1$.
\end{lemma}

The subspace $\mk{a}\subset\mk{m}$ spanned by the $n-1$ matrices
\beq
\bpm
E_k&0\\0&E_k\epm,
\quad
E_k=\diag (\ub{0,\ldots,0}_{k-1},\i,-\i,\ub{0,\ldots,0}_{n-k+1}),
\quad k=1,\ldots n-1
\eeq
is a Cartan subspace.
A special choice of an element of $\mk{a}$ is given by
\beq
\ee:=\bpm E&0\\0&E\epm\in\mk{m}=\mk{su}(2n)/\mk{sp}(n),\quad
E=\frac{1}{\sqrt{\chi}} \diag ((n-1)\i,\ub{-\i,\ldots,-\i}_{n-1}) =-\ov{E},
\eeq
which has the distinguishing property that the centralizer subspace
$\mk{c}(\ee)$ of $\ee$ in $\mk{g}=\mk{su}(2n)$ is of maximal dimension.
The corresponding linear operator $\ad(\ee)$
induces a direct sum decomposition of the vector spaces
$\mk{m}=\mk{su}(2n)/\mk{sp}(n)$ and $\mk{h}= \mk{sp}(n)$
into centralizer spaces $\mk{m}_{\pa}$ and $\mk{h}_{\pa}$
and their orthogonal complements (perp spaces)
$\mk{m}_{\pe}$ and $\mk{h}_{\pe}$
with respect to the Cartan-Killing form.
Through the Lie bracket relation \eqref{inclusion.three},
this operator $\ad(\ee)$ maps $\mk{h}_{\pe}$ into $\mk{m}_{\pe}$,
and vice versa,
whence $\ad(\ee)^2$ is well-defined as a linear mapping of
each subspace $\mk{h}_{\pe}$ and $\mk{m}_{\pe}$ into itself.
The eigenvalues of this linear map can be normalized
relative to the Cartan-Killing form by choosing the factor $\chi$ so that
$\ee$ has unit norm,
\beq
-1= \<\ee,\ee\> = 8n \tr(E^2)= -8(n-1)n^2/\chi,
\eeq
which determines
\beq
\chi = 8(n-1)n^2 .
\eeq

\begin{lemma}\hfil \newline
1. The matrix representations of $\mk{m}_{\pa}$ and $\mk{m}_{\pe}$
in $\mk{m}=\mk{su}(2n)/\mk{sp}(n)$ are given by
\beq\label{su.m.pa.pe}
(\mb{A}_{\pa},\mb{B}_{\pa}):=
\bpm A_{\pa}& B_{\pa}\\\ov{B}_{\pa}&-\ov{A}_{\pa}\epm
\in\mk{m}_{\pa},\quad
(\mb{a}_{\pe},\mb{b}_{\pe}):=
\bpm A_{\pe}&B_{\pe}\\\ov{B}_{\pe}&-\ov{A}_{\pe}\epm
\in\mk{m}_{\pe},
\eeq
in which
\begin{align*}
& A_{\pa}=\bpm -\tr\mb{A}_{\pa} &0\\0&\mb{A}_{\pa}\epm,
\quad
B_{\pa}=\bpm 0&0\\0&\mb{B}_{\pa}\epm,
\quad
\mb{A}_{\pa}^\t=-\ov{\mb{A}}_{\pa},
\quad
\mb{B}_{\pa}^\t=-\mb{B}_{\pa},
\\
& A_{\pe}=\bpm 0&\mb{a}_{\pe}\\-\ov{\mb{a}}_{\pe}^\t&0\epm,
\quad
B_{\pe}=\bpm 0&\mb{b}_{\pe}\\-\mb{b}_{\pe}^\t&0\epm,
\end{align*}
where $\mb{A}_{\pa}\in\mk{u}(n-1)$, $\mb{B}_{\pa}\in\mk{so}(n-1,\Cnum)$,
$\mb{a}_{\pe},\mb{b}_{\pe}\in\Cnum^{n-1}$.
\hfil \newline
2. The matrix representations of $\mk{h}_{\pa}$ and $\mk{h}_{\pe}$
in $\mk{h}=\mk{sp}(n)$ are given by
\beq\label{su.h.pa.pe}
((\mr{c}_{\pa},\mr{d}_{\pa}),(\mb{C}_{\pa},\mb{D}_{\pa})):=
\bpm C_{\pa}&D_{\pa}\\-\ov{D}_{\pa}&\ov{C}_{\pa}\epm
\in\mk{h}_{\pa},
\quad
(\mb{c}_{\pe},\mb{d}_{\pe}):=
\bpm C_{\pe}& D_{\pe}\\-\ov{D}_{\pe}&\ov{C}_{\pe}\epm
\in\mk{h}_{\pe},
\eeq
in which
\begin{align*}
& C_{\pa}=\bpm \mr{c}_{\pa}&0\\0&\mb{C}_{\pa}\epm,
\quad
D_{\pa}=\bpm \mr{d}_{\pa}&0\\0&\mb{D}_{\pa}\epm,
\quad
\mb{C}_{\pa}^\t=-\ov{\mb{C}}_{\pa},
\quad
\mb{D}_{\pa}^\t=\mb{D}_{\pa},\\
& C_{\pe}=\bpm 0&\mb{c}_{\pe}\\-\ov{\mb{c}}_{\pe}^\t&0\epm,
\quad
D_{\pe}=\bpm 0&\mb{d}_{\pe}\\\mb{d}_{\pe}^\t&0\epm,
\end{align*}
where $\mb{C}_{\pa}\in\mk{u}(n-1)$, $\mb{D}_{\pa}\in\mk{s}(n-1,\Cnum)$,
$\mb{c}_{\pe},\mb{d}_{\pe}\in \Cnum^{n-1}$,
$\mr{c}_{\pa}\in\i\Rnum$, $\mr{d}_{\pa}\in\Cnum$.
\hfil \newline
3. $\dim\mk{m}_{\pa}=(n-1)(2n-3)$, $\dim\mk{m}_{\pe}=\dim\mk{h}_{\pe}=4(n-1)$,
$\dim\mk{h}_{\pa}=(n-1)(2n-1)+3$.
\hfil \newline
4. The linear operator $\ad(\ee)$ acts on $\mk{m}_{\pe}$ and $\mk{h}_{\pe}$ by
\beq\label{su.ad.e.h.pe}
\ad(\ee)(\mb{a}_{\pe},\mb{b}_{\pe})
=\frac{1}{\sqrt{\rho}} (\i\mb{a}_{\pe},\i\mb{b}_{\pe})
\in\mk{h}_{\pe},
\quad
\ad(\ee)(\mb{c}_{\pe},\mb{d}_{\pe})
=\frac{1}{\sqrt{\rho}} (\i\mb{c}_{\pe},\i\mb{d}_{\pe})
\in\mk{m}_{\pe}
\eeq
where $\rho= \chi/n^2=8(n-1)$.
\end{lemma}

To write out the explicit Lie bracket relations on
$\mk{m}=\mk{m}_{\pa}\oplus\mk{m}_{\pe}$
and $\mk{h}=\mk{h}_{\pa}\oplus\mk{h}_{\pe}$,
we introduce the following inner products and outer products.
For $\mb{x},\mb{y}\in\Cnum^{n-1}$,
let
\begin{align}
&\C(\mb{x},\mb{y}):=\mb{x}\ov{\mb{y}}^\t-\mb{y}\ov{\mb{x}}^\t
=\i 2{\rm Im}<\mb{x},\mb{y}> \in\i\Rnum,
\label{com.vec}\\
&\A(\mb{x},\mb{y}):=\mb{x}\ov{\mb{y}}^\t+\mb{y}\ov{\mb{x}}^\t
=2{\rm Re}<\mb{x},\mb{y}>\in\Rnum,
\label{anti.com.vec}\\
&\S(\mb{x},\mb{y}):=\mb{x}\mb{y}^\t+\mb{y}\mb{x}^\t
=<\mb{x},\ov{\mb{y}}>+<\mb{y},\ov{\mb{x}}>
\in\Cnum,
\label{anti.conj.vec}
\end{align}
where 
\beq\label{Herm.inner.prod}
<\mb{x},\mb{y}>=\mb{x}\ov{\mb{y}}^\t=\ha\A(\mb{x},\mb{y}) + \ha\C(\mb{x},\mb{y})
\eeq
is the Hermitian inner product,
and where
\beq\label{Eucl.inner.prod}
<\mb{x},\ov{\mb{y}}>= \mb{x}\mb{y}^\t=\mb{y}\mb{x}^\t =\ha\S(\mb{x},\mb{y})
\eeq
is the standard Euclidean inner product. 
Also let
\begin{align}
&\CC(\mb{x},\mb{y})=\ov{\mb{x}}^\t\mb{y}-\ov{\mb{y}}^\t\mb{x}
\in\mk{u}(n-1),
\label{com.not.mat}\\
&\AA(\mb{x},\mb{y})=\mb{x}^\t\ov{\mb{y}}-\ov{\mb{y}}^\t\mb{x}
\in\mk{so}(n-1,\Cnum),
\label{anti.not.mat}\\
&\SS(\mb{x},\mb{y})=\mb{x}^\t\ov{\mb{y}}+\ov{\mb{y}}^\t\mb{x}
\in\mk{s}(n-1,\Cnum) .
\label{anti.conj.mat}
\end{align}
The inner products \eqref{com.vec}--\eqref{anti.conj.vec} have the following
symmetry properties
\beq
\C(\mb{y},\mb{x}) = - \C(\mb{x},\mb{y}) ,
\quad
\A(\mb{y},\mb{x}) = \A(\mb{x},\mb{y}) ,
\quad
\S(\mb{y},\mb{x}) = \S(\mb{x},\mb{y}) ,
\eeq
while the outer products \eqref{com.not.mat}--\eqref{anti.conj.mat}
obey the following transpose, symmetry, and trace properties
\begin{subequations}
\beq
\CC(\mb{x},\mb{y})^\t
=-\ov{\CC(\mb{x},\mb{y})},\quad \CC(\mb{y},\mb{x})=-\CC(\mb{x},\mb{y}),
\eeq\beq
\AA(\mb{x},\mb{y})^\t
=-\AA(\mb{x},\mb{y}),\quad \AA(\mb{y},\mb{x})
=-\ov{\AA(\mb{x},\mb{y})},
\eeq\beq
\SS(\mb{x},\mb{y})^\t
=\SS(\mb{x},\mb{y}),\quad
\SS(\mb{y},\mb{x})
=\ov{\SS(\mb{x},\mb{y})},
\eeq\beq
\tr(\CC(\mb{x},\mb{y}))=\C(\mb{y},\mb{x}),
\quad
\tr(\AA(\mb{x},\mb{y}))=0,
\quad
\tr(\SS(\mb{x},\mb{y}))=2<\mb{x},\mb{y}>.
\eeq
\end{subequations}

\begin{proposition}\hfil \newline
1. The Lie brackets \eqref{inclusion.one}--\eqref{inclusion.three}
are given by
\begin{subequations}
\begin{align}
& [(\mb{A}_{1\pa},\mb{B}_{1\pa}),(\mb{A}_{2\pa},\mb{B}_{2\pa})]
\nonumber\\&
=\big((0,0),([\mb{A}_{1\pa},\mb{A}_{2\pa}]-\mb{B}_{2\pa}\ov{\mb{B}}_{1\pa}+\mb{B}_{1\pa}\ov{\mb{B}}_{2\pa},
\mb{A}_{1\pa}\mb{B}_{2\pa}+\mb{B}_{2\pa}\ov{\mb{A}}_{1\pa}-\mb{B}_{1\pa}\ov{\mb{A}}_{2\pa}-\mb{A}_{2\pa}\mb{B}_{1\pa})\big)
\in\mk{h}_{\pa},
\label{su.inclu.1}\\
& [((\mr{c}_{1\pa},\mr{d}_{1\pa}),(\mb{C}_{1\pa},\mb{D}_{1\pa})),(\mb{A}_{2\pa},\mb{B}_{2\pa})]
\nonumber\\&
=([\mb{C}_{1\pa},\mb{A}_{2\pa}]+\mb{D}_{1\pa}\ov{\mb{B}}_{2\pa}+\mb{B}_{2\pa}\ov{\mb{D}}_{1\pa},
\mb{C}_{1\pa}\mb{B}_{2\pa}-\mb{B}_{2\pa}\ov{\mb{C}}_{1\pa}-\mb{D}_{1\pa}\ov{\mb{A}}_{2\pa}-\mb{A}_{2\pa}\ov{\mb{D}}_{1\pa})
\in\mk{m}_{\pa},
\label{su.inclu.2}\\
& [((\mr{c}_{1\pa},\mr{d}_{1\pa}),(\mb{C}_{1\pa},\mb{D}_{1\pa})),
((\mr{c}_{2\pa},\mr{d}_{2\pa}),(\mb{C}_{2\pa},\mb{D}_{2\pa}))]
\nonumber\\&=
\big((\mr{d}_{2\pa}\ov{\mr{d}}_{1\pa}-\mr{d}_{1\pa}\ov{\mr{d}}_{2\pa},
\mr{c}_{1\pa}\mr{d}_{2\pa}+\mr{d}_{2\pa}\mr{c}_{1\pa}-\mr{d}_{1\pa}\mr{c}_{2\pa}-\mr{c}_{2\pa}\mr{d}_{1\pa}),
\nonumber\\&\qquad
([\mb{C}_{1\pa},\mb{C}_{2\pa}]+\mb{D}_{2\pa}\ov{\mb{D}}_{1\pa}-\mb{D}_{1\pa}\ov{\mb{D}}_{2\pa},
\mb{C}_{1\pa}\mb{D}_{2\pa}-\mb{D}_{2\pa}\ov{\mb{C}}_{1\pa}+\mb{D}_{1\pa}\ov{\mb{C}}_{2\pa}-\mb{C}_{2\pa}\mb{D}_{1\pa})\big)
\in\mk{h}_{\pa},
\label{su.inclu.3}
\end{align}
\end{subequations}
\begin{subequations}
\begin{align}
& [((\mr{c}_{1\pa},\mr{d}_{1\pa}),(\mb{C}_{1\pa},\mb{D}_{1\pa})),
(\mb{a}_{2\pe},\mb{b}_{2\pe})]
\nonumber\\&
=(-\mb{a}_{2\pe}\mb{C}_{1\pa}+\mr{c}_{1\pa}\mb{a}_{2\pe}+\mb{b}_{2\pe}\mb{D}_{1\pa}+\mr{d}_{1\pa}\ov{\mb{b}}_{2\pe},
-\mb{a}_{2\pe}\mb{D}_{1\pa}-\mr{d}_{1\pa}\ov{\mb{a}}_{2\pe}-\mb{b}_{2\pe}\ov{\mb{C}}_{1\pa}+\mr{c}_{1\pa}\mb{b}_{2\pe})
\in\mk{m}_{\pe},
\label{su.inclu.6}\\
& [((\mr{c}_{1\pa},\mr{d}_{1\pa}),(\mb{C}_{1\pa},\mb{D}_{1\pa})),
(\mb{c}_{2\pe},\mb{d}_{2\pe})]
\nonumber\\&
=(\mr{c}_{1\pa}\mb{c}_{2\pe}-\mb{c}_{2\pe}\mb{C}_{1\pa}-\mr{d}_{1\pa}\ov{\mb{d}}_{2\pe}+\mb{d}_{2\pe}\ov{\mb{D}}_{1\pa},
\mr{c}_{1\pa}\mb{d}_{2\pe}-\mb{d}_{2\pe}\ov{\mb{C}}_{1\pa}+\mr{d}_{1\pa}\ov{\mb{c}}_{2\pe}-\mb{c}_{2\pe}\mb{D}_{1\pa})
\in\mk{h}_{\pe},
\label{su.inclu.7}
\end{align}
\end{subequations}
\begin{subequations}
\begin{align}
& [(\mb{A}_{1\pa},\mb{B}_{1\pa}),(\mb{a}_{2\pe},\mb{b}_{2\pe})]
\nonumber\\&
=(-(\tr\mb{A}_{1\pa})\mb{a}_{2\pe}-\mb{a}_{2\pe}\mb{A}_{1\pa}-\mb{b}_{2\pe}\ov{\mb{B}}_{1\pa},
-(\tr\mb{A}_{1\pa})\mb{b}_{2\pe}+\mb{b}_{2\pe}\ov{\mb{A}}_{1\pa}-\mb{a}_{2\pe}\mb{B}_{1\pa})
\in\mk{h}_{\pe},
\label{su.inclu.4}\\
& [(\mb{A}_{1\pa},\mb{B}_{1\pa}),(\mb{c}_{2\pe},\mb{d}_{2\pe})]
\nonumber\\&
=(-(\tr\mb{A}_{1\pa})\mb{c}_{2\pe}-\mb{c}_{2\pe}\mb{A}_{1\pa}-\mb{d}_{2\pe}\ov{\mb{B}}_{1\pa},
-(\tr\mb{A}_{1\pa})\mb{d}_{2\pe}+\mb{d}_{2\pe}\ov{\mb{A}}_{1\pa}-\mb{c}_{2\pe}\mb{B}_{1\pa})
\in\mk{m}_{\pe}.
\label{su.inclu.5}
\end{align}
\end{subequations}
2. The remaining Lie brackets \eqref{inclusion.gen} are given by
\begin{subequations}
\begin{align}
&[(\mb{a}_{1\pe},\mb{b}_{1\pe}),(\mb{a}_{2\pe},\mb{b}_{2\pe})]
\nonumber\\&
=\big((\C(\mb{a}_{2\pe},\mb{a}_{1\pe})-\C(\mb{b}_{1\pe},\mb{b}_{2\pe}),
-\S(\mb{a}_{1\pe},\mb{b}_{2\pe})+\S(\mb{b}_{1\pe},\mb{a}_{2\pe})),
\nonumber\\&\qquad
(\CC(\mb{a}_{2\pe},\mb{a}_{1\pe})+\CC(\mb{b}_{1\pe},\mb{b}_{2\pe})^\t,
-\SS(\mb{b}_{2\pe},\mb{a}_{1\pe})+\SS(\mb{b}_{1\pe},\mb{a}_{2\pe}))\big)
\in\mk{h}_{\pa},
\label{su.inclu.8}\\
&[(\mb{c}_{1\pe},\mb{d}_{1\pe}),(\mb{c}_{2\pe},\mb{d}_{2\pe})]
\nonumber\\&
=\big((\C(\mb{c}_{2\pe},\mb{c}_{1\pe})+\C(\mb{d}_{2\pe},\mb{d}_{1\pe}),
\S(\mb{c}_{1\pe},\mb{d}_{2\pe})-\S(\mb{d}_{1\pe},\mb{c}_{2\pe})),
\nonumber\\&\qquad
(\CC(\mb{c}_{2\pe},\mb{c}_{1\pe})+\CC(\mb{d}_{1\pe},\mb{d}_{2\pe})^\t,
-\SS(\mb{d}_{2\pe},\mb{c}_{1\pe})+\SS(\mb{d}_{1\pe},\mb{c}_{2\pe}))\big)
\in\mk{h}_{\pa},
\label{su.inclu.10}\\
&[(\mb{a}_{1\pe},\mb{b}_{1\pe}),(\mb{c}_{2\pe},\mb{d}_{2\pe})]
\nonumber\\&
=\big(\CC(\mb{c}_{2\pe},\mb{a}_{1\pe})+\CC(\mb{d}_{2\pe},\mb{b}_{1\pe})^\t,
\AA(\mb{d}_{2\pe},\mb{a}_{1\pe})-\AA(\mb{b}_{1\pe},\mb{c}_{2\pe})\big)
\in\mk{m}_{\pa}.
\label{su.inclu.12}
\end{align}
\end{subequations}
3. The Cartan-Killing form on $\mk{m}_{\pe}$ is given by
\beq\label{su.killing.h}
\<(\mb{a}_{1\pe},\mb{b}_{1\pe}),(\mb{a}_{2\pe},\mb{b}_{2\pe})\>=
-8n\big(\A(\mb{a}_{1\pe},\mb{a}_{2\pe})+\A(\mb{b}_{1\pe},\mb{b}_{2\pe})\big) .
\eeq
\end{proposition}

The adjoint action of the Lie subalgebra
$\mk{h}_{\pa}\subset \mk{h}=\mk{sp}(n)$ on $\mk{g}=\mk{su}(n)$
generates the linear transformation group $H^*_{\pa}\subset H^* = \Ad(H)$
that preserves the element $\ee$ in the Cartan subspace
$\mk{a}\subset\mk{m}=\mk{su}(n)/\mk{sp}(n)$.
This group $H^*_{\pa}$ can be identified with the adjoint action of
a symplectic group $Sp(1)\times Sp(n-1)\subset Sp(n)$
whose matrix representation is given by
\beq\label{su.isotr.H}
\bpm C&D\\ -\ov{D}&\ov{C} \epm
\in Sp(1)\times Sp(n-1) \simeq H^*_{\pa},
\quad
C= \bpm c&0\\0&\bs{C}\epm,
\quad
D= \bpm d&0\\0&\bs{D}\epm,
\eeq
where
\begin{align}
& \bs{C}^\t\ov{\bs{C}}+ \bs{D}^\t\ov{\bs{D}} = I_{n-1} ,
\quad
\bs{C}^\t\ov{\bs{D}} -\ov{\bs{D}}^\t\bs{C} = 0,
\\
& c\ov{c}+d\ov{d} =1 .
\end{align}
In particular,
the subgroup $Sp(n-1)\subset H^*_{\pa}$ acts on $\mk{m}_{\pe}$
by right multiplication,
\beq
\label{SP-n.action.su}
\Ad(\bs{C},\bs{D})(\mb{a}_{\pe},\mb{b}_{\pe})
=(\mb{a}_{\pe}\ov{\bs{C}}^\t+\mb{b}_{\pe}\ov{\bs{D}}^\t,
-\mb{a}_{\pe}\bs{D}^\t+\mb{b}_{\pe}\bs{C}^\t)
\in\mk{m}_{\pe},
\eeq
where $(\bs{C},\bs{D})\in Sp(n-1)$ is defined to be
the matrix \eqref{su.isotr.H} with $c=1$ and $d=0$,
and the subgroup $Sp(1)\subset H^*_{\pa}$ acts similarly by 
\beq\label{SP-1.action.su}
\Ad(c,d)(\mb{a}_{\pe},\mb{b}_{\pe})
=(c\mb{a}_{\pe}+d\ov{\mb{b}}_{\pe},
c\mb{b}_{\pe}-d\ov{\mb{a}}_{\pe})
\in\mk{m}_{\pe}.
\eeq
where $(c,d)\in Sp(1)$ is defined to be
the matrix \eqref{su.isotr.H} with $\bs{C}=I_{n-1}$ and $\bs{D}=0$.
Composition of these subgroups \eqref{SP-n.action.su} and \eqref{SP-1.action.su}
yields the group $H^*_{\pa}=\Ad(Sp(1)\times Sp(n-1)) \subset \Ad(Sp(n))$. 

\begin{proposition}
The vector space $\mk{m}_{\perp}\simeq \Cnum^{n-1}\oplus\Cnum^{n-1}$
is an irreducible representation of the group $H^*_{\pa}$ 
on which the linear map $\ad(\ee)^2$ is a multiple of the identity:
\beq\label{su.ad.e.2}
\ad(\ee)^2(\mb{a}_{\pe},\mb{b}_{\pe})
=-\frac{1}{\rho}(\mb{a}_{\pe},\mb{b}_{\pe})
\eeq
where
\beq
\rho=8(n-1).
\eeq
\end{proposition}

\subsection{The vector space $\mk{sp}(n+1)/\mk{sp}(1)\oplus\mk{sp}(n)$}
\label{g=sp}
The symplectic Lie algebra $\mk{sp}(n+1)$ consists of all matrices 
$\gg$ in $\mk{gl}(2(n+1),\Cnum)$ satisfying
\beq
\gg J+J\gg^\t=0,\quad 
\gg^\t=-\ov{\gg},\quad
J=\bpm 0 & I_{n+1}\\-I_{n+1} & 0\epm.
\eeq
There is an involutive automorphism of $\mk{gl}(2(n+1),\Cnum)$ given by
\beq
\sigma(\gg)=S\gg S,\quad S=\bpm I_{n,1}&0\\0&I_{n,1}\epm,\quad I_{n,1}=\bpm 1&0\\0&-I_n\epm
\eeq
preserving $\mk{sp}(n+1)\subset \mk{gl}(2(n+1),\Cnum)$.
The matrices in $\mk{sp}(n+1)$ that are invariant under $\sigma$ 
span the compact symplectic Lie algebra $\mk{sp}(1)\oplus\mk{sp}(n)$.
This leads to the orthogonal decomposition of $\mk{g}=\mk{sp}(n+1)$
as a symmetric Lie algebra given by the eigenspaces of $\sigma$,
\beq\label{sp.rep.eigen.h}
\mk{h}:= \mk{sp}(1)\oplus\mk{sp}(n)\subset \mk{g},
\quad
\sigma(\mk{h})=\mk{h}
\eeq
and
\beq\label{sp.rep.eigen.m}
\mk{m}:=\mk{sp}(n+1)/\mk{sp}(1)\oplus\mk{sp}(n)\subset \mk{g},
\quad
\sigma(\mk{m})=-\mk{m}.
\eeq

\begin{lemma}\hfil \newline
1. The matrix representation of the Lie algebra $\mk{g}=\mk{sp}(n+1)$ 
is given by
\beq
\bpm A&B\\-\ov{B}&\ov{A}\epm\in\mk{sp}(n+1),
\quad 
A^\t=-\ov{A},
\quad 
B^\t=B,
\eeq
where $A,B\in\mk{gl}(n+1,\Cnum)$.
The matrix representations of
the vector space $\mk{m}=\mk{sp}(n+1)/\mk{sp}(1)\oplus\mk{sp}(n)$
and the Lie subalgebra $\mk{h}= \mk{sp}(1)\oplus\mk{sp}(n)$ in $\mk{gl}(2(n+1),\Cnum)$ are respectively given by
\begin{align}
(\mb{a},\mb{b})&:=
\bpm A&B\\-\ov{B}&\ov{A}\epm
\in\mk{m},\quad  A^\t=-\ov{A},\quad B^\t=B
\label{sp.m}\\
((\mr{c},\mr{d}),(\mb{C},\mb{D}))&:= \bpm C&D\\-\ov{D}&\ov{C}\epm \in\mk{h},\quad
C^\t=-\ov{C},\quad D^\t=D, 
\label{sp.h}
\end{align}
in which
\begin{align}
A&=\bpm 0&\mb{a}\\-\ov{\mb{a}}^\t&0\epm,\quad B=\bpm 0&\mb{b}\\\mb{b}^\t&0\epm\\
C&=\bpm \mr{c}&0\\0&\mb{C}\epm,\quad D=\bpm \mr{d}&0\\0&\mb{D}\epm
\end{align}
where 
$\mb{a},\mb{b}\in\Cnum^{n}$, $\mr{c}\in\i\Rnum$, $\mr{d}\in\Cnum$, 
$\mb{C}\in\mk{u}(n,\Cnum)$, $\mb{D}\in\mk{s}(n,\Cnum)$. 
\hfil \newline
2. The Lie bracket relations \eqref{ber.rel} have the matrix representation
\begin{subequations}
\begin{align}
&[(\mb{a}_1,\mb{b}_1),(\mb{a}_2,\mb{b}_2)] =
((\mb{a}_2\ov{\mb{a}}_1^\t -\mb{a}_1\ov{\mb{a}}_2^\t
+ \mb{b}_2\ov{\mb{b}}_1^\t -\mb{b}_1\ov{\mb{b}}_2^\t, 
\mb{a}_1\mb{b}_2^\t + \mb{b}_2\mb{a}_1^\t - \mb{b}_1\mb{a}_2^\t -\mb{a}_2\mb{b}_1^\t),\nonumber\\&\qquad
(\ov{\mb{a}}_2^\t \mb{a}_1 - \ov{\mb{a}}_1^\t\mb{a}_2
+ \mb{b}_2^\t\ov{\mb{b}}_1 - \mb{b}_1^\t\ov{\mb{b}}_2, 
-\mb{b}_2^\t\ov{\mb{a}}_1 -\ov{\mb{a}}_1^\t\mb{b}_2
+\mb{b}_1^\t\ov{\mb{a}}_2 +\ov{\mb{a}}_2^\t\mb{b}_1))
\in\mk{h}, 
\label{sp.bracket.m.m}\\
&[(\mb{a}_1,\mb{b}_1),((\mr{c}_1,\mr{d}_1),(\mb{C}_1,\mb{D}_1))] =
(\mb{a}_1\mb{C}_1-\mr{c}_1\mb{a}_1-\mb{b}_1\ov{\mb{D}}_1+\mr{d}_1\mb{b}_1,\nonumber\\&\qquad
\mb{a}_1\mb{D}_1-\mr{d}_1\ov{\mb{a}}_1+\mb{b}_1\ov{\mb{C}}_1-\mr{c}_1\mb{b}_1)
\in\mk{m},
\label{sp.bracket.h.m}\\
&[((\mr{c}_1,\mr{d}_1),(\mb{C}_1,\mb{D}_1)),((\mr{c}_2,\mr{d}_2),(\mb{C}_2,\mb{D}_2))] =
((-\mr{d}_1\ov{\mr{d}}_2+\mr{d}_2\ov{\mr{d}}_1,\mr{c}_1\mr{d}_2-\mr{d}_2\ov{\mr{c}}_1+\mr{d}_1\ov{\mr{c}}_2-\mr{c}_2\mr{d}_1),\nonumber\\&\qquad
(\mb{C}_1\mb{C}_2-\mb{C}_2\mb{C}_1-\mb{D}_1\ov{\mb{D}}_2+\mb{D}_2\ov{\mb{D}}_1,\mb{C}_1\mb{D}_2-\mb{D}_2\ov{\mb{C}}_1+\mb{D}_1\ov{\mb{C}}_2-\mb{C}_2\mb{D}_1))
\in\mk{h}.
\label{sp.bracket.h.h}
\end{align}
\end{subequations}
3. The restriction of the Cartan-Killing form on $\mk{g}=\mk{sp}(n+1)$
to $\mk{m}=\mk{sp}(n+1)/\mk{sp}(1)\oplus\mk{sp}(n)$
yields a negative-definite inner product
\beq\label{sp.kill.on.m}
\<(\mb{a}_1,\mb{b}_1),(\mb{a}_2,\mb{b}_2)\>
=-4(n+2)\big(\mb{a}_1\ov{\mb{a}}_2^\t + \mb{a}_2\ov{\mb{a}}_1^\t 
+ \mb{b}_1\ov{\mb{b}}_2^\t + \mb{b}_2\ov{\mb{b}}_1^\t\big) . 
\eeq
3. The (real) dimension of $\mk{m}=\mk{sp}(n+1)/\mk{sp}(1)\oplus\mk{sp}(n)$ is $4n$
and its rank  is $1$.
\end{lemma}

The one-dimensional subspace $\mk{a}\subset\mk{m}=\mk{sp}(n+1)/\mk{sp}(1)\oplus\mk{sp}(n)$ 
spanned by the matrix
\beq
(\mb{e}_1,0):=\bpm E_1& 0\\0&E_1\epm\in\mk{m},
\quad 
E_1=\bpm 0&\mb{e}_1\\-\mb{e}_1^\t&0\epm,
\quad 
\mb{e}_1=(1,\underbrace{0,...,0}_{n-1})
\eeq
is a Cartan subspace.
The element 
\beq
\ee:=\frac{1}{\sqrt{\chi}}(\mb{e}_1,0)\in\mk{a}
\eeq
in this subspace has unit norm,
where 
\beq
-1= \<\ee,\ee\> = -8(n+2)/\chi
\eeq
determines
 \beq
\chi = 8(n+2).
\eeq
The corresponding linear operator $\ad(\ee)$
induces a direct sum decomposition of the vector spaces
$\mk{m}=\mk{sp}(n+1)/\mk{sp}(1)\oplus\mk{sp}(n)$ and $\mk{h}= \mk{sp}(n)$
into centralizer spaces $\mk{m}_{\pa}$ and $\mk{h}_{\pa}$
and their orthogonal complements (perp spaces)
$\mk{m}_{\pe}$ and $\mk{h}_{\pe}$
with respect to the Cartan-Killing form.
From the Lie bracket relation \eqref{inclusion.three},
$\mk{h}_{\pe}$ is mapped into $\mk{m}_{\pe}$, and vice versa,
under $\ad(\ee)$. 
Hence $\ad(\ee)^2$ defines a linear mapping of 
each subspace $\mk{h}_{\pe}$ and $\mk{m}_{\pe}$ into itself.

\begin{lemma}\hfil \newline
1. The matrix representations of $\mk{m}_{\pa}$ and $\mk{m}_{\pe}$
in $\mk{m}=\mk{sp}(n+1)/\mk{sp}(1)\oplus\mk{sp}(n)$ are given by
\beq\label{m.pa.pe.sp}
(\mr{a}_{\pa}):=\bpm A_{\pa}&B_{\pa}\\-\ov{B}_{\pa}&\ov{A}_{\pa}\epm\in\mk{m}_{\pa},\quad
((\mr{a}_{\pe},\mr{b}_{\pe}),(\mb{a}_{\pe},\mb{b}_{\pe})):=\bpm A_{\pe}&B_{\pe}\\-\ov{B}_{\pe}&\ov{A}_{\pe}\epm
\in\mk{m}_{\pe},
\eeq
in which
\begin{align*}
&A_{\pa}=\bpm 0&\mr{a}_{\pa}&0\\-\mr{a}_{\pa}&0&0\\0&0&0\epm ,\quad B_{\pa}=\bpm 0&0&0\\0&0&0\\0&0&0\epm,\\
&A_{\pe}=\bpm 0&\mr{a}_{\pe}&\mb{a}_{\pe}\\\mr{a}_{\pe}&0&0\\-\ov{\mb{a}}^\t_{\pe}&0&0 \epm,\quad
B_{\pe}=\bpm 0&\mr{b}_{\pe}&\mb{b}_{\pe}\\\mr{b}_{\pe}&0&0\\\mb{b}_{\pe}^\t&0&0 \epm
\end{align*}
where $\mr{a}_{\pa}\in\Rnum,\mr{a}_{\pe}\in\i\Rnum$, $\mr{b}_{\pe}\in\Cnum$, 
$\mb{b}_{\pe},\mb{a}_{\pe}\in\Cnum^{n-1}$. 
\hfil \newline
2. The matrix representations of $\mk{h}_{\pa}$ and $\mk{h}_{\pe}$
in $\mk{h}=\mk{sp}(1)\oplus\mk{sp}(n)$ are given by
\beq\label{h.pa.pe.sp}
((\mr{c}_{\pa},\mr{d}_{\pa}),(\mb{C}_{\pa},\mb{D}_{\pa})):=
\bpm C_{\pa}&D_{\pa}\\-\ov{D}_{\pa}&\ov{C}_{\pa}\epm
\in\mk{h}_{\pa},
\quad
((\mr{c}_{\pe},\mr{d}_{\pe}),(\mb{c}_{\pe},\mb{d}_{\pe})):=
\bpm C_{\pe}& D_{\pe}\\-\ov{D}_{\pe}&\ov{C}_{\pe}\epm
\in\mk{h}_{\pe},
\eeq
in which
\begin{align*}
& C_{\pa}=\bpm \mr{c}_{\pa}&0&0\\0&\mr{c}_{\pa}&0\\0&0&\mb{C}_{\pa}\epm,
\quad
D_{\pa}=\bpm \mr{d}_{\pa}&0&0\\0&\mr{d}_{\pa}&0\\0&0&\mb{D}_{\pa}\epm,
\quad
\mb{C}_{\pa}^\t=-\ov{\mb{C}}_{\pa},
\quad
\mb{D}_{\pa}^\t=\mb{D}_{\pa},\\
& C_{\pe}=\bpm \mr{c}_{\pe}&0&0\\0&-\mr{c}_{\pe}&\mb{c}_{\pe}\\0&-\ov{\mb{c}}_{\pe}^\t&0\epm,
\quad
D_{\pe}=\bpm \mr{d}_{\pe}&0&0\\0&-\mr{d}_{\pe}&\mb{d}_{\pe}\\0&\mb{d}_{\pe}^\t&0\epm,
\end{align*}
where $\mb{C}_{\pa}\in\mk{u}(n-1)$, $\mb{D}_{\pa}\in\mk{s}(n-1,\Cnum)$,
$\mb{c}_{\pe},\mb{d}_{\pe}\in \Cnum^{n-1}$,
$\mr{c}_{\pa},\mr{c}_{\pe}\in\i\Rnum$, $\mr{d}_{\pa},\mr{d}_{\pe}\in\Cnum$. 
\hfil \newline
3. $\dim\mk{m}_{\pa}=1$, $\dim\mk{m}_{\pe}=\dim\mk{h}_{\pe}=2n+1$,
$\dim\mk{h}_{\pa}=2(n-1)^2+n+2$.
\hfil \newline
4. The linear operator $\ad(\ee)$ acts on $\mk{m}_{\pe}$ and $\mk{h}_{\pe}$ by
\begin{subequations}
\begin{align}
&\ad(\ee)((\mr{a}_{\pe},\mr{b}_{\pe}),(\mb{a}_{\pe},\mb{b}_{\pe}))
=\frac{1}{\sqrt{\chi}}((2\mr{a}_{\pe},2\mr{b}_{\pe}),(-\mb{a}_{\pe},-\mb{b}_{\pe}))
\in\mk{h}_{\pe},\label{ad.e.m.pe.sp}\\&
\ad(\ee)((\mr{c}_{\pe},\mr{d}_{\pe}),(\mb{c}_{\pe},\mb{d}_{\pe}))
=\frac{1}{\sqrt{\chi}}((-2\mr{c}_{\pe},-2\mr{d}_{\pe}),(\mb{c}_{\pe},\mb{d}_{\pe}))
\in\mk{m}_{\pe}\label{ad.e.h.pe.sp}. 
\end{align}
\end{subequations}
\end{lemma}

We use the inner products \eqref{com.vec}--\eqref{anti.conj.vec}
and outer products \eqref{com.not.mat}--\eqref{anti.conj.mat}
to write out the explicit Lie bracket relations on
$\mk{m}=\mk{m}_{\pa}\oplus\mk{m}_{\pe}$
and $\mk{h}=\mk{h}_{\pa}\oplus\mk{h}_{\pe}$. 

\begin{proposition}\hfil \newline
1. The Lie brackets \eqref{inclusion.one}--\eqref{inclusion.three}
are given by
\begin{subequations}
\begin{align}
& [(\mr{a}_{1\pa}),(\mr{a}_{1\pa})]=0\in\mk{h}_{\pa},
\label{inclu.1.sp}\\
& [((\mr{c}_{1\pa},\mr{d}_{1\pa}),(\mb{C}_{1\pa},\mb{D}_{1\pa})),(\mr{a}_{2\pa})]=0\in\mk{m}_{\pa},
\label{inclu.2.sp}\\
& [((\mr{c}_{1\pa},\mr{d}_{1\pa}),(\mb{C}_{1\pa},\mb{D}_{1\pa})),
((\mr{c}_{2\pa},\mr{d}_{2\pa}),(\mb{C}_{2\pa},\mb{D}_{2\pa}))]
\nonumber\\&=
\big((\mr{d}_{2\pa}\ov{\mr{d}}_{1\pa}-\mr{d}_{1\pa}\ov{\mr{d}}_{2\pa},
\mr{c}_{1\pa}\mr{d}_{2\pa}+\mr{d}_{2\pa}\mr{c}_{1\pa}-\mr{d}_{1\pa}\mr{c}_{2\pa}-\mr{c}_{2\pa}\mr{d}_{1\pa}),
\nonumber\\&\qquad
([\mb{C}_{1\pa},\mb{C}_{2\pa}]+\mb{D}_{2\pa}\ov{\mb{D}}_{1\pa}-\mb{D}_{1\pa}\ov{\mb{D}}_{2\pa},
\mb{C}_{1\pa}\mb{D}_{2\pa}-\mb{D}_{2\pa}\ov{\mb{C}}_{1\pa}+\mb{D}_{1\pa}\ov{\mb{C}}_{2\pa}-\mb{C}_{2\pa}\mb{D}_{1\pa})\big)
\in\mk{h}_{\pa},
\label{inclu.3.sp}
\end{align}
\end{subequations}
\begin{subequations}
\begin{align}
& [((\mr{c}_{1\pa},\mr{d}_{1\pa}),(\mb{C}_{1\pa},\mb{D}_{1\pa})),
((\mr{a}_{2\pe},\mr{b}_{2\pe}),(\mb{a}_{2\pe},\mb{b}_{2\pe})]
\nonumber\\&
=\big((-\mr{d}_{1\pa}\ov{\mr{b}}_{2\pe}+\mr{b}_{2\pe}\ov{\mr{d}}_{1\pa},
\mr{c}_{1\pa}\mr{b}_{2\pe}+\mr{b}_{2\pe}\mr{c}_{1\pa}-\mr{d}_{1\pa}\mr{a}_{2\pe}-\mr{a}_{2\pe}\mr{d}_{1\pa}),
\nonumber\\&\qquad
(\mr{c}_{1\pa}\mb{a}_{2\pe}-\mb{a}_{2\pe}\mb{C}_{1\pa}-\mr{d}_{1\pa}\ov{\mb{b}}_{2\pe}+\mb{b}_{2\pe}\ov{\mb{D}}_{1\pa},
\mr{c}_{1\pa}\mb{b}_{2\pe}-\mb{b}_{2\pe}\ov{\mb{C}}_{1\pa}+\mr{d}_{1\pa}\ov{\mb{a}}_{2\pe}-\mb{a}_{2\pe}\mb{D}_{1\pa})\big)
\in\mk{m}_{\pe},
\label{inclu.6.sp}\\
& [((\mr{c}_{1\pa},\mr{d}_{1\pa}),(\mb{C}_{1\pa},\mb{D}_{1\pa})),
((\mr{c}_{2\pe},\mr{d}_{2\pe}),(\mb{c}_{2\pe},\mb{d}_{2\pe})]
\nonumber\\&
=\big((-\mr{d}_{1\pa}\ov{\mr{d}}_{2\pe}+\mr{d}_{2\pe}\ov{\mr{d}}_{1\pa},
\mr{c}_{1\pa}\mr{d}_{2\pe}+\mr{d}_{2\pe}\mr{c}_{1\pa}-\mr{d}_{1\pa}\mr{c}_{2\pe}-\mr{c}_{2\pe}\mr{d}_{1\pa}),
\nonumber\\&\qquad
(\mr{c}_{1\pa}\mb{c}_{2\pe}-\mb{c}_{2\pe}\mb{C}_{1\pa}-\mr{d}_{1\pa}\ov{\mb{d}}_{2\pe}+\mb{d}_{2\pe}\ov{\mb{D}}_{1\pa},
\mr{c}_{1\pa}\mb{d}_{2\pe}-\mb{d}_{2\pe}\ov{\mb{C}}_{1\pa}+\mr{d}_{1\pa}\ov{\mb{c}}_{2\pe}-\mb{c}_{2\pe}\mb{D}_{1\pa})\big)
\in\mk{h}_{\pe},
\label{inclu.7.sp}
\end{align}
\end{subequations}
\begin{subequations}
\begin{align}
& [(\mr{a}_{1\pa}),((\mr{a}_{2\pe},\mr{b}_{2\pe}),(\mb{a}_{2\pe},\mb{b}_{2\pe}))]
=((2\mr{a}_{1\pa}\mr{a}_{2\pe},2\mr{a}_{1\pa}\mr{b}_{2\pe}),(-\mr{a}_{1\pa}\mb{a}_{2\pe},-\mr{a}_{1\pa}\mb{b}_{2\pe}))
\in\mk{h}_{\pe},
\label{inclu.4.sp}\\
& [(\mr{a}_{1\pa}),((\mr{c}_{2\pe},\mr{d}_{2\pe}),(\mb{c}_{2\pe},\mb{d}_{2\pe}))]
=((-2\mr{a}_{1\pa}\mr{c}_{2\pe},-2\mr{a}_{1\pa}\mr{d}_{2\pe}),(\mr{a}_{1\pa}\mb{c}_{2\pe},\mr{a}_{1\pa}\mb{d}_{2\pe}))
\in\mk{m}_{\pe}.
\label{inclu.5.sp}
\end{align}
\end{subequations}
2. The remaining Lie brackets \eqref{inclusion.gen} are given by
\begin{subequations}
\begin{align}
&[((\mr{a}_{1\pe},\mr{b}_{1\pe}),(\mb{a}_{1\pe},\mb{b}_{1\pe})),
((\mr{a}_{2\pe},\mr{b}_{2\pe}),(\mb{a}_{2\pe},\mb{b}_{2\pe}))]_{\mk{h}_{\pa}}
\nonumber\\&
=\big((-\ha\C(\mb{a}_{1\pe},\mb{a}_{2\pe})-\ha\C(\mb{b}_{1\pe},\mb{b}_{2\pe})-\mr{b}_{1\pe}\ov{\mr{b}}_{2\pe}+\mr{b}_{2\pe}\ov{\mr{b}}_{1\pe},\nonumber\\&\qquad\qquad
2\mr{a}_{1\pe}\mr{b}_{2\pe}-2\mr{a}_{2\pe}\mr{b}_{1\pe}+\ha\S(\mb{a}_{1\pe},\mb{b}_{2\pe})-\ha\S(\mb{b}_{1\pe},\mb{a}_{2\pe})),
\nonumber\\&\qquad
(-\CC(\mb{a}_{1\pe},\mb{a}_{2\pe})+\CC(\mb{b}_{1\pe},\mb{b}_{2\pe})^\t,
-\SS(\mb{b}_{2\pe},\mb{a}_{1\pe})+\SS(\mb{b}_{1\pe},\mb{a}_{2\pe}))\big)
\in\mk{h}_{\pa},
\label{inclu.8.sp}\\
&[((\mr{a}_{1\pe},\mr{b}_{1\pe}),(\mb{a}_{1\pe},\mb{b}_{1\pe})),
((\mr{a}_{2\pe},\mr{b}_{2\pe}),(\mb{a}_{2\pe},\mb{b}_{2\pe}))]_{\mk{h}_{\pe}}
\nonumber\\&
=\big((-\ha\C(\mb{a}_{1\pe},\mb{a}_{2\pe})-\ha\C(\mb{b}_{1\pe},\mb{b}_{2\pe}),\ha\S(\mb{a}_{1\pe},\mb{b}_{2\pe})-\ha\S(\mb{b}_{1\pe},\mb{a}_{2\pe})),\nonumber\\&\qquad
(\mr{a}_{1\pe}\mb{a}_{2\pe}-\mr{a}_{2\pe}\mb{a}_{1\pe}-\mr{b}_{1\pe}\ov{\mb{b}}_{2\pe}+\mr{b}_{2\pe}\ov{\mb{b}}_{1\pe},\mr{a}_{1\pe}\mb{b}_{2\pe}-\mr{b}_{2\pe}\ov{\mb{a}}_{1\pe}+\mr{b}_{1\pe}\ov{\mb{a}}_{2\pe}-\mr{a}_{2\pe}\mb{b}_{1\pe})\big)
\in\mk{h}_{\pe},
\end{align}
\end{subequations}
\begin{subequations}
\begin{align}
&[((\mr{c}_{1\pe},\mr{d}_{1\pe}),(\mb{c}_{1\pe},\mb{d}_{1\pe})),
((\mr{c}_{2\pe},\mr{d}_{2\pe}),(\mb{c}_{2\pe},\mb{d}_{2\pe}))]_{\mk{h}_{\pa}}
\nonumber\\&
=\big((-\mr{d}_{1\pe}\ov{\mr{d}}_{2\pe}+\mr{d}_{2\pe}\ov{\mr{d}}_{1\pe}-\ha\C(\mb{c}_{1\pe},\mb{c}_{2\pe})-\ha\C(\mb{d}_{1\pe},\mb{d}_{2\pe}),
\nonumber\\&\qquad\qquad
2\mr{c}_{1\pe}\mr{d}_{2\pe}-2\mr{d}_{1\pe}\mr{c}_{2\pe}+\ha\S(\mb{c}_{1\pe},\mb{d}_{2\pe})-\ha\S(\mb{d}_{1\pe},\mb{c}_{2\pe})),\nonumber\\&\qquad
(-\CC(\mb{c}_{1\pe},\mb{c}_{2\pe})+\CC(\mb{d}_{1\pe},\mb{d}_{2\pe})^\t,-\SS(\mb{d}_{2\pe},\mb{c}_{1\pe})+\SS(\mb{d}_{1\pe},\mb{c}_{2\pe}))\big)
\in\mk{h}_{\pa},
\label{inclu.10.sp.1}\\&
[((\mr{c}_{1\pe},\mr{d}_{1\pe}),(\mb{c}_{1\pe},\mb{d}_{1\pe})),
((\mr{c}_{2\pe},\mr{d}_{2\pe}),(\mb{c}_{2\pe},\mb{d}_{2\pe}))]_{\mk{h}_{\pe}}
\nonumber\\&
=\big((\ha\C(\mb{c}_{1\pe},\mb{c}_{2\pe})+\ha\C(\mb{d}_{1\pe},\mb{d}_{2\pe}),-\ha\S(\mb{c}_{1\pe},\mb{d}_{2\pe})+\ha\S(\mb{d}_{1\pe},\mb{c}_{2\pe})),\nonumber\\&\qquad
(-\mr{c}_{1\pe}\mb{c}_{2\pe}+\mr{c}_{2\pe}\mb{c}_{1\pe}+\mr{d}_{1\pe}\ov{\mb{d}}_{2\pe}-\mr{d}_{2\pe}\ov{\mb{d}}_{1\pe},
-\mr{c}_{1\pe}\mb{d}_{2\pe}+\mr{d}_{2\pe}\ov{\mb{c}}_{1\pe}-\mr{d}_{1\pe}\ov{\mb{c}}_{2\pe}+\mr{c}_{2\pe}\mb{d}_{1\pe})\big)\in\mk{h}_{\pe},
\label{inclu.10.sp.2}
\end{align}
\end{subequations}
\begin{subequations}
\begin{align}
&[((\mr{a}_{1\pe},\mr{b}_{1\pe}),(\mb{a}_{1\pe},\mb{b}_{1\pe})),
((\mr{c}_{2\pe},\mr{d}_{2\pe}),(\mb{c}_{2\pe},\mb{d}_{2\pe}))]_{\mk{m}_{\pa}}
\nonumber\\&
=\big(\mr{b}_{1\pe}\ov{\mr{d}}_{2\pe}+\mr{d}_{2\pe}\ov{\mr{b}}_{1\pe}-\ha\A(\mb{b}_{1\pe},\mb{d}_{2\pe})-2\mr{a}_{1\pe}\mr{c}_{2\pe}-\ha\A(\mb{a}_{1\pe},\mb{c}_{2\pe})\big)
\in\mk{m}_{\pa},
\label{inclu.12.sp.1}\\&
[((\mr{a}_{1\pe},\mr{b}_{1\pe}),(\mb{a}_{1\pe},\mb{b}_{1\pe})),
((\mr{c}_{2\pe},\mr{d}_{2\pe}),(\mb{c}_{2\pe},\mb{d}_{2\pe}))]_{\mk{m}_{\pe}}\nonumber\\&
=\big((-\ha\C(\mb{b}_{1\pe},\mb{d}_{2\pe})-\ha\C(\mb{a}_{1\pe},\mb{c}_{2\pe}),
\ha\S(\mb{a}_{1\pe},\mb{d}_{2\pe})-\ha\S(\mb{b}_{1\pe},\mb{c}_{2\pe})),\nonumber\\&\quad
(\mr{a}_{1\pe}\mb{c}_{2\pe}-\mr{c}_{2\pe}\mb{a}_{1\pe}-\mr{b}_{1\pe}\ov{\mb{d}}_{2\pe}+\mr{d}_{2\pe}\ov{\mb{b}}_{1\pe},
\mr{a}_{1\pe}\mb{d}_{2\pe}-\mr{d}_{2\pe}\ov{\mb{a}}_{1\pe}+\mr{b}_{1\pe}\ov{\mb{c}}_{2\pe}-\mr{c}_{2\pe}\mb{b}_{1\pe})\big)\in\mk{m}_{\pe}.
\label{inclu.12.sp.2}
\end{align}
\end{subequations}
3. The Cartan-Killing form on $\mk{m}_{\pe}$ is given by
\begin{align}\label{killing.h.sp}
&\<((\mr{a}_{1\pe},\mr{b}_{1\pe}),(\mb{a}_{1\pe},\mb{b}_{1\pe})),
((\mr{a}_{2\pe},\mr{b}_{2\pe}),(\mb{a}_{2\pe},\mb{b}_{2\pe}))\>\nonumber\\&
=-4(n+2)\big(\A(\mr{a}_{1\pe},\mr{a}_{2\pe})+\A(\mb{a}_{1\pe},\mb{a}_{2\pe})
+\A(\mr{b}_{1\pe},\mr{b}_{2\pe})+\A(\mb{b}_{1\pe},\mb{b}_{2\pe})\big).
\end{align}
\end{proposition}

The adjoint action of the Lie subalgebra
$\mk{h}_{\pa}\subset \mk{h}=\mk{sp}(1)\oplus\mk{sp}(n)$ on $\mk{g}=\mk{sp}(n+1)$
generates the linear transformation group $H^*_{\pa}\subset H^* = \Ad(H)$
that preserves the element $\ee$ in the Cartan subspace
$\mk{a}\subset\mk{m}=\mk{sp}(n+1)/\mk{sp}(1)\oplus\mk{sp}(n)$.
This group $H^*_{\pa}$ can be identified with the adjoint action of
a symplectic group $Sp(1)\times Sp(n-1)\subset Sp(1)\times Sp(n)$
whose matrix representation is given by
\beq\label{isotr.H.sp}
\bpm C&D\\ -\ov{D}&\ov{C} \epm
\in Sp(1)\times Sp(n-1) \simeq H^*_{\pa},
\quad
C= \bpm c&0&0\\0&c&0\\0&0&\bs{C}\epm,
\quad
D= \bpm d&0&0\\0&d&0\\0&0&\bs{D}\epm,
\eeq
where
\begin{align}
& \bs{C}^\t\ov{\bs{C}}+ \bs{D}^\t\ov{\bs{D}} = I_{n-1} ,
\quad
\bs{C}^\t\ov{\bs{D}} -\ov{\bs{D}}^\t\bs{C} = 0,
\\
& c\ov{c}+d\ov{d} =1 .
\end{align}
In particular,
the subgroup $Sp(n-1)\subset H^*_{\pa}$ acts on $\mk{m}_{\pe}$
by right multiplication,
\beq\label{SP-n.action.sp}
\Ad(\bs{C},\bs{D})((\mr{a}_{\pe},\mr{b}_{\pe}),(\mb{a}_{\pe},\mb{b}_{\pe}))
=
((\mr{a}_{\pe},\mr{b}_{\pe}),(\mb{a}_{\pe}\ov{\bs{C}}^\t-\mb{b}_{\pe}\bs{D}^\t,
\mb{a}_{\pe}\ov{\bs{D}}^\t+\mb{b}_{\pe}\bs{C}^\t))
\in\mk{m}_{\pe},
\eeq
where $(\bs{C},\bs{D})\in Sp(n-1)$ is defined to be
the matrix \eqref{isotr.H.sp} with $c=1$ and $d=0$,
while the subgroup $Sp(1)\subset H^*_{\pa}$ has a non-standard action on
$\mk{m}_{\pe}$ given by
\beq\label{SP-1.action.sp}
\begin{aligned}
&\Ad(c,d)((\mr{a}_{\pe},\mr{b}_{\pe}),(\mb{a}_{\pe},\mb{b}_{\pe}))=\nonumber\\&
((c\mr{a}_{\pe}\ov{c}-d\ov{\mr{b}}_{\pe}\ov{c}-c\mr{b}_{\pe}d+d\mr{a}_{\pe}d,
c\mr{a}_{\pe}\ov{d}-d\ov{\mr{b}}_{\pe}\ov{d}+c\mr{b}_{\pe}c-d\mr{a}_{\pe}c),
(c\mb{a}_{\pe}-d\ov{\mb{b}}_{\pe},
c\mb{b}_{\pe}+d\ov{\mb{a}}_{\pe}))
\in\mk{m}_{\pe},
\end{aligned}
\eeq
where $(c,d)\in Sp(1)$ is defined to be
the matrix \eqref{isotr.H.sp} with $\bs{C}=I_{n-1}$ and $\bs{D}=0$.
Composition of these subgroups \eqref{SP-n.action.sp} and \eqref{SP-1.action.sp}
yields the group $H^*_{\pa}=\Ad(Sp(1)\times Sp(n-1)) \subset \Ad(Sp(1)\times Sp(n))$. 

\begin{proposition}
\label{sp.prop.reduc}
The vector space $\mk{m}_{\perp}\simeq \i\Rnum\oplus\Cnum\oplus\Cnum^{n-1}\oplus\Cnum^{n-1}$ 
is a reducible representation of the group $H^*_{\pa}$ 
such that the linear map $\ad(\ee)^2$ is given by
\beq\label{ad.e.2.sp}
\ad(\ee)^2((\mr{a}_{\pe},\mr{b}_{\pe}),(\mb{a}_{\pe},\mb{b}_{\pe}))
=\frac{1}{\chi}((-4\mr{a}_{\pe},-4\mr{b}_{\pe}),(-\mb{a}_{\pe},-\mb{b}_{\pe})).
\eeq
The irreducible subspaces in this representation consist of 
$((\mr{a}_{\pe},\mr{b}_{\pe}),(0,0)) \simeq \i\Rnum\oplus\Cnum$ 
and $((0,0),(\mb{a}_{\pe},\mb{b}_{\pe})) \simeq \Cnum^{n-1}\oplus\Cnum^{n-1}$ 
on which $\ad(\ee)^2$ is a multiple of the identity with respective eigenvalues
$-4/\chi$ and $-1/\chi$. 
\end{proposition}

\section{Bi-Hamiltonian soliton equations in $SU(2n)/Sp(n)$}
\label{SU.curveflows}

Employing the notation and preliminaries
in \secref{construction} and \secref{g=su},
we consider a non-stretching curve flow $\map(t,x)$ 
in $M=SU(2n)/Sp(n)$
having a $Sp(n)$-parallel framing 
as expressed in terms of the variables
\begin{align}
&
e_x=\frac{1}{\sqrt{\chi}}( -\i I_{n-1},0)
\in\mk{u}(n-1)\oplus\mk{so}(n-1,\Cnum) \simeq \mk{m}_{\pa},
\quad
\chi=8(n-1)n^2,
\label{e}\\
&
\conx_x=(\tbu_{1},\tbu_{2})
\in\Cnum^{n-1}\oplus\Cnum^{n-1}\simeq \mk{h}_{\pe},
\label{u.pe}
\end{align}
and
\begin{align}
&
h_{\pa}=(\tbHH_{1\pa},\tbHH_{2\pa})
\in\mk{u}(n-1)\oplus \mk{so}(n-1,\Cnum)\simeq\mk{m}_{\pa},
\label{h.pa}\\
&
h_{\pe}=(\tbh_{1\pe},\tbh_{2\pe})
\in\Cnum^{n-1}\oplus\Cnum^{n-1}\simeq \mk{m}_{\pe},
\label{h.pe}\\
&\varpi^{\pa}=((\ww^{1\pa},\ww^{2\pa}),(\tbWW^{1\pa},\tbWW^{2\pa}))
\in\mk{sp}(1)\oplus\mk{sp}(n-1)\simeq \mk{h}_{\pa},
\label{w.pa}\\
& \varpi^{\pe}=(\tbw^{1\pe},\tbw^{2\pe})
\in\Cnum^{n-1}\oplus\Cnum^{n-1}\simeq \mk{h}_{\pe},
\label{w.pe}
\end{align}
using the matrix identifications \eqref{su.m.pa.pe}--\eqref{su.h.pa.pe},
where
$\ww^{1\pa}\in\i\Rnum$ is an imaginary (complex) scalar variable,
$\ww^{2\pa}\in\Cnum$ is a complex scalar variable,
$\tbu_{1}$, $\tbu_{2}$, $\tbw^{1\pe}$, $\tbw^{2\pe}$,
$\tbh_{1\pe}$, $\tbh_{2\pe}$ $\in\Cnum^{n-1}$ are complex vector variables,
$\tbWW^{1\pa}$, $\tbHH_{1\pa}\in\mk{u}(n-1)$ are anti-Hermitian matrix variables,
$\tbWW^{2\pa}\in\mk{s}(n-1,\Cnum)$ is a complex symmetric matrix variable,
and $\tbHH_{2\pa}\in\mk{so}(n-1,\Cnum)$ is a complex anti-symmetric matrix variable.
For later use, through property \eqref{su.ad.e.h.pe}
we also introduce the variable
\beq
h^{\pe}=(\tbh^{1\pe},\tbh^{2\pe})= \ad(e_x)h_{\pe}
=\frac{1}{\sqrt{\rho}}(\i\tbh_{1\pe},\i\tbh_{2\pe})
\in\Cnum^{n-1}\oplus\Cnum^{n-1}\simeq \mk{h}_{\pe},
\quad
\rho=8(n-1),
\label{h.pe.up.su}
\eeq
where $\tbh^{1\pe}$, $\tbh^{2\pe}$ $\in\Cnum^{n-1}$ are complex vector variables.

Up to the rigid ($x$-independent) action of the equivalence group
$H^*_{\pa} = \Ad(Sp(1)\times Sp(n-1)) \subset \Ad(Sp(n))$,
a $Sp(n)$-parallel linear coframe $e$ along $\map$ is then determined by
the variables \eqref{e} and \eqref{u.pe}
via the transport equation
\beq\label{transport}
\nabla_x e=-\ad(\conx_x)e
\eeq
together with the soldering relation
\beq
e\rfloor \map_x = e_x .
\eeq
The resulting coframe $e$ defines an isomorphism between
$T_\map M$ and $\mk{m} \simeq \mk{u}(n-1)\oplus \mk{so}(n-1,\Cnum)\oplus\Cnum^{n-1}\oplus\Cnum^{n-1}$,
which yields a correspondence between the set of frames for $T_\map M$
and the set of basis vectors for $\mk{m}$, as follows.
Let $e_\pa$ and $e_\pe$ be the respective projections of $e$ into
$\mk{m}_{\pa}$ and $\mk{m}_{\pe}$ given in terms of
the matrix identifications \eqref{su.m.pa.pe}--\eqref{su.h.pa.pe} by
\begin{align}
& e_\pa= (\mb{A}_{\pa}(\cdot),\mb{B}_{\pa}(\cdot))
\label{epa}\\
& e_\pe= (\mb{a}_{\pe}(\cdot),\mb{b}_{\pe}(\cdot))
\label{epe}
\end{align}
where $\mb{A}_{\pa}(\cdot)$ and $\mb{B}_{\pa}(\cdot)$ are linear maps
from $T_x M$ into $\mk{u}(n-1)$ and $\mk{so}(n-1,\Cnum)$ respectively,
and where both $\mb{a}_{\pe}(\cdot)$ and $\mb{b}_{\pe}(\cdot)$ are linear maps
from $T_x M$ into $\Cnum^{n-1}$.
Let $(T_\map M)_\pa$ and $(T_\map M)_\pe$ be the orthogonal subspaces of
$T_\map M$ respectively defined by the kernels of $e_\pa$ and $e_\pe$,
so thus
\[
e_\pa\rfloor (T_\map M)_\pe = e_\pe\rfloor (T_\map M)_\pa = 0
\]
and hence
\begin{align}
& e\rfloor (T_\map M)_\pa = e_\pa\rfloor T_\map M
= \mk{m}_{\pa}\simeq \mk{u}(n-1)\oplus\mk{so}(n-1,\Cnum) ,
\\
& e\rfloor (T_\map M)_\pe = e_\pe\rfloor T_\map M
= \mk{m}_{\pe}\simeq \Cnum^{n-1}\oplus\Cnum^{n-1} .
\end{align}
Note that, in this notation,
\begin{align}
& e_\pa\rfloor \map_x = e_x, \quad
e_\pe\rfloor \map_x = 0 ,\\
& e_\pa\rfloor \map_t = h_\pa, \quad
e_\pe\rfloor \map_t = h_\pe .
\end{align}
Now if $\{\mb{M}_{\pa\mk{u}}^{(i)}\}$, $i=1,\ldots,(n-1)^2$,
is a matrix basis for $\mk{u}(n-1)$ viewed as a real vector space, 
and $\{\mb{M}_{\pa\mk{so}}^{(j)}\}$, $j=1,\ldots,(n-1)(n-2)$,
is a matrix basis for $\mk{so}(n-1,\Cnum)$ viewed as a real vector space, 
then $e_\pa$ determines a corresponding basis
$\{X_{\pa\mk{u}}^{(i)}, X_{\pa\mk{so}}^{(j)}\}$,
$i=1,\ldots,(n-1)^2$ and $j=1,\ldots,(n-1)(n-2)$,
for the vector space $(T_\map M)_\pa$ given by
\[
(\mb{A}_{\pa}(X_{\pa\mk{u}}^{(i)}),\mb{B}_{\pa}(X_{\pa\mk{u}}^{(i)}))
= (\mb{M}_{\pa\mk{u}}^{(i)},0) ,
\quad
(\mb{A}_{\pa}(X_{\pa\mk{so}}^{(j)}),\mb{B}_{\pa}(X_{\pa\mk{so}}^{(j)}))
= (0,\mb{M}_{\pa\mk{so}}^{(j)}) .
\]
Similarly if $\{\mb{m}_{\pe\Cnum}^{(k)}\}$, $k=1,\ldots,2(n-1)$,
is a basis for $\Cnum^{n-1}$ viewed as a real vector space, 
then $e_\pe$ determines a corresponding basis
$\{X_{\pe\Cnum}^{(k)}, X_{\pe\Cnum'}^{(k')}\}$, $k,k'=1,\ldots,2(n-1)$,
for the vector space $(T_\map M)_\pe$ given by
\[
(\mb{a}_{\pe}(X_{\pe\Cnum}^{(k)}),\mb{b}_{\pe}(X_{\pe\Cnum}^{(k)}))
= (\mb{m}_{\pe\Cnum}^{(k)},0) ,
\quad
(\mb{a}_{\pe}(X_{\pe\Cnum'}^{(k')}),\mb{b}_{\pe}(X_{\pe\Cnum'}^{(k')}))
= (0,\mb{m}_{\pe\Cnum}^{(k')}) .
\]
In addition, if each basis
$\{\mb{M}_{\pa\mk{u}}^{(i)}\}$, $\{\mb{M}_{\pa\mk{so}}^{(j)}\}$, $\{\mb{m}_{\pe\Cnum}^{(k)}\}$
is normalized such that
\[
\<\mb{M}_{\pa\mk{u}}^{(i)},\mb{M}_{\pa\mk{u}}^{(i')}\> =-\delta_{ii'} ,
\quad
\<\mb{M}_{\pa\mk{so}}^{(j)},\mb{M}_{\pa\mk{so}}^{(j')}\> =-\delta_{jj'} ,
\quad
\<\mb{m}_{\pe\Cnum}^{(k)},\mb{m}_{\pe\Cnum}^{(k')}\> =-\delta_{kk'} ,
\]
then the basis for $T_\map M=(T_\map M)_\pa\oplus (T_\map M)_\pe$
has the corresponding normalization
\[
g(X_{\pa\mk{u}}^{(i)},X_{\pa\mk{u}}^{(i')}) =\delta_{ii'} ,
\quad
g(X_{\pa\mk{so}}^{(j)},X_{\pa\mk{so}}^{(j')}) =\delta_{jj'} ,
\quad
g(X_{\pe\Cnum}^{(k)},X_{\pe\Cnum}^{(k')}) = g(X_{\pe\Cnum'}^{(k)},X_{\pe\Cnum'}^{(k')}) =\delta_{kk'} .
\]
Consequently, the resulting orthonormal frame
\beq
\{X_{\pa\mk{u}}^{(i)}, X_{\pa\mk{so}}^{(j)}, X_{\pe\Cnum}^{(k)}, X_{\pe\Cnum'}^{(k')}\}
\eeq
can be shown to satisfy the Frenet equations
\beq
\begin{aligned}
& \nabla_x X_{\pa\mk{u}}^{(i)} =
\sum_{k} U_{\mk{u},\Cnum}^{(i,k)} X_{\pe\Cnum}^{(k)}
+ \sum_{k'} U_{\mk{u},\Cnum'}^{(i,k')} X_{\pe\Cnum'}^{(k')}
\\
& \nabla_x X_{\pa\mk{so}}^{(j)} =
\sum_{k} U_{\mk{so},\Cnum}^{(j,k)} X_{\pe\Cnum}^{(k)}
+ \sum_{k'} U_{\mk{so},\Cnum'}^{(j,k')} X_{\pe\Cnum'}^{(k')}
\\
& \nabla_x X_{\pe\Cnum}^{(k)} =
-\sum_{i} U_{\mk{u},\Cnum}^{(i,k)} X_{\pa\mk{u}}^{(i)}
- \sum_{j} U_{\mk{so},\Cnum}^{(j,k)} X_{\pa\mk{so}}^{(j)}
\\
& \nabla_x X_{\pe\Cnum'}^{(k')} =
-\sum_{i} U_{\mk{u},\Cnum'}^{(i,k')} X_{\pa\mk{u}}^{(i)}
- \sum_{j} U_{\mk{so},\Cnum'}^{(j,k')} X_{\pa\mk{so}}^{(j)}
\end{aligned}
\eeq
obtained from the transport equation \eqref{transport}
combined with the Lie brackets \eqref{su.inclu.5} and \eqref{su.inclu.12},
where 
\beq
\begin{aligned}
& U_{\mk{u},\Cnum}^{(i,k)} 
=\<[(\mb{M}_{\pa\mk{u}}^{(i)},0),(\tbu_{1},\tbu_{2})],(\mb{m}_{\pe\Cnum}^{(k)},0)\>
= 8n \A((\tr\mb{M}_{\pa\mk{u}}^{(i)})\tbu_{1}+\tbu_{1}\mb{M}_{\pa\mk{u}}^{(i)},
\mb{m}_{\pe\Cnum}^{(k)})
\\
& U_{\mk{u},\Cnum'}^{(i,k')} 
=\<[(\mb{M}_{\pa\mk{u}}^{(i)},0),(\tbu_{1},\tbu_{2})],(0,\mb{m}_{\pe\Cnum}^{(k')})\>
=8n \A((\tr\mb{M}_{\pa\mk{u}}^{(i)})\tbu_{2}-\tbu_{2}\ov{\mb{M}}_{\pa\mk{u}}^{(i)},
\mb{m}_{\pe\Cnum}^{(k')})
\\
& U_{\mk{so},\Cnum}^{(j,k)} 
=\<[(0,\mb{M}_{\pa\mk{so}}^{(j)}),(\tbu_{1},\tbu_{2})],(\mb{m}_{\pe\Cnum}^{(k)},0)\>
=8n \A(\tbu_{2}\ov{\mb{M}}_{\pa\mk{so}}^{(j)},\mb{m}_{\pe\Cnum}^{(k)})
\\
& U_{\mk{so},\Cnum'}^{(j,k')} 
=\<[(0,\mb{M}_{\pa\mk{so}}^{(j)}),(\tbu_{1},\tbu_{2})],(0,\mb{m}_{\pe\Cnum}^{(k')})\>
=8n \A(\tbu_{1}\mb{M}_{\pa\mk{so}}^{(j)},\mb{m}_{\pe\Cnum}^{(k')})
\end{aligned}
\eeq
denote the Cartan matrix components of the underlying
$Sp(n)$-parallel linear connection \eqref{u.pe}
projected into the tangent space of the curve.

The geometrical meaning of this linear connection is seen through
looking at the frame components of the principal normal vector
\beq\label{prin.nor.vec}
N:=\nabla_x X=\<e^*,\ad(e_x)\conx_x\>
\eeq
given by
\beq
e\rfloor N=-\ad(e_x)\conx_x=
\frac{-1}{\sqrt{\rho}}(\i\tbu_{1},\i\tbu_{2})
\in\Cnum^{n-1}\oplus\Cnum^{n-1}\simeq \mk{m}_{\pe},\quad
\rho = 8(n-1)
\eeq
again using the relation \eqref{su.ad.e.h.pe}.
These components $(\i\tbu_{1},\i\tbu_{2})$ are invariantly defined by
the curve $\map$ up to the rigid ($x$-independent) action of
the equivalence group $H^*_{\pa}=\Ad(Sp(1)\times Sp(n-1))\subset \Ad(Sp(n))$
that preserves the framing at each point $x$.
Hence, in geometrical terms,
the complex vector pair $(\i\tbu_{1},\i\tbu_{2})$
describes a \emph{covariant} of the curve $\map$ relative to the group $H^*_{\pa}$. 
Moreover, $x$-derivatives of the pair $(\i\tbu_{1},\i\tbu_{2})$
describe \emph{differential covariants} of $\map$ relative to $H^*_{\pa}$,
which arise geometrically from the frame components of
$x$-derivatives of the principal normal vector $N$.
We thus note that the geometric invariants of $\map$
as defined by Riemannian inner products of the tangent vector $X=\map_x$
and its derivatives $N=\nabla_x\map_x,\nabla_xN=\nabla_x^2\map_x$, etc.\
along the curve $\map$
can be expressed as scalars formed from Cartan-Killing inner products of
the covariant $(\i\tbu_{1},\i\tbu_{2})$
and differential covariants $(\i\tbu_{1\,x},\i\tbu_{2\,x})$, $(\i\tbu_{1\,xx},\i\tbu_{2\,xx})$, etc.;
for example
\[
g(N,N)=-g(X,\nabla_x^2 X)
=\frac{2n}{n-1} (|\tbu_{1}|^2+|\tbu_{2}|^2)
\]
yields the square of the classical curvature invariant of the curve $\map$.
In particular,
the set of invariants given by
$\{g(X,\nabla_x^{2l}X)\}$, $l=1,\ldots,2n^2-n-2(=\dim \mk{m} -1)$, 
generates the components of the connection matrix of
a classical Frenet frame \cite{Guggenheimer} determined by $\map_x$.

Since $T_\map M$ has rank $n-1\geq 1$, 
if $n=2$ then all non-stretching curve flows belong to the same algebraic equivalence class, 
corresponding to the element \eqref{e},
while if $n>2$ then the element \eqref{e} determines 
one particular algebraic equivalence class of non-stretching curve flows.

\subsection{Hamiltonian operators and flows}
\label{SU.ops.hier}
The Cartan structure equations \eqref{pull.1} and \eqref{pull.2}
for the $Sp(n)$-parallel framing of $\map$
expressed in terms of the variables \eqref{e}--\eqref{w.pe}
are respectively given by
\beq\label{tors.w.pe.su}
\begin{aligned}
&
\frac{-\i}{\sqrt{\rho}}\tbw^{1\pe}=
D_x\tbh_{1\pe}+(\tr\tbHH_{1\pa})\tbu_{1}+\tbu_{1}\tbHH_{1\pa}+\tbu_{2}\btbHH_{2\pa},
\\
&
\frac{-\i}{\sqrt{\rho}}\tbw^{2\pe}=
D_x\tbh_{2\pe}+(\tr\tbHH_{1\pa})\tbu_{2}-\tbu_{2}\btbHH_{1\pa}+\tbu_{1}\tbHH_{2\pa},
\end{aligned}
\eeq
\beq\label{tors.h.pa.su}
\begin{aligned}
&
D_x\tbHH_{1\pa}=
\CC(\tbu_{1},\tbh_{1\pe})-\ov{\CC(\tbu_{2},\tbh_{2\pe})},
\\
&
D_x\tbHH_{2\pa}=
\AA(\tbu_{2},\tbh_{1\pe}) +\ov{\AA(\tbu_{1},\tbh_{2\pe})},
\end{aligned}
\eeq
and
\beq\label{curv.u.su}
\begin{aligned}
&
\tbu_{1\,t}=D_x\tbw^{1\pe}-\ww^{1\pa}\tbu_{1} +\ww^{2\pa}\btbu_{2}
+\tbu_{1}\tbWW^{1\pa} -\tbu_{2}\btbWW^{2\pa}+\frac{\i}{\sqrt{\rho}}\tbh_{1\pe},
\\
&
\tbu_{2\,t}=D_x\tbw^{2\pe}-\ww^{1\pa}\tbu_{2}-\ww^{2\pa}\btbu_{1}
+\tbu_{2}\btbWW^{1\pa}+\tbu_{1}\tbWW^{2\pa}+\frac{\i}{\sqrt{\rho}}\tbh_{2\pe},
\end{aligned}
\eeq
\beq\label{curv.w.pa.su}
\begin{aligned}
&
D_x\ww^{1\pa}=\C(\tbu_{1},\tbw^{1\pe})+\C(\tbu_{2},\tbw^{2\pe}),
\\
&
D_x\ww^{2\pa}=-\S(\tbu_{1},\tbw^{2\pe})+\S(\tbu_{2},\tbw^{1\pe}),
\\
&
D_x\tbWW^{1\pa}=\CC(\tbu_{1},\tbw^{1\pe})-\ov{\CC(\tbu_{2},\tbw^{2\pe})},
\\
&
D_x\tbWW^{2\pa}=\ov{\SS(\tbu_{1},\tbw^{2\pe})} -\ov{\SS(\tbu_{2},\tbw^{1\pe})}.
\end{aligned}
\eeq
As stated by Theorem~\ref{biHam.flow.eqn},
these equations \eqref{tors.w.pe.su}--\eqref{curv.w.pa.su} directly encode
a pair of compatible Hamiltonian operators.
To display the operators explicitly,
we first define the following operator notations in terms of
the inner products \eqref{com.vec}--\eqref{anti.conj.vec}
and outer products \eqref{com.not.mat}--\eqref{anti.conj.mat}. 
For $\mr{x}\in\Cnum$,
$\mb{x},\mb{y}\in\Cnum^{n-1}$,
$\mb{X}\in\mk{gl}(n-1,\Cnum)$,
let
\beq\label{CASinnerops}
\begin{aligned}
&
\C_{\mb{x}}\mb{y}:=\C(\mb{x},\mb{y})\in\i\Rnum ,
\\
&
\A_{\mb{x}}\mb{y}:= \A(\mb{x},\mb{y})\in\Rnum ,
\\
&
\S_{\mb{x}}\mb{y}:=\S(\mb{x},\mb{y})\in\Cnum ,
\end{aligned}
\eeq
\beq\label{CASouterops}
\begin{aligned}
& \CC_{\mb{x}}\mb{y}:=\CC(\mb{x},\mb{y})\in\mk{u}(n-1), 
\\
& 
\AA_{\mb{x}}\mb{y}:=\AA(\mb{x},\mb{y})\in\mk{so}(n-1,\Cnum), 
\\
&
\SS_{\mb{x}}\mb{y}:=\SS(\mb{x},\mb{y})\in\mk{s}(n-1,\Cnum), 
\end{aligned}
\eeq
and
\beq\label{Rop}
R_{\mb{y}}\mr{x}:=\mr{x}\mb{y}\in\Cnum^{n-1} ,
\eeq
\beq\label{Lop}
L_{\mb{y}}\mb{X}:=\mb{y}\mb{X}\in\Cnum^{n-1} ,
\eeq
\beq\label{ccop}
\cc\mb{X}:=\ov{\mb{X}} \in\Cnum^{n-1} .
\eeq
Next we eliminate $\tbHH_{1\pa}$, $\tbHH_{2\pa}$
through the torsion equation \eqref{tors.h.pa.su},
and also eliminate $\ww^{1\pa}$, $\ww^{2\pa}$, $\tbWW^{1\pa}$, $\tbWW^{2\pa}$
through the curvature equation \eqref{curv.w.pa.su}.
We also replace $\tbh_{1\pe}$, $\tbh_{2\pe}$ respectively in terms of
$\tbh^{1\pe}$, $\tbh^{2\pe}$ from equation \eqref{h.pe.up.su}, 
which leads to the following main result.

\begin{theorem}\label{g=su.theorem}
The flow equations given by \eqref{tors.w.pe.su}--\eqref{curv.w.pa.su}
for the pair of complex vector variables
$\tbu_{1}(t,x),\tbu_{2}(t,x)\in\Cnum^{n-1}$
have the operator form
\beq\label{g=su.flow.eq}
\bpm\tbu_{1}\\\tbu_{2}\epm _t
=\mathcal{H}\bpm\tbw^{1\pe}\\\tbw^{2\pe}\epm
+\bpm\tbh^{1\pe}\\\tbh^{2\pe}\epm ,
\quad
\bpm\tbw^{1\pe}\\\tbw^{2\pe}\epm
=8(n-1)\mathcal{J}\bpm\tbh^{1\pe}\\\tbh^{2\pe}\epm ,
\eeq
where
\beq\label{su.H.op}
\mathcal{H}=\bpm
\begin{aligned}
& D_x-R_{\tbu_{1}}D_x^{-1}\C_{\tbu_{1}}+R_{\btbu_{2}}D_x^{-1}\S_{\tbu_{2}}
\\& \quad
+L_{\tbu_{1}}D_x^{-1}\CC_{\tbu_{1}}+L_{\tbu_{2}}D_x^{-1}\SS_{\tbu_{2}}
\\&
\end{aligned}
&
\begin{aligned}
& -R_{\tbu_{1}}D_x^{-1}\C_{\tbu_{2}}-R_{\btbu_{2}}D_x^{-1}\S_{\tbu_{1}}
\\&\quad
-L_{\tbu_{1}}D_x^{-1}\cc\CC_{\tbu_{2}} -L_{\tbu_{2}}D_x^{-1}\SS_{\tbu_{1}}
\\&
\end{aligned}
\\
\begin{aligned}
& -R_{\tbu_{2}}D_x^{-1}\C_{\tbu_{1}}-R_{\btbu_{1}}D_x^{-1}\S_{\tbu_{2}}
\\&\quad
+L_{\tbu_{2}}D_x^{-1}\cc\CC_{\tbu_{1}}-L_{\tbu_{1}}D_x^{-1}\cc\SS_{\tbu_{2}}
\end{aligned}
&
\begin{aligned}
& D_x-R_{\tbu_{2}}D_x^{-1}\C_{\tbu_{2}} + R_{\btbu_{1}}D_x^{-1}\S_{\tbu_{1}}
\\&\quad
-L_{\tbu_{2}}D_x^{-1}\CC_{\tbu_{2}}+L_{\tbu_{1}}D_x^{-1}\cc\SS_{\tbu_{1}}
\end{aligned}
\epm
\eeq
and
\beq\label{su.J.op}
\mathcal{J}=\bpm
\begin{aligned}
& D_x + R_{\tbu_{1}}D_x^{-1}\A_{\tbu_{1}} +L_{\i\tbu_{2}}D_x^{-1}\cc\AA_{\i\tbu_{2}}
\\&\quad
+L_{\i\tbu_{1}}D_x^{-1}\CC_{\i\tbu_{1}}
\\&
\end{aligned}
&
\begin{aligned}
& R_{\tbu_{1}}D_x^{-1}\A_{\tbu_{2}} +L_{\i\tbu_{2}}D_x^{-1}\AA_{\i\tbu_{1}}
\\&\quad
-L_{\i\tbu_{1}}D_x^{-1}\cc\CC_{\i\tbu_{2}}
\\&
\end{aligned}
\\
\begin{aligned}
& R_{\tbu_{2}}D_x^{-1}\A_{\tbu_{1}}+L_{\i\tbu_{1}}D_x^{-1}\AA_{\i\tbu_{2}}
\\&\quad
-L_{\i\tbu_{2}}D_x^{-1}\cc\CC_{\i\tbu_{1}}
\end{aligned}
&
\begin{aligned}
& D_x +R_{\tbu_{2}}D_x^{-1}\A_{\tbu_{2}}+L_{\i\tbu_{1}}D_x^{-1}\cc\AA_{\i\tbu_{1}}
\\&\quad
+L_{\i\tbu_{2}}D_x^{-1}\CC_{\i\tbu_{2}}
\end{aligned}
\epm
\eeq
are compatible Hamiltonian cosymplectic and symplectic operators on the $x$-jet space of $(\tbu_{1},\tbu_{2})$.
\end{theorem}

We now explain some details about this Hamiltonian structure.
Let $J^\infty$ denote the $x$-jet space of the variables $(\tbu_{1},\tbu_{2})$,
and let subscripts $l,l'=1,2$ denote the $2\times 2$ components of
$\Hop$ and $\Jop$.

Associated to the operator $\Hop$ is the Poisson bracket
\beq\label{su.poisson}
\{\mk{H}_{1},\mk{H}_{2}\}_{\Hop} :=
\int \sum_{\substack{l=1,2\\l'=1,2}}
\A(\delta \mk{H}_{1}/\delta\tbu_{l}, \Hop_{ll'}(\delta \mk{H}_{2}/\delta\tbu_{l'})) dx
\eeq
where $\mk{H}_{1},\mk{H}_{2}$ are real-valued functionals on $J^\infty$.
The cosymplectic property of $\Hop$ means that this bracket is skew-symmetric 
\beq
\{\mk{H}_{1},\mk{H}_{2}\}_{\Hop} + \{\mk{H}_{2},\mk{H}_{1}\}_{\Hop} =0
\eeq
and obeys the Jacobi identity
\beq
\{\mk{H}_{1},\{\mk{H}_{2},\mk{H}_{3}\}_{\Hop}\}_{\Hop} + \text{ cyclic} =0. 
\eeq
A dual of the Poisson bracket is the symplectic $2$-form
associated to the operator $\Jop$, 
\beq\label{su.sympform}
\bs{\omega}(\X_{1},\X_{2})_{\Jop} :=
\int \sum_{\substack{l=1,2\\l'=1,2}}
\A( \X_{1}\tbu_{l}, \Jop_{ll'}(\X_{2}\tbu_{l'})) dx
\eeq
where $\X_{1},\X_{2}$ are vector fields
$\X=\tbh^{1\pe}\cdot \p/\p\tbu_{1} + \tbh^{2\pe}\cdot \p/\p\tbu_{2}$
defined in terms of vector function pairs 
$(\tbh^{1\perp},\tbh^{2\perp})\in\Cnum^{n-1}\oplus \Cnum^{n-1}$ on $J^\infty$
(with ``$\cdot$'' standing for summation with respect to vector components).
The symplectic property of $\Jop$ corresponds to $\bs{\omega}$ 
being skew-symmetric 
\beq\label{su.symp.skew}
\bs{\omega}(\X_{1},\X_{2}) +  \bs{\omega}(\X_{2},\X_{1}) =0
\eeq
and closed
\beq\label{su.symp.cyclic}
\begin{aligned}
& \pr(\X_{1})\bs{\omega}(\X_{2},\X_{3})+ \text{ cyclic}\\
&=
\int \sum_{\substack{l=1,2\\l'=1,2}}
\A( \tbh_{2}^{l\perp}, \pr(\sum_{l''=1,2}\tbh^{l''\perp}_{1}\cdot\p/\p\tbu_{l''})
\Jop_{ll'}(\tbh_3^{l\perp}) )dx+\text{ cyclic}
=0 .
\end{aligned}
\eeq
Compatibility of the operators $\Hop$ and $\Jop$ is the statement that
every linear combination $c_{1}\Hop+c_{2}\Jop^{-1}$ is a cosymplectic Hamiltonian operator,
or equivalently that $c_{1}\Hop^{-1}+c_{2}\Jop$ is a symplectic operator,
where $\Hop^{-1}$ and $\Jop^{-1}$ denote formal inverse operators
defined on $J^\infty$.

The following result is a consequence of Theorem~\ref{biHam.hier}.

\begin{corollary}\label{g=su.hier}
The operator $\Rop=\Hop\Jop$ generates a hierarchy of
bi-Hamiltonian flows \eqref{g=su.flow.eq} on $(\tbu_{1}(t,x),\tbu_{2}(t,x))$,
given by
\beq\label{g=su.h.pe.hier}
\bpm\tbh_{(k)}^{1\pe}\\\tbh_{(k)}^{2\pe}\epm
=\Rop^k\bpm\tbu_{1\,x}\\\tbu_{2\,x}\epm ,
\quad
k=0,1,2,\ldots
\eeq
and
\beq\label{g=su.w.pe.hier}
\bpm\tbw_{(k)}^{1\pe}\\\tbw_{(k)}^{2\pe}\epm
= \bpm\delta H^{(k)}/\delta\tbu_{1}\\\delta H^{(k)}/\delta\tbu_{2}\epm
=\Rop^{*k}\bpm\tbu_{1}\\\tbu_{2}\epm ,
\quad
k=0,1,2,\ldots
\eeq
in terms of the Hamiltonians
\beq\label{g=su.H.hier}
H^{(k)}=\frac{1}{1+2k}\tr(\i\tbHH_{1\pa}^{(k)}), 
\quad
k=0,1,2,\ldots 
\eeq
with
\beq
\tr(\i\tbHH_{1\pa}^{(k)}) 
= D_x^{-1}\big(A(\tbu_{1},\tbh_{(k)}^{1\pe})+ A(\tbu_{2},\tbh_{(k)}^{2\pe})\big),
\eeq
where the operator $\Rop^*=\Jop\Hop$ is the adjoint of $\Rop$.
\end{corollary}

The $+k$ flow in this hierarchy \eqref{g=su.h.pe.hier} is scaling invariant 
under $(\tbu_{1},\tbu_{2}) \rightarrow \lambda^{-1} (\tbu_{1},\tbu_{2})$,
$x\rightarrow \lambda x$, $t\rightarrow \lambda^{1+2k}t$.

\subsection{mKdV flow}
\label{SU.mkdv.eqns}
After a scaling of $t\rightarrow t/\rho$, where $\rho=8(n-1)$,
the $+1$ flow in the hierarchy \eqref{g=su.h.pe.hier}
yields an integrable system of coupled vector mKdV equations
\beq\label{g=su.mkdv}
\begin{aligned}
\tbu_{1\,t}-\rho^{-1}\tbu_{1\,x}  & =
\tbu_{1\,xxx}+3(\tbu_{1}\cdot\btbu_{1} +\tbu_{2}\cdot\btbu_{2})\tbu_{1\,x}
+3(\tbu_{2}\cdot\tbu_{1\,x}-\tbu_{2\,x}\cdot\tbu_{1})\btbu_{2}\\&\qquad
+3(\tbu_{1\,x}\cdot\btbu_{1} +\tbu_{2\,x}\cdot\btbu_{2})\tbu_{1}
\\
\tbu_{2\,t}-\rho^{-1}\tbu_{2\,x} & =
\tbu_{2\,xxx}+3(\tbu_{1}\cdot\btbu_{1} +\tbu_{2}\cdot\btbu_{2})\tbu_{2\,x}
-3(\tbu_{2}\cdot\tbu_{1\,x}-\tbu_{2\,x}\cdot\tbu_{1})\btbu_{1}\\&\qquad
+3(\tbu_{1\,x}\cdot\btbu_{1} +\tbu_{2\,x}\cdot\btbu_{2})\tbu_{2}
\end{aligned}
\eeq
where a dot denotes the standard Euclidean inner product (cf \eqref{Herm.inner.prod}--\eqref{Eucl.inner.prod}).
This system is invariant under the symplectic group $Sp(1)\times Sp(n-1)$,
defined by the transformations \eqref{SP-n.action.su}--\eqref{SP-1.action.su}
on the vector pair $(\tbu_{1},\tbu_{2})$,
and has the following bi-Hamiltonian structure
\beq\label{g=su.mkdv.Ham}
\bpm\tbu_{1}\\\tbu_{2}\epm _t
- \rho^{-1}\bpm\tbu_{1}\\\tbu_{2}\epm _x
=\mathcal{H}\bpm\delta H^{(1)}/\delta\tbu_{1}\\\delta H^{(1)}/\delta\tbu_{2}\epm
= \mathcal{E}\bpm\delta H^{(0)}/\delta\tbu_{1}\\\delta H^{(0)}/\delta\tbu_{2}\epm
\eeq
in terms of the Hamiltonians
\begin{align}
& H^{(0)} = 
\tbu_{1}\cdot\btbu_{1} + \tbu_{2}\cdot\btbu_{2},
\\
& H^{(1)} = 
-\tbu_{1\,x}\cdot\btbu_{1\,x} - \tbu_{2\,x}\cdot\btbu_{2\,x}
+ (\tbu_{1}\cdot\btbu_{1}+\tbu_{2}\cdot\btbu_{2})^2 ,
\end{align}
where $\Eop=\Hop\Jop\Hop$ is a Hamiltonian cosymplectic operator
compatible with $\Hop$. 

We remark that the convective terms $\tbu_{1\,x}$, $\tbu_{2\,x}$
on the left-hand side in the system \eqref{g=su.mkdv} and \eqref{g=su.mkdv.Ham}
can be removed by the Galilean transformation
$t\rightarrow t$, $x\rightarrow x+\rho\inv t$.

\subsection{SG flow}
\label{SU.sg.eqns}
The $-1$ flow connected with the hierarchy \eqref{g=su.h.pe.hier}
is defined by
\beq\label{g=su.-1flow}
0=\bpm\tbw^{1\pe}\\\tbw^{2\pe}\epm
=\mathcal{J}\bpm\tbh^{1\pe}\\\tbh^{2\pe}\epm ,
\eeq
yielding the flow equation
\beq\label{g=su.sg}
\bpm\tbu_{1}\\\tbu_{2}\epm_{t}
=\bpm\tbh^{1\pe}\\\tbh^{2\pe}\epm
\eeq
with
\beq\label{su.sg.h.pe}
\begin{aligned}
& \i\sqrt{\rho}D_x\tbh^{1\pe}=
(\tr\tbHH_{1\pa})\tbu_{1}+\tbu_{1}\tbHH_{1\pa}+\tbu_{2}\btbHH_{2\pa},
\\
& \i\sqrt{\rho}D_x\tbh^{2\pe}=
(\tr\tbHH_{1\pa})\tbu_{2}-\tbu_{2}\btbHH_{1\pa}+\tbu_{1}\tbHH_{2\pa} , 
\end{aligned}
\eeq
and
\beq\label{su.sg.H.pa}
\begin{aligned}
& D_x\tbHH_{1\pa}=\sqrt{\rho}\big(\CC(\i\tbh^{1\pe},\tbu_{1})+\CC(\i\tbh^{2\pe},\tbu_{2})^\t\big),
\\
& D_x\tbHH_{2\pa}= \sqrt{\rho}\big(-\AA(\tbu_{2},\i\tbh^{1\pe})+\AA(\i\tbh^{2\pe},\tbu_{1})\big) .
\end{aligned}
\eeq
Note that equations \eqref{su.sg.h.pe}--\eqref{su.sg.H.pa} will determine 
the variables $\tbh^{1\pe}$, $\tbh^{2\pe}$, $\tbHH_{1\pa}$, $\tbHH_{2\pa}$
as nonlocal functions of $\tbu_{1}$, $\tbu_{2}$. 
Similarly to the method used to derive the SG flow in the case $SU(n)/SO(n)$ 
\cite{AncoSIGMA},
we will seek inverse local expressions for $\tbu_{1}$ and $\tbu_{2}$ 
arising from an algebraic reduction of the form
\begin{align}
& \tbHH_{1\pa} 
= \ha\alpha \big(\CC(\tbh^{1\pe},\i\tbh^{1\pe}) + \CC(\tbh^{2\pe},\i\tbh^{2\pe})^\t\big) +\beta\i I_{n-1} 
\label{su.red.H1.pa}\\
& \tbHH_{2\pa} 
= \map \AA(\tbh^{2\pe},\i\tbh^{1\pe})
\label{su.red.H2.pa}
\end{align}
for some expressions 
$\alpha(\mr{h}_{\pa}),\beta(\mr{h}_{\pa}),\map(\mr{h}_{\pa})\in\Rnum$, 
where it is convenient to introduce the variable 
\beq
\mr{h}_{\pa} := -\i\tr\tbHH_{1\pa}
\eeq
satisfying
\beq\label{su.sg.h.pa}
D_x\mr{h}_{\pa} =
-\sqrt{\rho}\big(\A(\tbu_{1},\tbh^{1\pe})+ \A(\tbu_{2},\tbh^{2\pe})\big) .
\eeq 

To proceed, we substitute expressions \eqref{su.red.H1.pa} and \eqref{su.red.H2.pa} into equation \eqref{su.sg.H.pa}
and use equations \eqref{su.sg.h.pe} and \eqref{su.sg.h.pa} to eliminate $x$ derivatives. 
This yields
\begin{align}
\map-\alpha=0
\label{gamma.eq}\\
D_x\beta =0 
\label{beta.eq}\\
D_x \alpha +(\alpha^2/\sqrt{\rho})\big(\A(\tbu_{1},\tbh^{1\pe})+ \A(\tbu_{2},\tbh^{2\pe})\big) =0
\label{Dx.alpha.eq}\\
\ha\big(\A(\tbh^{1\pe},\tbh^{1\pe})+ \A(\tbh^{2\pe},\tbh^{2\pe})\big) \alpha^2 
+ n\beta\alpha +\rho =0
\label{alpha.eq}
\end{align}
By applying $D_x$ to equation \eqref{alpha.eq} and using equation \eqref{beta.eq} together with equation \eqref{su.sg.h.pe}, we obtain equation \eqref{Dx.alpha.eq}. 
Therefore, we can just algebraically solve equation \eqref{alpha.eq} to get 
\beq\label{alpha.gamma}
\alpha=\map
= \frac{-n\beta\pm\sqrt{ n^2\beta^2 -4\rho(|\tbh^{1\pe}|^2+ |\tbh^{2\pe}|^2) }}{2(|\tbh^{1\pe}|^2+ |\tbh^{2\pe}|^2)}
\eeq
where
\beq
|\tbh^{1\pe}|^2 := \ha \A(\tbh^{1\pe},\tbh^{1\pe}) ,
\quad
|\tbh^{2\pe}|^2 := \ha \A(\tbh^{2\pe},\tbh^{2\pe}) .
\eeq
To determine $\beta$
we use the conservation law 
\beq
0 =D_{x}\big( |\tbh^{1\pe}|^2+ |\tbh^{2\pe}|^2 +\frac{1}{\rho}( \mr{h}_{\pa}^2 +|\tbHH_{1\pa}|^2 +|\tbHH_{2\pa}|^2 ) \big)
\label{su.sg.conslaw}
\eeq
admitted by the system of equations \eqref{su.sg.h.pe}, \eqref{su.sg.H.pa}, \eqref{su.sg.h.pa}, 
where
\beq
\mr{h}_{\pa} = \alpha( |\tbh^{1\pe}|^2+ |\tbh^{2\pe}|^2 ) +(n-1)\beta
\label{su.red.h.pa}
\eeq
and 
\beq\label{su.H.pa.norm}
\begin{aligned}
|\tbHH_{1\pa}|^2 := -\tr( \tbHH_{1\pa}^2 )
& = \alpha^2 ( |\tbh^{1\pe}|^4 + |\tbh^{2\pe}|^4 + \ha\S(\tbh^{1\pe},\tbh^{2\pe})\S(\btbh^{1\pe},\btbh^{2\pe}) ) 
\\&\qquad
+2\alpha \beta( |\tbh^{1\pe}|^2+ |\tbh^{2\pe}|^2 ) +\beta^2 (n-1) 
\\
|\tbHH_{2\pa}|^2 := -\tr( \btbHH_{2\pa}\tbHH_{2\pa} )
& = \alpha^2 ( |\tbh^{1\pe}|^2 |\tbh^{2\pe}|^2 - \ha\S(\tbh^{1\pe},\tbh^{2\pe})\S(\btbh^{1\pe},\btbh^{2\pe}) ) 
\end{aligned}
\eeq
are obtained from equations \eqref{su.red.H1.pa}--\eqref{su.red.h.pa}. 
Substitution of the expressions \eqref{su.red.h.pa} and \eqref{su.H.pa.norm} 
into the conservation law \eqref{su.sg.conslaw},
followed by use of the algebraic equation \eqref{alpha.eq}, 
gives 
\beq
|\tbh^{1\pe}|^2+ |\tbh^{2\pe}|^2 +\frac{1}{\rho}\big( \mr{h}_{\pa}^2 +|\tbHH_{1\pa}|^2 +|\tbHH_{2\pa}|^2 \big)
= \ha n(n-1)\beta^2/\rho = n(\beta/4)^2 . 
\label{beta.const}
\eeq
Through equations \eqref{beta.const} and \eqref{su.sg.conslaw}, 
we see that a conformal scaling of $t$ can be used to make $\beta$ equal to a constant.
We will put 
\beq
\beta = - 2\sqrt{\rho}/n
\eeq
which simplifies the expression \eqref{alpha.gamma} for $\alpha$ and $\map$,
\beq\label{alpha.gamma.normalized}
\alpha=\map
= \sqrt{\rho}\;
 \frac{1\pm\sqrt{ 1 -|\tbh^{1\pe}|^2 -|\tbh^{2\pe}|^2 }}{|\tbh^{1\pe}|^2+ |\tbh^{2\pe}|^2} . 
\eeq

Local expressions for $\tbu_{1}$ and $\tbu_{2}$ now arise directly 
from substitution of expressions \eqref{su.red.H1.pa}, \eqref{su.red.H2.pa}, \eqref{su.red.h.pa} 
into equation \eqref{su.sg.h.pe} to get 
\begin{align}
& \tbh^{1\pe}_{x} = 
-\frac{\sqrt{\rho}}{\alpha} \tbu_{1}
+\frac{\alpha}{\sqrt{\rho}}( \S(\tbu_{1},\btbh^{1\pe}) + \S(\tbu_{2},\btbh^{2\pe}) ) \tbh^{1\pe}
+\frac{\alpha}{2\sqrt{\rho}}( \S(\tbu_{1},\tbh^{2\pe}) - \S(\tbu_{2},\tbh^{1\pe}) ) \btbh^{2\pe} , 
\label{su.sg.h1x.pe}\\
& \tbh^{2\pe}_{x} =
-\frac{\sqrt{\rho}}{\alpha} \tbu_{2}
+\frac{\alpha}{\sqrt{\rho}}( \S(\tbu_{1},\btbh^{1\pe}) + \S(\tbu_{2},\btbh^{2\pe}) ) \tbh^{2\pe}
-\frac{\alpha}{2\sqrt{\rho}}( \S(\tbu_{1},\tbh^{2\pe}) - \S(\tbu_{2},\tbh^{1\pe}) ) \btbh^{1\pe} \big) .
\label{su.sg.h2x.pe}
\end{align}
Algebraically combining equations \eqref{su.sg.h1x.pe} and \eqref{su.sg.h2x.pe}, we obtain 
\beq\label{su.sg.u.rels}
\begin{aligned}
& \tbu_{1}= 
-\frac{\alpha}{\sqrt{\rho}}\tbh^{1\pe}_{x} 
+ \frac{\alpha^2}{\rho}(a \tbh^{1\pe} + b \btbh^{2\pe}) , 
\\
& \tbu_{2}= 
-\frac{\alpha}{\sqrt{\rho}}\tbh^{2\pe}_{x} 
+ \frac{\alpha^2}{\rho}(a \tbh^{2\pe} - b \btbh^{1\pe}) ,
\end{aligned}
\eeq
where, after using expression \eqref{alpha.gamma.normalized}, we have
\beq\label{su.sg.u.rels2}
a = 
\frac{ \btbh^{1\pe}\cdot\tbh^{1\pe}_{x} + \btbh^{2\pe}\cdot\tbh^{2\pe}_{x} }
{ \pm 2\sqrt{1 -|\tbh^{1\pe}|^2 -|\tbh^{2\pe}|^2} } ,
\quad
 b = 
\frac{ \tbh^{2\pe}\cdot\tbh^{1\pe}_{x} - \tbh^{1\pe}\cdot\tbh^{2\pe}_{x} }
{ \pm 2\sqrt{1 -|\tbh^{1\pe}|^2 -|\tbh^{2\pe}|^2} } ,
\eeq
with a dot denoting the standard Euclidean inner product (cf \eqref{Herm.inner.prod}--\eqref{Eucl.inner.prod}).

Finally, we express the flow equation \eqref{g=su.sg} 
entirely in terms of $\tbu_{1}$, $\tbu_{2}$, and their $t$ derivatives. 
Substitution of $\tbh^{1\pe}= \tbu_{1\,t}$ and $\tbh^{2\pe}= \tbu_{2\,t}$ 
into equations \eqref{alpha.gamma.normalized}--\eqref{su.sg.h2x.pe} 
directly yields the nonlocal evolution equation 
\beq\label{su.-1floweq}
\bpm\tbu_{1}\\\tbu_{2}\epm_{t}
=D_x^{-1}\bpm 
-\dfrac{\sqrt{\rho}}{\alpha} \tbu_{1}
+\dfrac{\alpha}{\sqrt{\rho}}\big( (\tbu_{1}\cdot\btbu_{1\,t} +\tbu_{2}\cdot\btbu_{2\,t})\tbu_{1\,t}
+ (\tbu_{1}\cdot\tbu_{2\,t} -\tbu_{2}\cdot\tbu_{1\,t}) \btbu_{2\,t} \big) \\
-\dfrac{\sqrt{\rho}}{\alpha} \tbu_{2}
+\dfrac{\alpha}{\sqrt{\rho}}\big( (\tbu_{1}\cdot\btbu_{1\,t} +\tbu_{2}\cdot\btbu_{2\,t})\tbu_{2\,t}
-(\tbu_{1}\cdot\tbu_{2\,t} -\tbu_{2}\cdot\tbu_{1\,t}) \btbu_{1\,t} \big) 
\epm
\eeq
with 
\beq\label{su.sgflow.alpha}
\frac{\alpha}{\sqrt{\rho}}
= \frac{1\pm\sqrt{ 1 -|\tbu_{1\,t}|^2 - |\tbu_{2\,t}|^2 }}{|\tbu_{1\,t}|^2+ |\tbu_{2\,t}|^2} 
\eeq
and 
\beq\label{su.sgflow.alpha.inv}
\frac{\sqrt{\rho}}{\alpha}
= 1\mp\sqrt{ 1 -|\tbu_{1\,t}|^2 - |\tbu_{2\,t}|^2 } .
\eeq
This $-1$ flow equation \eqref{su.-1floweq} is equivalent to 
a hyperbolic system of coupled vector SG equations
\beq\label{su.sgeq}
\begin{aligned}
\tbu_{1\,tx} = &
\frac{1\pm\sqrt{ 1 -|\tbu_{1\,t}|^2 - |\tbu_{2\,t}|^2 }}{|\tbu_{1\,t}|^2+ |\tbu_{2\,t}|^2} 
\big( (\tbu_{1}\cdot\btbu_{1\,t} +\tbu_{2}\cdot\btbu_{2\,t}) \tbu_{1\,t} 
+ ( \tbu_{1}\cdot\tbu_{2\,t} -\tbu_{2}\cdot\tbu_{1\,t} ) \btbu_{2\,t} \big) \\&\qquad
-\big( 1\mp\sqrt{ 1 -|\tbu_{1\,t}|^2 - |\tbu_{2\,t}|^2 } \big)
 \tbu_{1}
\\
\tbu_{2\,tx} = &
\frac{1\pm\sqrt{ 1 -|\tbu_{1\,t}|^2 - |\tbu_{2\,t}|^2 }}{|\tbu_{1\,t}|^2+ |\tbu_{2\,t}|^2} 
\big( (\tbu_{1}\cdot\btbu_{1\,t} +\tbu_{2}\cdot\btbu_{2\,t}) \tbu_{2\,t}  
- ( \tbu_{1}\cdot\tbu_{2\,t}-\tbu_{2}\cdot\tbu_{1\,t} ) \btbu_{1\,t} \big) \\&\qquad
-\big( 1\mp\sqrt{ 1 -|\tbu_{1\,t}|^2 - |\tbu_{2\,t}|^2 } \big)
 \tbu_{2}
\end{aligned}
\eeq
which is invariant under the symplectic group $Sp(1)\times Sp(n-1)$,
defined by the transformations \eqref{SP-n.action.su}--\eqref{SP-1.action.su}
on the vector pair $(\tbu_{1},\tbu_{2})$. 

Alternatively, from the flow equation \eqref{g=su.sg} combined with
the relations \eqref{su.sg.u.rels}--\eqref{su.sg.u.rels2}, 
the variables $\tbh^{1\pe}$ and $\tbh^{2\pe}$
are found to obey coupled vector SG equations
\beq\label{su.h.pe.sgeq}
\begin{aligned}
& \big( -\frac{\alpha}{\sqrt{\rho}}\tbh^{1\pe}_{x} 
+ \frac{\alpha^2}{\rho}(a \tbh^{1\pe} + b \btbh^{2\pe}) \big)_{t} = \tbh^{1\pe},  
\\
& \big( -\frac{\alpha}{\sqrt{\rho}}\tbh^{2\pe}_{x} 
+ \frac{\alpha^2}{\rho}(a \tbh^{2\pe} - b \btbh^{1\pe}) \big)_{t} = \tbh^{2\pe},
\end{aligned}
\eeq
with 
\beq\label{su.sg.u.rels3}
\frac{\alpha}{\sqrt{\rho}} = 
\frac{1\pm\sqrt{ 1 -|\tbh^{1\pe}|^2 -|\tbh^{2\pe}|^2 }}{|\tbh^{1\pe}|^2+ |\tbh^{2\pe}|^2} . 
\eeq

A Hamiltonian structure for the system \eqref{su.sgeq} is given by 
\beq\label{g=su.sg.Ham}
\bpm\tbu_{1}\\\tbu_{2}\epm _t
=\mathcal{H}\bpm\delta H^{(-1)}/\delta\tbu_{1}\\\delta H^{(-1)}/\delta\tbu_{2}\epm
\eeq
in terms of 
\beq
H^{(-1)} = \pm 8\sqrt{1 -|\tbh^{1\pe}|^2 -|\tbh^{2\pe}|^2}
\eeq
where $\tbh^{1\pe}$ and $\tbh^{2\pe}$ are implicitly determined as 
nonlocal functions of the variables $(\tbu_{1},\tbu_{2})$ 
(and their $x$-derivatives) through expressions
\eqref{su.sg.u.rels}, \eqref{su.sg.u.rels2}, \eqref{su.sg.u.rels3}.

\subsection{Geometric curve flows}
\label{SU.curve.flows}
From Theorem~\ref{geom.curve.flows}, 
the flows in the hierarchy \eqref{g=su.h.pe.hier} and \eqref{g=su.-1flow}
for $(\tbu_{1}(t,x),\tbu_{2}(t,x))$ $\in\Cnum^{n-1}\oplus\Cnum^{n-1}$
correspond to $SU(2n)$-invariant non-stretching geometric curve flows for 
$\map(t,x)$ $\in M=SU(2n)/Sp(n)$.
The resulting equations of motion can be expressed covariantly in terms of 
$X=\map_x$, $N=\nabla_x\map_x$, and $\nabla_x$-derivatives of $N$,
in addition to the Riemannian metric and curvature tensors on $M$. 

The SG flow \eqref{su.sgeq} is given by $\tbw^{1\pe}=\tbw^{2\pe}=0$, 
which implies $\ww^{1\pa}=\ww^{2\pa}=0$ and $\tbWW^{1\pa}=\tbWW^{2\pa}=0$ 
as a consequence of the structure equation \eqref{curv.w.pa.su}. 
This determines
\beq
\varpi^{\pe}=\varpi^{\pa}=0 . 
\eeq
Hence the corresponding flow vector $\map_t=Y_{(-1)}$ satisfies 
\beq
e\rfloor \nabla_x \map_t
= D_x e_t + [\conx_x,e_t] = [\conx_t,e_x] 
=-\ad(e_x)\varpi^{\pe}
=0
\eeq
yielding the $SU(2n)$-invariant curve flow equation
\beq\label{g=su.wavemap}
0=\nabla_t\map_x, \quad 
|\map_x|=1, 
\eeq
which is called the {\it non-stretching wave map} on $M=SU(2n)/Sp(n)$.
In addition to satisfying the non-stretching property $\nabla_t|\map_x|=0$, 
this equation \eqref{g=su.wavemap} possesses the conservation law 
$\nabla_x|\map_t|=0$, 
corresponding to equation \eqref{su.sg.conslaw}. 
Thus, up to a conformal scaling of $t$, 
the wave map equation describes a flow with unit speed, $|\map_t|=1$. 

The mKdV flow \eqref{g=su.mkdv}, after $t$ has been rescaled, 
is given by 
$\tbh_{1\pe} =-\i\sqrt{\rho} \tbu_{1\,x}$
and $\tbh_{2\pe} =-\i\sqrt{\rho} \tbu_{2\,x}$,
from which 
$\tbHH_{1\pa} = -\i\sqrt{\rho}( \btbu_{1}^\t\tbu_{1} + \btbu_{2}^\t\tbu_{2} )$
and 
$\tbHH_{2\pa} = -\i\sqrt{\rho}( \btbu_{1}^\t\tbu_{2} - \tbu_{2}^\t\btbu_{1} )$
are obtained by the structure equation \eqref{tors.h.pa.su}. 
This determines
\begin{align}
& (e_t)_{\pe} = 
-\sqrt{\rho} ( \i\tbu_{1\,x}, \i\tbu_{2\,x} ) \in\mk{m}_{\pe} , 
\label{su.et.pe}\\
& (e_t)_{\pa} = 
\sqrt{\rho} ( \ha\CC(\i\tbu_{1},\tbu_{1}) -\ha\ov{\CC(\i\tbu_{2},\tbu_{2})}, 
\AA(\i\tbu_{2},\tbu_{1}) ) \in\mk{m}_{\pa} . 
\label{su.et.pa}
\end{align}
Then $e_t=e\rfloor\map_t$ can be expressed as follows in terms of
\begin{align}
& e\rfloor N =[\conx_x,e_x] 
= -\frac{1}{\sqrt{\rho}} ( \i\tbu_{1}, \i\tbu_{2} ) \in\mk{m}_{\pe}, 
\\
& (e\rfloor \nabla_x N)_{\pe} =(e\rfloor N)_x
= -\frac{1}{\sqrt{\rho}} ( \i\tbu_{1\,x}, \i\tbu_{2\,x} ) \in\mk{m}_{\pe}, 
\label{su.gradN.pe}\\
& (e\rfloor \nabla_x N)_{\pa} =[\conx_x,e\rfloor N] 
= -\frac{1}{\sqrt{\rho}} ( \CC(\i\tbu_{1},\tbu_{1}) -\ov{\CC(\i\tbu_{2},\tbu_{2})},2\AA(\i\tbu_{2},\tbu_{1}) ) \in\mk{m}_{\pa}.
\label{su.gradN.pa}
\end{align}
Consider
\beq
\ad(e\rfloor N) e_x
= -\frac{1}{\rho} ( \tbu_{1}, \tbu_{2} ) \in\mk{h}_{\pe}
\eeq
which leads to 
\beq\label{su.adsq.ex}
\ad(e\rfloor N)^2 e_x
= -\frac{1}{\sqrt{\rho}^3} ( \CC(\i\tbu_{1},\tbu_{1}) -\ov{\CC(\i\tbu_{2},\tbu_{2})},2\AA(\i\tbu_{2},\tbu_{1}) ) \in\mk{m}_{\pa}
\eeq
by means of the Lie brackets \eqref{su.inclu.4} and \eqref{su.inclu.12}.
Comparing equations \eqref{su.et.pe}--\eqref{su.et.pa} with equations \eqref{su.gradN.pe}--\eqref{su.gradN.pa} and \eqref{su.adsq.ex}, 
we see that 
\begin{align}
& (e_t)_{\pe} - 2(e_t)_{\pa} = \rho e\rfloor \nabla_x N , 
\\
& 2(e_t)_{\pa} = -\rho \ad(e\rfloor N)^2 e_x . 
\end{align}
This yields
\beq
e\rfloor\map_t = (e_t)_{\pe} + (e_t)_{\pa} 
= \rho e\rfloor \nabla_x N  -\frac{3}{2}\rho^2 e\rfloor \ad(N)^2 X
\eeq
where 
\beq
\ad(N)^2 = -R(\cdot,N)N
\eeq
is a linear map on $T_\map M$. 
Hence the flow vector $\map_t=Y_{(1)}$ satisfies 
\beq\label{g=su.mkdvmap}
\map_t = \nabla_x^2\map_x -\rho \frac{3}{2}\ad(\nabla_x \map_x)^2 \map_x, \quad 
|\map_x|=1, 
\eeq
which is a $SU(2n)$-invariant curve flow equation
called the {\it non-stretching mKdV map} on $M=SU(2n)/Sp(n)$.
The simple form of the nonlinearities in this equation is due to 
the algebraic property that $\ad(e_x)^2$ is a multiple of the identity 
on the vector spaces $\mk{m}_{\pe}\simeq\mk{h}_{\pe}$, 
as explained by the general results in \Ref{AncoJGP}.

\section{Bi-Hamiltonian soliton equations in $Sp(n+1)/Sp(1)\times Sp(n)$}
\label{Sp.curveflows}

Employing the notation and preliminaries
in \secref{construction} and \secref{g=sp},
we consider a non-stretching curve flow $\map(t,x)$
in $M=Sp(n+1)/Sp(1)\times Sp(n)$
having a $Sp(n)\times Sp(1)$-parallel framing 
as expressed in terms of the variables
\begin{align}
&
e_x=\frac{1}{\sqrt{\chi}}(1)\in\Rnum\simeq\mk{m}_{\pa},\quad \chi=8(n+2)
\label{e.sp}\\
&
\conx_x=((\u_1,\u_2),(\tbu_{1},\tbu_{2}))
\in\i\Rnum\oplus\Cnum\oplus\Cnum^{n-1}\oplus\Cnum^{n-1}\simeq \mk{h}_{\pe},
\label{u.pe.sp}
\end{align}
and
\begin{align}
&
h_{\pa}=(\hh_{\pa})
\in\Rnum\simeq\mk{m}_{\pa},
\label{h.pa.sp}\\
&
h_{\pe}=((\hh_{1\pe},\hh_{2\pe}),(\tbh_{1\pe},\tbh_{2\pe}))
\in\i\Rnum\oplus\Cnum\oplus\Cnum^{n-1}\oplus\Cnum^{n-1}\simeq \mk{m}_{\pe},
\label{h.pe.sp}\\
&\varpi^{\pa}=((\ww^{1\pa},\ww^{2\pa}),(\tbWW^{1\pa},\tbWW^{2\pa}))
\in\mk{sp}(1)\oplus\mk{sp}(n-1)\simeq \mk{h}_{\pa},
\label{w.pa.sp}\\
& \varpi^{\pe}=((\ww^{1\pe},\ww^{2\pe}),(\tbw^{1\pe},\tbw^{2\pe}))
\in\Rnum\oplus\Cnum\oplus\Cnum^{n-1}\oplus\Cnum^{n-1}\simeq \mk{h}_{\pe},
\label{w.pe.sp}
\end{align}
using the matrix identifications \eqref{m.pa.pe.sp}--\eqref{h.pa.pe.sp},
where $\hh_{1\pa}$ $\in\Rnum$ is a real variable, 
$\ww^{1\pe}, \ww^{1\pa},\hh_{1\pe}$ $\in\i\Rnum$ are imaginary (complex) scalar variables,
$\ww^{2\pe}, \ww^{2\pa},\hh_{2\pe}\in\Cnum$ are complex scalar variables,
$\tbu_{1},\tbu_{2},\tbw^{1\pe},\tbw^{2\pe},\tbh_{1\pe},\tbh_{2\pe}$ $\in\Cnum^{n-1}$ are complex vector variables,
$\tbWW^{1\pa}$ $\in\mk{u}(n-1)$ is an anti-Hermitian matrix variable, 
and $\tbWW^{2\pa}$ $\in\mk{s}(n-1,\Cnum)$ is a complex symmetric matrix variable.
For later use, through properties \eqref{ad.e.m.pe.sp}--\eqref{ad.e.h.pe.sp}
we also introduce the variable
\begin{align}
h^{\pe}&=((\hh^{1\pe},\hh^{2\pe}),(\tbh^{1\pe},\tbh^{2\pe}))= \ad(e_x)h_{\pe}\nonumber\\
&=\frac{1}{\sqrt{\chi}}((2\hh_{1\pe},2\hh_{2\pe}),(-\tbh_{1\pe},-\tbh_{2\pe}))
\in\i\Rnum\oplus\Cnum\oplus\Cnum^{n-1}\oplus\Cnum^{n-1}\simeq \mk{h}_{\pe},\quad 
\chi=8(n+2),
\label{h.pe.up.sp}
\end{align}
where $\hh^{1\pe}$ $\in\i\Rnum$ is an imaginary (complex) scalar variable, 
$\hh^{2\pe}$ $\in\Cnum$ is a complex scalar variable, 
and $\tbh^{1\pe},\tbh^{2\pe}$ $\in\Cnum^{n-1}$ are complex vector variables.

Up to the rigid ($x$-independent) action of the equivalence group
$H^*_{\pa} = \Ad(Sp(1)\times Sp(n-1)) \subset \Ad(Sp(n)\times Sp(1))$,
a $Sp(n)\times Sp(1)$-parallel linear coframe $e$ along $\map$ 
is then determined by the variables \eqref{e.sp} and \eqref{u.pe.sp}
via the transport equation
\beq\label{transport.sp}
\nabla_x e=-\ad(\conx_x)e
\eeq
together with the soldering relation
\beq
e\rfloor \map_x = e_x .
\eeq
The resulting coframe $e$ defines an isomorphism between
$T_\map M$ and $\mk{m}\simeq \Rnum\oplus\i\Rnum\oplus\Cnum\oplus\Cnum^{n-1}\oplus\Cnum^{n-1}$, 
which yields the following correspondence between 
the set of frames for $T_\map M$ and the set of bases for $\mk{m}$. 
Let $e_\pa$ and $e_\pe$ be the respective projections of $e$ into
$\mk{m}_{\pa}$ and $\mk{m}_{\pe}$ given in terms of
the matrix identifications \eqref{m.pa.pe.sp}--\eqref{h.pa.pe.sp} by
\begin{align}
& e_\pa= (\mr{a}_{\pa}(\cdot))
\label{epa.sp}\\
& e_\pe= ((\mr{a}_{\pe}(\cdot),\mr{b}_{\pe}(\cdot)),(\mb{a}_{\pe}(\cdot),\mb{b}_{\pe}(\cdot)))
\label{epe.sp}
\end{align}
where $\mr{a}_{\pa}(\cdot)$, $\mr{a}_{\pe}(\cdot)$, $\mr{b}_{\pe}(\cdot)$ 
are linear maps from $T_x M$ into $\Rnum$, $\i\Rnum$, $\Cnum$, respectively, 
and where both $\mb{a}_{\pe}(\cdot)$ and $\mb{b}_{\pe}(\cdot)$ are linear maps
from $T_x M$ into $\Cnum^{n-1}$.
Let $(T_\map M)_\pa$ and $(T_\map M)_\pe$ be the orthogonal subspaces of
$T_\map M$ respectively defined by the kernels of $e_\pa$ and $e_\pe$,
so thus
\beq
e_\pa\rfloor (T_\map M)_\pe = e_\pe\rfloor (T_\map M)_\pa = 0
\eeq
and hence
\begin{align}
& e\rfloor (T_\map M)_\pa = e_\pa\rfloor T_\map M
= \mk{m}_{\pa}\simeq \Rnum ,
\label{TM.sp.pa}\\
& e\rfloor (T_\map M)_\pe = e_\pe\rfloor T_\map M
= \mk{m}_{\pe}\simeq \i\Rnum\oplus\Cnum\oplus\Cnum^{n-1}\oplus\Cnum^{n-1}.
\label{TM.sp.pe}
\end{align}
Note that, in this notation,
\begin{align}
& e_\pa\rfloor \map_x = e_x, \quad
e_\pe\rfloor \map_x = 0 ,\\
& e_\pa\rfloor \map_t = h_\pa, \quad
e_\pe\rfloor \map_t = h_\pe .
\end{align}
Now if $\{\mr{m}_{\pa\Rnum}\}$ is a basis for $\Rnum$,
then $e_\pa$ determines a corresponding basis $\{X_{\pa\Rnum}\}$
for $(T_\map M)_\pa$ given by
\[
\mr{a}_{\pa}(X_{\pa\Rnum})
= \mr{m}_{\pa\Rnum}.
\]
Similarly if $\{\mr{m}_{\pe\i\Rnum}\}$ and $\{\mr{m}_{\pe\Cnum},\mr{m'}_{\pe\Cnum}\}$
are respectively a basis for $\i\Rnum$ and $\Cnum$ viewed as real vector spaces,
and if
$\{\mb{m}_{\pe\Cnum}^{(k)}\}$, $k=1,\ldots,2(n-1)$,  
is a basis for $\Cnum^{n-1}$ viewed as a real vector space, 
then $e_\pe$ determines a corresponding basis
$\{X_{\pe\i\Rnum},X_{\pe\Cnum},X'{}_{\pe\Cnum},X_{\pe\Cnum}^{(k)}, X_{\pe\Cnum'}^{(k')}\}$, $k,k'=1,\ldots,2(n-1)$,
for the vector space $(T_\map M)_\pe$ given by
\begin{align*}
&((\mr{a}_{\pe}(X_{\pe\i\Rnum}),\mr{b}_{\pe}(X_{\pe\i\Rnum})),
(\mb{a}_{\pe}(X_{\pe\i\Rnum}),\mb{b}_{\pe}(X_{\pe\i\Rnum})))
= ((\mr{m}_{\pe\i\Rnum},0),(0,0)),
\\
&((\mr{a}_{\pe}(X_{\pe\Cnum}),\mr{b}_{\pe}(X_{\pe\Cnum})),
(\mb{a}_{\pe}(X_{\pe\Cnum}),\mb{b}_{\pe}(X_{\pe\Cnum})))
= ((0,\mr{m}_{\pe\Cnum}),(0,0)),
\\
&((\mr{a}_{\pe}(X'{}_{\pe\Cnum}),\mr{b}_{\pe}(X'{}_{\pe\Cnum})),
(\mb{a}_{\pe}(X'{}_{\pe\Cnum}),\mb{b}_{\pe}(X'{}_{\pe\Cnum})))
= ((0,\mr{m'}_{\pe\Cnum}),(0,0)),
\\
&((\mr{a}_{\pe}(X_{\pe\Cnum}^{(k)}),\mr{b}_{\pe}(X_{\pe\Cnum}^{(k)})),
(\mb{a}_{\pe}(X_{\pe\Cnum}^{(k)}),\mb{b}_{\pe}(X_{\pe\Cnum}^{(k)})))
= ((0,0),(\mb{m}_{\pe\Cnum}^{(k)},0)) ,
\\
&((\mr{a}_{\pe}(X_{\pe\Cnum}^{(k')}),\mr{b}_{\pe}(X_{\pe\Cnum}^{(k')})),
(\mb{a}_{\pe}(X_{\pe\Cnum'}^{(k')}),\mb{b}_{\pe}(X_{\pe\Cnum'}^{(k')})))
= ((0,0),(0,\mb{m}_{\pe\Cnum}^{(k')})) .
\end{align*}
In addition, if each basis
$\{\mr{m}_{\pa\Rnum}\}$, $\{\mr{m}_{\pe\i\Rnum}\}$,
$\{\mr{m}_{\pe\Cnum}, \mr{m'}_{\pe\Cnum}\}$,
$\{\mb{m}_{\pe\Cnum}^{(k)}\}$
is normalized such that
\begin{align*}
& \<\mr{m}_{\pa\Rnum},\mr{m}_{\pa\Rnum}\> =\<\mr{m}_{\pe\i\Rnum},\mr{m}_{\pe\i\Rnum}\>=-1 ,
\\
& \<\mr{m}_{\pe\Cnum},\mr{m}_{\pe\Cnum}\>=\<\mr{m'}_{\pe\Cnum},\mr{m'}_{\pe\Cnum}\>=-1 ,
\quad 
\<\mr{m}_{\pe\Cnum},\mr{m'}_{\pe\Cnum}\>=0 ,
\\
& 
\<\mb{m}_{\pe\Cnum}^{(k)},\mb{m}_{\pe\Cnum}^{(k')}\> =-\delta_{kk'} ,
\end{align*}
then the basis for $T_\map M=(T_\map M)_\pa\oplus (T_\map M)_\pe$
has the corresponding normalization
\begin{align*}
& g(X_{\pe\Rnum},X_{\pe\Rnum}) = g(X_{\pe\i\Rnum},X_{\pe\i\Rnum}) =1 ,
\\
& g(X_{\pe\Cnum},X_{\pe\Cnum}) =g(X'{}_{\pe\Cnum},X'{}_{\pe\Cnum}) =1 ,
\quad
g(X_{\pe\Cnum},X'{}_{\pe\Cnum}) =0, 
\\
& g(X_{\pe\Cnum}^{(k)},X_{\pe\Cnum}^{(k')}) 
= g(X_{\pe\Cnum'}^{(k)},X_{\pe\Cnum'}^{(k')}) =\delta_{kk'} .
\end{align*}
Consequently, from the transport equation \eqref{transport.sp} together with
the Lie brackets \eqref{inclu.5.sp}, \eqref{inclu.12.sp.1} and \eqref{inclu.12.sp.2}, 
the resulting orthonormal frame
\beq
\{X_{\pa\Rnum},X_{\pe\i\Rnum},X_{\pe\Cnum},X'{}_{\pe\Cnum},X_{\pe\Cnum}^{(k)}, X_{\pe\Cnum'}^{(k')}\},
\eeq
can be shown to satisfy the Frenet equations 
\begin{subequations}
\beq
\begin{aligned}
& \nabla_x X_{\pa\Rnum}
=U_{\Rnum,\i\Rnum}X_{\pe\i\Rnum}+U_{\Rnum,\Cnum}X_{\pe\Cnum}+U'{}_{\Rnum,\Cnum}X'{}_{\pe\Cnum}
+\sum_{k} U_{\Rnum,\Cnum}^{(k)}X_{\pe\Cnum}^{(k)}
+\sum_{k'} U_{\Rnum,\Cnum'}^{(k')}X_{\pe\Cnum'}^{(k')}
\\
& \nabla_x X_{\pe\i\Rnum}
= -U_{\Rnum,\i\Rnum}X_{\pa\Rnum}
+ \sum_{k} U_{\i\Rnum,\Cnum}^{(k)}X_{\pe\Cnum}^{(k)}
+\sum_{k'} U_{\i\Rnum,\Cnum'}^{(k')}X_{\pe\Cnum'}^{(k')}
\\
&\nabla_x X_{\pe\Cnum}
= -U_{\Rnum,\Cnum}X_{\pa\Rnum}
+ \sum_{k} U_{\Cnum,\Cnum}^{(k)}X_{\pe\Cnum}^{(k)}
+\sum_{k'} U_{\Cnum,\Cnum'}^{(k')}X_{\pe\Cnum'}^{(k')}
\\
&\nabla_x X'{}_{\pe\Cnum}
= -U'{}_{\Rnum,\Cnum}X_{\pa\Rnum}
+\sum_{k} U'{}_{\Cnum,\Cnum}^{(k)}X_{\pe\Cnum}^{(k)}
+\sum_{k'} U'{}_{\Cnum,\Cnum'}^{(k')}X_{\pe\Cnum'}^{(k')}
\end{aligned}
\eeq
\beq
\begin{aligned}
\nabla_x X_{\pe\Cnum}^{(k)}
& =-U^{(k)}_{\Rnum,\Cnum}X_{\pa\Rnum}
-U^{(k)}_{\i\Rnum,\Cnum}X_{\pe\i\Rnum}
-U^{(k)}_{\Cnum,\Cnum}X_{\pe\Cnum} -U^{'(k)}_{\Cnum,\Cnum}X'{}_{\pe\Cnum} \\&\qquad
+\sum_{j} U_{\Cnum,\Cnum}^{(k,j)}X_{\pe\Cnum}^{(j)}
+\sum_{j'} U_{\Cnum,\Cnum'}^{(k,j')}X_{\pe\Cnum'}^{(j')}
\\
\nabla_x X_{\pe\Cnum'}^{(k')}
& =-U^{(k')}_{\Rnum,\Cnum'}X_{\pa\Rnum}
-U^{(k')}_{\i\Rnum,\Cnum'}X_{\pe\i\Rnum}
-U^{(k')}_{\Cnum,\Cnum'}X_{\pe\Cnum} -U^{'(k')}_{\Cnum,\Cnum'}X'{}_{\pe\Cnum} \\&\qquad
-\sum_{j} U_{\Cnum,\Cnum'}^{(j,k')}X_{\pe\Cnum}^{(j)}
+\sum_{j'} U_{\Cnum',\Cnum'}^{(k',j')}X_{\pe\Cnum'}^{(j')}
\end{aligned}
\eeq
\end{subequations}
where 
\begin{subequations}
\beq
\begin{aligned}
U_{\Rnum,\i\Rnum}
&= \<[((\mr{m}_{\pa\Rnum}),((\u_1,\u_2),(\tbu_{1},\tbu_{2}))],
((\mr{m}_{\pe\i\Rnum},0),(0,0))\> \\
&=8(n+2)\A(\mr{m}_{\pa\Rnum}\u_1,\mr{m}_{\pe\i\Rnum})
\\
U_{\Rnum,\Cnum}
&= \<[((\mr{m}_{\pa\Rnum}),((\u_1,\u_2),(\tbu_{1},\tbu_{2}))],
((0,\mr{m}_{\pe\Cnum}),(0,0))\> \\
& =8(n+2)\A(\mr{m}_{\pa\Rnum}\u_2,\mr{m}_{\pe\Cnum})
\\
U'{}_{\Rnum,\Cnum}
&= \<[((\mr{m}_{\pa\Rnum}),((\u_1,\u_2),(\tbu_{1},\tbu_{2}))],
((0,\mr{m'}_{\pe\Cnum}),(0,0))\> \\
& =8(n+2)\A(\mr{m}_{\pa\Rnum}\u_2,\mr{m'}_{\pe\Cnum})
\end{aligned}
\eeq
\beq
\begin{aligned}
U_{\Rnum,\Cnum}^{(k)}
&= \<[((\mr{m}_{\pa\Rnum})),((\u_1,\u_2),(\tbu_{1},\tbu_{2}))],
((0,0),(\mb{m}_{\pe\Cnum}^{(k)},0))\> \\
& =-4(n+2)\A(\mr{m}_{\Rnum}\tbu_1,\mb{m}_{\pe\Cnum}^{(k)})
\\
U_{\Rnum,\Cnum'}^{(k')}
&= \<[((\mr{m}_{\pa\Rnum}),((\u_1,\u_2),(\tbu_{1},\tbu_{2}))],
((0,0),(0,\mb{m}_{\pe\Cnum}^{(k')}))\> \\
& =-4(n+2)\A(\mr{m}_{\pa\Rnum}\tbu_2,\mb{m}_{\pe\Cnum}^{(k')})
\\
U_{\i\Rnum,\Cnum}^{(k)}
&= \<[((\mr{m}_{\pe\i\Rnum},0),(0,0)),((\u_1,\u_2),(\tbu_{1},\tbu_{2}))],
((0,0),(\mb{m}_{\pe\Cnum}^{(k)},0))\> \\
& =-4(n+2)\A(\mr{m}_{\pe\i\Rnum}\tbu_1,\mb{m}_{\pe\Cnum}^{(k)})
\\
U_{\i\Rnum,\Cnum'}^{(k')}
&= \<[((\mr{m}_{\pe\i\Rnum},0),(0,0)),((\u_1,\u_2),(\tbu_{1},\tbu_{2}))],
((0,0),(0,\mb{m}_{\pe\Cnum}^{(k')}))\> \\
&=-4(n+2)\A(\mr{m}_{\pe\i\Rnum}\tbu_2,\mb{m}_{\pe\Cnum}^{(k')})
\\
U_{\Cnum,\Cnum}^{(k)}
&= \<[((0,\mr{m}_{\pe\Cnum}),(0,0)),((\u_1,\u_2),(\tbu_{1},\tbu_{2}))],
((0,0),(\mb{m}_{\pe\Cnum}^{(k)},0))\> \\
&=4(n+2)\A(\mr{m}_{\pe\Cnum}\btbu_2,\mb{m}_{\pe\Cnum}^{(k)})
\\
U'{}_{\Cnum,\Cnum}^{(k)}
&= \<[((0,\mr{m'}_{\pe\Cnum}),(0,0)),((\u_1,\u_2),(\tbu_{1},\tbu_{2}))],
((0,0),(\mb{m}_{\pe\Cnum}^{(k)},0))\> \\
&=4(n+2)\A(\mr{m'}_{\pe\Cnum}\btbu_2,\mb{m}_{\pe\Cnum}^{(k)})
\\
U_{\Cnum,\Cnum'}^{(k')}
&= \<[((0,\mr{m}_{\pe\Cnum}),(0,0)),((\u_1,\u_2),(\tbu_{1},\tbu_{2}))],
((0,0),(0,\mb{m}_{\pe\Cnum}^{(k')}))\> \\
&=-4(n+2)\A(\mr{m}_{\pe\Cnum}\btbu_1,\mb{m}_{\pe\Cnum}^{(k')})
\\
U'{}_{\Cnum,\Cnum'}^{(k')}
&= \<[((0,\mr{m'}_{\pe\Cnum}),(0,0)),((\u_1,\u_2),(\tbu_{1},\tbu_{2}))],
((0,0),(0,\mb{m}_{\pe\Cnum}^{(k')}))\> \\
&=-4(n+2)\A(\mr{m'}_{\pe\Cnum}\btbu_1,\mb{m}_{\pe\Cnum}^{(k')})
\end{aligned}
\eeq
\beq
\begin{aligned}
U_{\Cnum,\Cnum}^{(k,j)}
&= \<[((0,0),(\mb{m}_{\pe\Cnum}^{(k)},0)),((\u_1,\u_2),(\tbu_{1},\tbu_{2}))],
((0,0),(\mb{m}_{\pe\Cnum}^{(j)},0))\> \\
&=4(n+2)\A(\u_1 \mb{m}_{\pe\Cnum}^{(k)},\mb{m}_{\pe\Cnum}^{(j)})
= -4(n+2)\A(\u_1 \mb{m}_{\pe\Cnum}^{(j)},\mb{m}_{\pe\Cnum}^{(k)})
\\
U_{\Cnum,\Cnum'}^{(k,j')}
&= \<[((0,0),(\mb{m}_{\pe\Cnum}^{(k)},0)),((\u_1,\u_2),(\tbu_{1},\tbu_{2}))],
((0,0),(0,\mb{m}_{\pe\Cnum}^{(j')}))\> \\
& =4(n+2)\A(\u_2\ov{\mb{m}}_{\pe\Cnum}^{(k)},\mb{m}_{\pe\Cnum}^{(j')})
\\
U_{\Cnum',\Cnum'}^{(k',j')}
&= \<[((0,0),(0,\mb{m}_{\pe\Cnum}^{(k')})),((\u_1,\u_2),(\tbu_{1},\tbu_{2}))],
((0,0),(0,\mb{m}_{\pe\Cnum}^{(j')}))\> \\
&=4(n+2)\A(\u_1 \mb{m}_{\pe\Cnum}^{(k')},\mb{m}_{\pe\Cnum}^{(j')})
=-4(n+2)\A(\u_1 \mb{m}_{\pe\Cnum}^{(j')},\mb{m}_{\pe\Cnum}^{(k')})
\end{aligned}
\eeq
\end{subequations}
denote the Cartan matrix components of the underlying
$Sp(n)\times Sp(1)$-parallel linear connection \eqref{u.pe.sp}
projected into the tangent space of the curve.

In this frame, the components of the principal normal vector
\beq\label{prin.nor.vec.sp}
N:=\nabla_x X=\<e^*,\ad(e_x)\conx_x\>
\eeq
are given by
\beq
e\rfloor N=-\ad(e_x)\conx_x=
\frac{1}{\sqrt{\chi}}((-2\u_1,-2\u_2),(\tbu_{1},\tbu_{2}))
\in\i\Rnum\oplus\Cnum\oplus\Cnum^{n-1}\oplus\Cnum^{n-1}\simeq \mk{m}_{\pe}
\eeq
through relation \eqref{ad.e.h.pe.sp}.
These components $((-2\u_1,-2\u_2),(\tbu_{1},\tbu_{2}))$ 
are invariantly defined by the curve $\map$ 
up to the rigid ($x$-independent) action of 
the equivalence group $H^*_{\pa}=\Ad(Sp(1)\times Sp(n-1))\subset \Ad(Sp(n))$
that preserves the framing at each point $x$.
Moreover, the pair of scalars $(\u_1,\u_2)$ and the pair of vectors $(\tbu_{1},\tbu_{2})$  
belong to separate irreducible representations of this group. 
Hence, in geometrical terms,
the complex scalar-vector pair $((-2\u_1,-2\u_2),(\tbu_{1},\tbu_{2}))$
describes \emph{covariants} of the curve $\map$ relative to the group $H^*_{\pa}$, 
while $x$-derivatives of this pair describe \emph{differential covariants} 
which arise geometrically from the frame components of
$x$-derivatives of the principal normal vector $N$.
We thus note that the geometric invariants of $\map$
as defined by Riemannian inner products of the tangent vector $X=\map_x$
and its derivatives $N=\nabla_x\map_x$, $\nabla_xN=\nabla_x^2\map_x$, etc.\
along the curve $\map$
can be expressed as scalars formed from Cartan-Killing inner products of
the covariants $((-2\u_1,-2\u_2),(\tbu_{1},\tbu_{2}))$
and the differential covariants $((-2\u_{1\,x},-2\u_{2\,x}),(\tbu_{1\,x},\tbu_{2\,x}))$, etc.;
for example
\[
g(N,N)=-g(X,\nabla_x^2X)
=\frac{1}{\chi}\big(4(|\u_1|^2+|\u_2|^2)+|\tbu_1|^2+|\tbu_2|^2\big)
\]
yields the square of the classical curvature invariant of the curve $\map$.
In particular,
the set of invariants given by
$\{g(X,\nabla_x^{2l}X)\}$, $l=1,\ldots,4n-1(=\dim \mk{m} -1)$
generates the components of the connection matrix of
a classical Frenet frame \cite{Guggenheimer} determined by $\map_x$.

Since $T_\map M$ has rank $1$, 
all non-stretching curve flows belong to the same algebraic equivalence class,
as determined by the element \eqref{e.sp}.

\subsection{Hamiltonian operators and flows}
\label{Sp.ops.hier}
The Cartan structure equations \eqref{pull.1} and \eqref{pull.2}
for the $Sp(n)\times Sp(1)$-parallel framing of $\map$
expressed in terms of the variables \eqref{e.sp}--\eqref{w.pe.sp}
are respectively given by 
\beq\label{tors.w.pe.sp}
\begin{aligned}
&\ww^{1\pe}
=\frac{\sqrt{\chi}}{2}\big( D_x\hh_{1\pe} -\ha\C(\tbu_2,\tbh_{2\pe})-\ha\C(\tbu_1,\tbh_{1\pe}) +2\hh_{\pa}\u_1\big),
\\
&\ww^{2\pe}
=\frac{\sqrt{\chi}}{2}\big(D_x\hh_{2\pe}-\ha\S(\tbu_2,\tbh_{1\pe}) +\ha\S(\tbu_1,\tbh_{2\pe}) +2\hh_{\pa}\u_2\big),
\\
&\tbw^{1\pe}
=-\sqrt{\chi}\big(D_x\tbh_{1\pe}-\hh_{1\pe}\tbu_{1}+\u_1\tbh_{1\pe} +\hh_{2\pe}\btbu_2-\u_2\btbh_{2\pe} -\hh_{\pa}\tbu_1\big),
\\
&\tbw^{2\pe}
=-\sqrt{\chi}\big(D_x\tbh_{2\pe}-\hh_{1\pe}\tbu_2 +\u_2\btbh_{1\pe} -\hh_{2\pe}\btbu_1 +\u_1\tbh_2 -\hh_{\pa}\tbu_2\big),
\end{aligned}
\eeq
\beq\label{tors.h.pa.sp}
D_x\hh_{\pa}
= \A(\u_1,\hh_{1\pe}) -\ha\A(\tbu_1,\tbh_{1\pe})
+ \A(\u_2,\hh_{2\pe}) -\ha\A(\tbu_2,\tbh_{2\pe}), 
\eeq
and
\beq\label{curv.u.sp}
\begin{aligned}
\u_{1\,t}
& =D_x\ww^{1\pe}+\ha\C(\tbu_1,\tbw^{1\pe})+\ha\C(\tbu_2,\tbw^{2\pe})
+\bu_2 \ww^{2\pa} -\u_2 \bww^{2\pa}+\hh^{1\pe}, 
\\
\u_{2\,t}
& =D_x\ww^{2\pe}-\ha\S(\tbu_1,\tbw^{2\pe}) +\ha\S(\tbu_2,\tbw^{1\pe})
-2\u_2\ww^{1\pa} +2\u_1\ww^{2\pa} +\hh^{2\pe}, 
\\
\tbu_{1\,t}
& =D_x\tbw^{1\pe} -\u_1\tbw^{1\pe}+\ww^{1\pe}\tbu_1 +\u_2\btbw^{2\pe}-\ww^{2\pe}\btbu_2 -\ww^{1\pa}\tbu_1 +\ww^{2\pa}\btbu_2 
\\&\qquad
+\tbu_1\tbWW^{1\pa} -\tbu_2\btbWW^{2\pa} +\tbh^{1\pe}, 
\\
\tbu_{2\,t}
& =D_x\tbw^{2\pe} -\u_1\tbw^{2\pe}+\ww^{2\pe}\btbu_1-\u_2\btbw^{1\pe}+\ww^{1\pe}\tbu_2 -\ww^{1\pa}\tbu_2 -\ww^{2\pa}\btbu_1
\\&\qquad
+\tbu_2\btbWW^{1\pa} +\tbu_1\tbWW^{2\pa} +\tbh^{2\pe}, 
\end{aligned}
\eeq
\beq\label{curv.w.pa.sp}
\begin{aligned}
&D_x\ww^{1\pa}
=-\bu_2\ww^{2\pe} +\u_2\bww^{2\pe} +\ha\C(\tbu_{1},\tbw^{1\pe}) +\ha\C(\tbu_{2},\tbw^{2\pe}),
\\
&D_x\ww^{2\pa}
=2 \u_2\ww^{1\pe} -2\u_1\ww^{2\pe} +\ha\S(\tbu_{2},\tbw^{1\pe}) -\ha\S(\tbu_{1},\tbw^{2\pe}),
\\
&D_x\tbWW^{1\pa}
=\CC(\tbu_{1},\tbw^{1\pe})+\ov{\CC(\tbu_{2},\tbw^{2\pe})}, 
\\
&D_x\tbWW^{2\pa}
=\ov{\SS(\tbu_{1},\tbw^{2\pe})} -\SS(\tbu_2,\tbw^{1\pe}) . 
\end{aligned}
\eeq
These equations \eqref{tors.w.pe.sp}--\eqref{curv.w.pa.sp} directly encode
a pair of compatible Hamiltonian operators as stated by Theorem~\ref{biHam.flow.eqn}. 
Using the operator notation \eqref{CASinnerops}--\eqref{ccop},
and eliminating $\hh_{\pa}$ through the torsion equation \eqref{tors.h.pa.sp}
and $\ww^{1\pa}$, $\ww^{2\pa}$, $\tbWW^{1\pa}$, $\tbWW^{2\pa}$
through the curvature equation \eqref{curv.w.pa.sp},
as well as replacing $\hh_{1\pe}$, $\hh_{2\pe}$, $\tbh_{1\pe}$, $\tbh_{2\pe}$ 
respectively in terms of $\hh^{1\pe}$,$\hh^{2\pe}$,$\tbh^{1\pe}$, $\tbh^{2\pe}$,
we obtain the following main result.

\begin{theorem}\label{g=sp.theorem}
For the imaginary scalar variable $\u_1\in\i\Rnum$,
the complex scalar variable $\u_2\in\Cnum$, 
and the pair of complex vector variables $\tbu_{1}(t,x),\tbu_{2}(t,x)\in\Cnum^{n-1}$, 
the flow equations given by \eqref{tors.w.pe.sp}--\eqref{curv.u.sp}
have the operator form
\beq\label{g=sp.flow.eq}
\bpm\u_1\\\u_2\\\tbu_{1}\\\tbu_{2}\epm _t
=\mathcal{H}\bpm\ww^{1\pe}\\\ww^{2\pe}\\\tbw^{1\pe}\\\tbw^{2\pe}\epm
+\bpm\hh^{1\pe}\\\hh^{2\pe}\\\tbh^{1\pe}\\\tbh^{2\pe}\epm ,
\quad
\bpm\ww^{1\pe}\\\ww^{2\pe}\\\tbw^{1\pe}\\\tbw^{2\pe}\epm
=\frac{\chi}{4}\mathcal{J}\bpm\hh^{1\pe}\\\hh^{2\pe}\\ \tbh^{1\pe}\\\tbh^{2\pe}\epm ,
\eeq
where $\mathcal{H}$ and $\mathcal{J}$ 
are compatible Hamiltonian cosymplectic and symplectic operators 
on the $x$-jet space of $(\u_1,\u_2,\tbu_{1},\tbu_{2})$.
The $4\times 4$ components of these operators 
$\mathcal{H}=(\mathcal{H}_{ij})$ and $\mathcal{J}=(\mathcal{J}_{ij})$, 
with $i,j=1,2,3,4$, 
are given by 
\beq\label{Ham.op.1.sp}
\begin{aligned}
&\mathcal{H}_{11}=
D_x-\C_{\u_2}D_x^{-1}\S_{\u_2}
\\
&\mathcal{H}_{12}=
\C_{u_2}D_x^{-1}\S_{\u_1}
\\
&\mathcal{H}_{13}=
\ha\C_{\tbu_1}-\ha \C_{\u_2}D_x^{-1}\S_{\tbu_2}
\\
&\mathcal{H}_{14}=
\ha\C_{\tbu_2}+\ha \C_{\u_2}D_x^{-1}\S_{\tbu_1}
\end{aligned}
\eeq
\beq\label{Ham.op.2.sp}
\begin{aligned}
&\mathcal{H}_{21}=
\S_{\u_1}D_x^{-1}\S_{\u_2}
\\
&\mathcal{H}_{22}=
D_x -\S_{\u_2}D_x^{-1}\C_{\u_2}-\S_{\u_1}D_x^{-1}\S_{\u_1}
\\
&\mathcal{H}_{23}=
\ha\S_{\tbu_2}-\ha S_{\u_2}D_x^{-1}\C_{\tbu_1}+\ha \S_{\u_1}D_x^{-1}\S_{\tbu_2}
\\
&\mathcal{H}_{24}=
-\ha\S_{\tbu_1}-\ha\S_{\u_2}D_x^{-1}\C_{\tbu_2}-\ha \S_{\u_1}D_x^{-1}\S_{\tbu_1}
\end{aligned}
\eeq
\beq\label{Ham.op.3.sp}
\begin{aligned}
&\mathcal{H}_{31}=
R_{\tbu_1}+R_{\btbu_2}D_x^{-1}\S_{\u_2}
\\
&\mathcal{H}_{32}=
-R_{\btbu_2}-R_{\tbu_1}D_x^{-1}\C_{\u_2}-R_{\btbu_2}D_x^{-1}\S_{\u_1}
\\
&\mathcal{H}_{33}=
D_x-L_{\u_1}-\ha R_{\tbu_1}D_x^{-1}\C_{\tbu_1}+L_{\tbu_1}D_x^{-1}\CC_{\tbu_1}
+\ha R_{\btbu_2}D_x^{-1}\S_{\tbu_2} +L_{\tbu_2}D_x^{-1}\cc\SS_{\tbu_2}
\\
&\mathcal{H}_{34}=
L_{\u_2}\cc-\ha R_{\tbu_1}D_x^{-1}\C_{\tbu_2}+L_{\tbu_1}D_x^{-1}\cc\CC_{\tbu_2}
-R_{\btbu_2}D_x^{-1}\S_{\tbu_1} -L_{\tbu_2}D_x^{-1}\S_{\tbu_1}
\end{aligned}
\eeq
\beq\label{Ham.op.4.sp}
\begin{aligned}
&\mathcal{H}_{41}=
R_{\tbu_2}-R_{\btbu_1}D_x^{-1}\S_{\u_2}
\\
&\mathcal{H}_{42}=
R_{\btbu_1}-R_{\tbu_2}D_x^{-1}\C_{\u_2}+R_{\btbu_1}D_x^{-1}\S_{\u_1}
\\
&\mathcal{H}_{43}=
-L_{\u_2}\cc-\ha R_{\tbu_2}D_x^{-1}\C_{\tbu_1}+L_{\tbu_2}D_x^{-1}\cc\CC_{\tbu_1}
-\ha R_{\btbu_1}D_x^{-1}\S_{\tbu_2}-L_{\tbu_1}D_x^{-1}\SS_{\tbu_2}
\\
&\mathcal{H}_{44}=
D_x-L_{\u_1}-\ha R_{\tbu_2}D_x^{-1}\C_{\tbu_2}+L_{\tbu_2}D_x^{-1}\CC_{\tbu_2}
+\ha R_{\btbu_1}D_x^{-1}\S_{\tbu_1}
+L_{\tbu_1}D_x^{-1}\cc\SS_{\tbu_1}
\end{aligned}
\eeq
and
\beq\label{Sym.op.sp}
\begin{aligned}
&\mathcal{J}_{11}=
D_x+\S_{\u_1}D_x^{-1}\A_{\u_1}
\\
&\mathcal{J}_{12}=
\S_{\u_1}D_x^{-1}\A_{\u_2}
\\
&\mathcal{J}_{13}=
\C_{\tbu_1}+\S_{\u_1}D_x^{-1}\A_{\tbu_1}
\\
&\mathcal{J}_{14}=
\C_{\tbu_2}+\S_{\u_1}D_x^{-1}\A_{\tbu_2}
\end{aligned}
\eeq
\beq\label{Sym.op.sp.1}
\begin{aligned}
&\mathcal{J}_{21}=
\S_{\u_2}D_x^{-1}\A_{\u_1}\\
&\mathcal{J}_{22}=
D_x+\S_{\u_2}D_x^{-1}\A_{\u_2}\\
&\mathcal{J}_{23}=
\S_{\tbu_2}+\S_{\u_2}D_x^{-1}\A_{\tbu_1}\\
&\mathcal{J}_{24}=
-\S_{\tbu_1}+\S_{\u_2}D_x^{-1}\A_{\tbu_2}
\end{aligned}
\eeq
\beq\label{Sym.op.sp.2}
\begin{aligned}
&\mathcal{J}_{31}=
2R_{\tbu_1}+2R_{\tbu_1}D_x^{-1}\A_{\u_1}\\
&\mathcal{J}_{32}=
-2R_{\btbu_2}+2R_{\tbu_1}D_x^{-1}\A_{\u_2}\\
&\mathcal{J}_{33}=
4D_x+4L_{\u_1}+2R_{\tbu_1}D_x^{-1}\A_{\tbu_1}\\
&\mathcal{J}_{34}=
-4L_{\u_2}\cc+2R_{\tbu_1}D_x^{-1}\A_{\tbu_2}
\end{aligned}
\eeq
\beq\label{Sym.op.sp.3}
\begin{aligned}
&\mathcal{J}_{41}=
2R_{\tbu_2}+2R_{\tbu_2}D_x^{-1}\A_{\u_1}\\
&\mathcal{J}_{42}=
2R_{\btbu_1}+2R_{\tbu_2}D_x^{-1}\A_{\u_2}
\\
&\mathcal{J}_{43}=
4L_{\u_2}\cc+2R_{\tbu_2}D_x^{-1}\A_{\tbu_1}\\
&\mathcal{J}_{44}=
4D_x+4L_{\u_1}+2R_{\tbu_2}D_x^{-1}\A_{\tbu_2} .
\end{aligned}
\eeq
\end{theorem}

The properties of these operators are similar to the Hamiltonian structure
\eqref{su.poisson}--\eqref{su.symp.cyclic}. 
Let $J^\infty$ denote the $x$-jet space of the variables 
$(\u_1,\u_2,\tbu_{1},\tbu_{2})$. 
The cosymplectic property of $\Hop$ means that it defines
an associated Poisson bracket
\beq\label{poisson}
\{\mk{H}_{1},\mk{H}_{2}\}_{\Hop} :=
\int \sum_{\substack{l=1,2\\l'=1,2}} \A(\delta \mk{H}_{1}/\delta\u_{l}, \Hop_{ll'}(\delta \mk{H}_{2}/\delta\u_{l'})) 
+ \sum_{\substack{l=3,4\\l'=3,4}} \A(\delta \mk{H}_{1}/\delta\tbu_{l-2}, \Hop_{ll'}(\delta \mk{H}_{2}/\delta\tbu_{l'-2})) 
\;dx
\eeq
which is skew-symmetric and obeys the Jacobi identity,
for all real-valued functionals $\mk{H}$ on $J^\infty$.
The symplectic property of $\Jop$ means that it defines 
an associated symplectic $2$-form 
\beq
\bs{\omega}(\X_{1},\X_{2})_{\Jop} :=
\int \sum_{\substack{l=1,2\\l'=1,2}} \A( \X_{1}\u_{l}, \Jop_{ll'}(\X_{2}\u_{l'})) 
+ \sum_{\substack{l=3,4\\l'=3,4}} \A( \X_{1}\tbu_{l-2}, \Jop_{ll'}(\X_{2}\tbu_{l'-2})) 
\;dx
\eeq
which is skew-symmetric and closed
for all vector fields 
$\X=\hh^{1\pe}\cdot \p/\p\u_{1} + \hh^{2\pe}\cdot \p/\p\u_{2}
+\tbh^{1\pe}\cdot \p/\p\tbu_{1} + \tbh^{2\pe}\cdot \p/\p\tbu_{2}$
defined in terms of scalar-vector function pairs 
$(\hh^{1\perp},\hh^{2\perp})\in\i\Rnum\oplus \Cnum$, 
$(\tbh^{1\perp},\tbh^{2\perp})\in\Cnum^{n-1}\oplus \Cnum^{n-1}$ on $J^\infty$. 
Compatibility of the operators $\Hop$ and $\Jop$ means that
every linear combination $c_{1}\Hop+c_{2}\Jop^{-1}$ is a cosymplectic Hamiltonian operator,
or equivalently that $c_{1}\Hop^{-1}+c_{2}\Jop$ is a symplectic operator,
where $\Hop^{-1}$ and $\Jop^{-1}$ denote formal inverse operators
defined on $J^\infty$.

As a consequence of Theorem~\ref{biHam.hier}, we have the following result.

\begin{corollary}\label{g=sp.hier}
The operator $\Rop=\Hop\Jop$ generates a hierarchy of
bi-Hamiltonian flows \eqref{g=sp.flow.eq} on 
$(\u_1(t,x),\u_2(t,x),\tbu_{1}(t,x),\tbu_{2}(t,x))$
given by
\beq\label{g=sp.h.pe.hier}
\bpm\hh_{(k)}^{1\pe}\\\hh_{(k)}^{2\pe}\\\tbh_{(k)}^{1\pe}\\\tbh_{(k)}^{2\pe}\epm
=\Rop^k\bpm\u_{1\,x}\\\u_{2\,x}\\\tbu_{1\,x}\\\tbu_{2\,x}\epm ,
\quad
k=0,1,2,\ldots
\eeq
and
\beq\label{g=sp.w.pe.hier}
\bpm\ww_{(k)}^{1\pe}\\\ww_{(k)}^{2\pe}\\
\tbw_{(k)}^{1\pe}\\\tbw_{(k)}^{2\pe}\epm
= \bpm\delta H^{(k)}/\delta\u_{1}\\\delta H^{(k)}/\delta\u_{2}\\
\delta H^{(k)}/\delta\tbu_{1}\\\delta H^{(k)}/\delta\tbu_{2}\epm
=\Rop^{*k}\bpm\u_{1}\\\u_{2}\\\tbu_{1}\\\tbu_{2}\epm ,
\quad
k=0,1,2,\ldots
\eeq
in terms of the Hamiltonians
\beq\label{g=sp.H.hier}
H^{(k)} =\frac{1}{1+2k}\hh_{\pa}^{(k)}, \quad
k=0,1,2,\ldots 
\eeq
with 
\beq
\hh_{\pa}^{(k)} = D_x^{-1}\big(
A(\u_{1},\hh_{(k)}^{1\pe})+ A(\u_{2},\hh_{(k)}^{2\pe})
+ A(\tbu_{1},\tbh_{(k)}^{1\pe})+ A(\tbu_{2},\tbh_{(k)}^{2\pe})
\big),
\eeq
where the operator $\Rop^*=\Jop\Hop$ is the adjoint of $\Rop$.
\end{corollary}

The $+k$ flow in this hierarchy \eqref{g=sp.h.pe.hier} is scaling invariant 
under $(\u_1,\u_2,\tbu_{1},\tbu_{2}) \rightarrow \lambda^{-1} (\u_1,\u_2,\tbu_{1},\tbu_{2})$, $x\rightarrow \lambda x$, $t\rightarrow \lambda^{1+2k}t$.

\subsection{mKdV flow}
\label{Sp.mkdv.eqns}
After a scaling of $t\rightarrow 4t/\chi$, where $\chi=8(n+2)$,
the $+1$ flow in the hierarchy \eqref{g=sp.h.pe.hier}
yields an integrable system of coupled scalar-vector mKdV equations
\beq\label{g=sp.scal.mkdv}
\begin{aligned}
\u_{1\,t}-\frac{1}{\chi}\u_{1\,x} & =
\u_{1\,xxx}-3( \btbu_{1\,xx}\cdot\tbu_{1} - \btbu_{1}\cdot\tbu_{1\,xx}
+ \btbu_{2\,xx}\cdot\tbu_{2} - \btbu_{2}\cdot\tbu_{2\,xx} )
+6\u_{1\,x}(|\u_1|^2+|\u_2|^2)\\&\qquad
+3\bu_{2} ( \tbu_{2}\cdot\tbu_{1\,x} -\tbu_{1}\cdot\tbu_{2\,x} )
-3\u_{2}( \btbu_{2}\cdot\btbu_{1\,x} -\btbu_{1}\cdot\btbu_{2\,x} )
\\
\u_{2\,t}-\frac{1}{\chi}\u_{2\,x} & =
\u_{2\,xxx}+3( \tbu_{1\,xx}\cdot\tbu_2 -\tbu_{2\,xx}\cdot\tbu_1 )
+6\u_{2\,x}(|\u_2|^2+|\u_1|^2)\\&\qquad
+6\u_{1}( \tbu_{1\,x}\cdot\tbu_{2} -\tbu_{2\,x}\cdot\tbu_{1} )
+6\u_{2}( \btbu_{1\,x}\cdot\tbu_{1} -\btbu_{1}\cdot\tbu_{1\,x} 
+ \btbu_{2\,x}\cdot\tbu_{2} -\btbu_{2}\cdot\tbu_{2\,x} )
\end{aligned}
\eeq
\beq\label{g=sp.vec.mkdv}
\begin{aligned}
\tbu_{1\,t}-\frac{1}{\chi}\tbu_{1\,x} & =
4\tbu_{1\,xxx} +\u_{1\,xx}\tbu_1 -3\u_{2\,xx}\btbu_2
+6( \u_{1\,x} +|\tbu_1|^2+|\tbu_2|^2+|\u_2|^2+|\u_1|^2 )\tbu_{1\,x}\\&\qquad
-6\u_{2x}\btbu_{2\,x}
+3\big( \btbu_{1\,x}\cdot\tbu_{1}- \btbu_{1}\cdot\tbu_{1\,x}
+ \btbu_{2\,x}\cdot\tbu_{2}- \btbu_{2}\cdot\tbu_{2\,x} 
+(|\u_2|^2+|\u_1|^2)_x \\&\qquad
+2\u_1(|\tbu_1|^2+|\tbu_2|^2) \big)\tbu_1
+3\big( \tbu_{1\,x}\cdot\tbu_2-\tbu_1\cdot\tbu_{2\,x} -2\u_2(|\tbu_2|^2+|\tbu_1|^2) \big)\btbu_2
\\
\tbu_{2\,t}-\frac{1}{\chi}\tbu_{2\,x}=&
4\tbu_{2\,xxx} +3\u_{2\,xx}\btbu_1+\u_{1\,xx}\tbu_2
+6( \u_{1\,x} +|\tbu_1|^2+|\tbu_2|^2+|\u_2|^2+|\u_1|^2 )\tbu_{2\,x}\\&\qquad
+6\u_{2\,x}\btbu_{1\,x}
+3\big( \btbu_{1\,x}\cdot\tbu_{1} -\btbu_{1}\cdot\tbu_{1\,x} 
+ \btbu_{2\,x}\cdot\tbu_{2} -\btbu_{2}\cdot\tbu_{2\,x} 
+(|\u_2|^2+|\u_1|^2)_x \\&\qquad
+2\u_1(|\tbu_1|^2+|\tbu_2|^2) \big)\tbu_2
+ 3\big( \tbu_1\cdot\tbu_{2\,x} - \tbu_{1\,x}\cdot\tbu_2 
+2\u_2(|\tbu_1|^2+|\tbu_2|^2) \big)\btbu_1
\end{aligned}
\eeq
where a dot denotes the standard Euclidean inner product (cf \eqref{Herm.inner.prod}--\eqref{Eucl.inner.prod}).

This system is invariant under the symplectic group $Sp(1)\times Sp(n-1)$,
defined by the transformations \eqref{SP-n.action.sp}--\eqref{SP-1.action.sp}
on the scalar-vector pair $((\u_1,\u_2),(\tbu_{1},\tbu_{2}))$,
and has the following bi-Hamiltonian structure
\beq\label{g=sp.mkdv.Ham}
\bpm\u_1\\\u_2\\\tbu_{1}\\\tbu_{2}\epm _t
- \chi^{-1}\bpm\u_1\\\u_2\\\tbu_{1}\\\tbu_{2}\epm _x
=\mathcal{H}
\bpm \delta H^{(1)}/\delta\u_{1}\\
\delta H^{(1)}/\delta\u_{2}\\
\delta H^{(1)}/\delta\tbu_{1}\\
\delta H^{(1)}/\delta\tbu_{2}\epm
= \mathcal{E}\bpm \delta H^{(0)}/\delta\u_{1}\\
\delta H^{(0)}/\delta\u_{2}\\
\delta H^{(0)}/\delta\tbu_{1}\\
\delta H^{(0)}/\delta\tbu_{2}\epm
\eeq
in terms of the Hamiltonians
\begin{align}
& H^{(0)} = 
4(|\u_1|^2+|\u_2|^2)+|\tbu_1|^2+|\tbu_2|^2 , 
\\
& \begin{aligned}
H^{(1)} & = 
-(|\u_{1\,x}|^2+|\u_{2\,x}|^2 + |\tbu_{1\,x}|^2+|\tbu_{2\,x}|^2)
+2\u_{1}( \tbu_{1\, x}\cdot\btbu_{1} - \tbu_{1}\cdot\btbu_{1\, x} ) \\&\qquad
+\u_2( \btbu_{1\,x}\cdot\btbu_{2} -\btbu_{2\,x}\cdot\btbu_{1} )
+ \bu_2( \tbu_{1\,x}\cdot\tbu_{2}-\tbu_{2\,x}\cdot\tbu_{1} ) \\&\qquad
+(|\u_1|^2+|\u_2|^2)^2 +(|\tbu_1|^2+|\tbu_2|^2)^2 
+2(|\tbu_1|^2+|\tbu_2|^2)(|\u_1|^2+|\u_2|^2)
\end{aligned}
\end{align}
where $\Eop=\Hop\Jop\Hop$ is a Hamiltonian cosymplectic operator
compatible with $\Hop$. 

We remark that the convective terms $\u_{1\,x}$, $\u_{2\,x}$,$\tbu_{1\,x}$, $\tbu_{2\,x}$
on the left-hand side in the system \eqref{g=sp.scal.mkdv}--\eqref{g=sp.vec.mkdv} and \eqref{g=sp.mkdv.Ham}
can be removed by the Galilean transformation
$t\rightarrow t$, $x\rightarrow x+\chi\inv t$.

\subsection{SG flow}
\label{Sp.sg.eqns}
The recursion operator $\mathcal{R}=\mathcal{H}\mathcal{J}$ yields a $-1$ flow defined by 
\beq\label{g=sp.-1flow}
0=
\bpm\ww^{1\pe}\\\ww^{2\pe}\\\tbw^{1\pe}\\\tbw^{2\pe}\epm
=\frac{\chi}{4}\mathcal{J}\bpm\hh^{1\pe}\\\hh^{2\pe}\\ \tbh^{1\pe}\\\tbh^{2\pe}\epm .
\eeq
The resulting flow equations \eqref{g=sp.flow.eq} have the form 
\beq\label{curv.u.sp.-1.flow}
\bpm\u_{1}\\\u_{2}\\\tbu_{1}\\\tbu_{2}\epm _t
=\bpm\hh^{1\pe}\\\hh^{2\pe}\\\tbh^{1\pe}\\\tbh^{2\pe}\epm, 
\eeq
with 
\beq\label{h.pe.sp.-1.flow}
\begin{aligned}
& D_x\hh^{1\pe}=
-\C(\tbu_{1},\tbh^{1\pe})-\C(\tbu_{2},\tbh^{2\pe})
-\frac{4}{\sqrt{\chi}}\hh_{\pa}\u_{1} , 
\\
& D_x\hh^{2\pe}=
-\S(\tbu_{2},\tbh^{1\pe}) +\S(\tbu_{1},\tbh^{2\pe})
-\frac{4}{\sqrt{\chi}}\hh_{\pa}\u_{2} , 
\end{aligned}
\eeq
\beq\label{hvec.pe.sp.-1.flow}
\begin{aligned}
& -D_x\tbh^{1\pe}=
\ha \hh^{1\pe}\tbu_{1}+\u_{1}\tbh^{1\pe}-\ha \hh^{2\pe}\btbu_{2}-\u_{2}\btbh^{2\pe}
+\frac{1}{\sqrt{\chi}}\hh_{\pa}\tbu_{1} , 
\\
& -D_x\tbh^{2\pe}=
\ha\hh^{1\pe}\tbu_{2}+\u_{2}\btbh^{1\pe}+\ha\hh^{2\pe}\btbu_{1}+\u_{1}\tbh^{2\pe}
+\frac{1}{\sqrt{\chi}}\hh_{\pa}\tbu_{2} , 
\end{aligned}
\eeq
and
\beq\label{h.pa.sp.-1.flow}
\frac{1}{\sqrt{\chi}} D_x\hh_{\pa} 
= \ha\A(\hh^{1\pe},\u_{1}) +\ha\A(\tbh^{1\pe},\tbu_{1})
+ \ha\A(\hh^{2\pe},\u_{2}) +\ha\A(\tbh^{2\pe},\tbu_{2}) . 
\eeq

This system of equations \eqref{h.pe.sp.-1.flow}--\eqref{h.pa.sp.-1.flow}
for the variables $\hh^{1\pe}$, $\hh^{2\pe}$, $\tbh^{1\pe}$, $\tbh^{2\pe}$, $\hh_{\pa}$
possesses the conservation law
\beq\label{sp.sg.conslaw}
D_x\big( \frac{1}{\chi}\hh_{\pa}^2+\frac{1}{4}|\hh^{2\pe}|^2+|\tbh^{2\pe}|^2+\frac{1}{4}|\hh^{1\pe}|^2+|\tbh^{1\pe}|^2 \big)
=0 . 
\eeq
Hence, after a conformal scaling of $t$, we can put 
\beq
\frac{1}{\chi}\hh_{\pa}^2+\frac{1}{4}|\hh^{2\pe}|^2+|\tbh^{2\pe}|^2+\frac{1}{4}|\hh^{1\pe}|^2+|\tbh^{1\pe}|^2
=1
\eeq
which yields the relation 
\beq\label{sp.red.h.pa}
\frac{1}{\sqrt{\chi}}\hh_{\pa}=
\pm \sqrt{1-\tfrac{1}{4}|\hh^{1\pe}|^2 -\tfrac{1}{4}|\hh^{2\pe}|^2 
-|\tbh^{1\pe}|^2 -|\tbh^{2\pe}|^2}.
\eeq
Substitution of $\hh^{1\pe}=\u_{1\,t}$, $\hh^{2\pe}=\u_{2\,t}$, $\tbh^{1\pe}=\tbu_{1\,t}$, $\tbh^{2\pe}=\tbu_{2\,t}$
into the equations \eqref{sp.red.h.pa} and \eqref{h.pe.sp.-1.flow}--\eqref{hvec.pe.sp.-1.flow}
produces a hyperbolic system of coupled scalar-vector SG equations 
\beq\label{g=sp.scal.sg}
\begin{aligned}
&\u_{1\,tx}=
\tbu_{1\,t}\cdot\btbu_1 - \tbu_1\cdot\btbu_{1\,t}
+ \tbu_{2\,t}\cdot\btbu_2 - \tbu_2\cdot\btbu_{2\,t}
\mp 4\u_1\sqrt{1-\tfrac{1}{4}(|\u_{1\,t}|^2+ |\u_{2\,t}|^2)-|\tbu_{1\,t}|^2-|\tbu_{2\,t}|^2}
\\
&\u_{2\,tx}=
\tbu_{2\,t}\cdot\tbu_1-\tbu_{1\,t}\cdot\tbu_2
\mp 4\u_2\sqrt{1-\tfrac{1}{4}(|\u_{1\,t}|^2+ |\u_{2\,t}|^2)
-|\tbu_{1\,t}|^2-|\tbu_{2\,t}|^2}
\end{aligned}
\eeq
\beq\label{g=sp.vec.sg}
\begin{aligned}
&\tbu_{1\,tx}=
\ha(\u_{2\,t}\btbu_{2}-\u_{1\,t}\tbu_{1}) +\u_2\btbu_{2\,t}-\u_1\tbu_{1\,t}
\mp \sqrt{1-\tfrac{1}{4}(|\u_{1\,t}|^2+ |\u_{2\,t}|^2)-|\tbu_{1\,t}|^2-|\tbu_{2\,t}|^2}\ \tbu_1
\\
&\tbu_{2\,tx}=
-\ha(\u_{1\,t}\tbu_2 + \u_{2\,t}\btbu_1) -\u_1\tbu_{2\,t}-\u_2\btbu_{1\,t}
\mp\sqrt{1-\tfrac{1}{4}(|\u_{1\,t}|^2+ |\u_{2\,t}|^2)-|\tbu_{1\,t}|^2-|\tbu_{2\,t}|^2}\ \tbu_2 
\end{aligned}
\eeq
with a dot denoting the standard Euclidean inner product (cf \eqref{Herm.inner.prod}--\eqref{Eucl.inner.prod}).
This system is invariant under the symplectic group $Sp(1)\times Sp(n-1)$,
defined by the transformations \eqref{SP-n.action.sp}--\eqref{SP-1.action.sp}
on the scalar-vector pair $((\u_1,\u_2),(\tbu_{1},\tbu_{2}))$. 

Alternatively, we can algebraically combine equations \eqref{h.pe.sp.-1.flow}--\eqref{hvec.pe.sp.-1.flow} 
to express $\u_{1}$, $\u_{2}$, $\tbu_{1}$, $\tbu_{2}$ entirely in terms of 
$\hh^{1\pe}$, $\hh^{2\pe}$, $\tbh^{1\pe}$, $\tbh^{2\pe}$, and their $x$ derivatives. 
Substitution of the resulting expressions into the flow equation \eqref{curv.u.sp.-1.flow} 
yields coupled scalar-vector SG equations for the variables 
$\hh^{1\pe}$, $\hh^{2\pe}$, $\tbh^{1\pe}$, $\tbh^{2\pe}$. 

\subsection{Geometric curve flows}
\label{Sp.curve.flows}
From Theorem~\ref{geom.curve.flows}, 
the flows in the hierarchy \eqref{g=sp.h.pe.hier} and \eqref{g=sp.-1flow}
for $(\u_{1}(t,x),\u_{2}(t,x),\tbu_{1}(t,x),\tbu_{2}(t,x))$ 
$\in\i\Rnum\oplus\Cnum\oplus\Cnum^{n-1}\oplus\Cnum^{n-1}$
correspond to $Sp(n+1)$-invariant non-stretching geometric curve flows for 
$\map(t,x)$ $\in M=Sp(n+1)/Sp(1)\times Sp(n)$.
The resulting equations of motion can be expressed covariantly in terms of 
$X=\map_x$, $N=\nabla_x\map_x$, and $\nabla_x$-derivatives of $N$,
in addition to the Riemannian metric and curvature tensors on $M$. 

The SG flow \eqref{g=sp.scal.sg}--\eqref{g=sp.vec.sg} is given by 
$\ww^{1\pe}=\ww^{2\pe}=0$ and $\tbw^{1\pe}=\tbw^{2\pe}=0$, 
which implies $\ww^{1\pa}=\ww^{2\pa}=0$ and $\tbWW^{1\pa}=\tbWW^{2\pa}=0$ 
from the structure equation \eqref{curv.w.pa.sp}. 
This determines
\beq
\varpi^{\pe}=\varpi^{\pa}=0 
\eeq
and consequently the corresponding flow vector $\map_t=Y_{(-1)}$ satisfies 
\beq
e\rfloor \nabla_x \map_t
= D_x e_t + [\conx_x,e_t] = [\conx_t,e_x] 
=-\ad(e_x)\varpi^{\pe}
=0
\eeq
Hence we obtain the $Sp(n+1)$-invariant curve flow equation
\beq\label{g=sp.wavemap}
0=\nabla_t\map_x, \quad 
|\map_x|=1, 
\eeq
which is called the {\it non-stretching wave map} on $M=Sp(n+1)/Sp(1)\times Sp(n)$.
This equation \eqref{g=sp.wavemap} satisfies 
the non-stretching property $\nabla_t|\map_x|=0$ 
and possesses the conservation law $\nabla_x|\map_t|=0$, 
corresponding to equation \eqref{sp.sg.conslaw}. 
Thus, up to a conformal scaling of $t$, 
the wave map equation describes a flow with unit speed, $|\map_t|=1$. 

The mKdV flow \eqref{g=sp.scal.mkdv}--\eqref{g=sp.vec.mkdv}, 
after $t$ has been rescaled, 
is given by 
$\hh_{1\pe}= \ha\sqrt{\chi}\u_{1\,x}$, 
$\hh_{2\pe}= \ha\sqrt{\chi}\u_{2\,x}$, 
$\tbh_{1\pe} =-\sqrt{\chi} \tbu_{1\,x}$, 
$\tbh_{2\pe} =-\sqrt{\chi} \tbu_{2\,x}$,
which gives $\hh_{\pa}= \ha\sqrt{\chi}( |\u_{1}|^2 + |\u_{2}|^2 + |\tbu_{1}|^2 + |\tbu_{2}|^2 )$
from the structure equation \eqref{tors.h.pa.sp}. 
This determines
\beq
(e_t)_{\pe} 
= \sqrt{\chi} \big( (\ha\u_{1\,x}, \ha\u_{2\,x}), (-\tbu_{1\,x}, -\tbu_{2\,x}) \big)
\in\mk{m}_{\pe} , 
\label{sp.et.pe}
\eeq
and
\beq\label{sp.et.pa} 
\begin{aligned}
(e_t)_{\pa} 
& = 
\frac{\sqrt{\chi}}{4} \big( \A(\u_{1},\u_{1}) + \A(\u_{2},\u_{2}) + \A(\tbu_{1},\tbu_{1}) + \A(\tbu_{2},\tbu_{2}) \big)
\in\mk{m}_{\pa}\\
& = \frac{\chi}{4} ( \A(\u_{1},\u_{1}) + \A(\u_{2},\u_{2}) + \A(\tbu_{1},\tbu_{1}) + \A(\tbu_{2},\tbu_{2}) ) e_x .
\end{aligned}
\eeq
Then $e_t=e\rfloor\map_t$ can be expressed as follows in terms of
\begin{align}
e\rfloor N 
& =[\conx_x,e_x] 
=\frac{1}{\sqrt{\chi}} \big( (-2\u_{1},-2\u_{2}), (\tbu_{1},\tbu_{2}) \big) 
\in\mk{m}_{\pe}, 
\\
(e\rfloor \nabla_x N)_{\pe} 
& =(e\rfloor N)_x + [\conx_x,e\rfloor N]_{\pe}
\nonumber\\& 
= \frac{1}{\sqrt{\chi}} \big( (-2\u_{1\,x},-2\u_{2\,x}), 
(\tbu_{1\,x}+3\u_{1}\tbu_{1}-3\u_{2}\btbu_{2},\tbu_{2\,x}+3\u_{1}\tbu_{2}+3\u_{2}\btbu_{1}) \big) 
\in\mk{m}_{\pe}, 
\label{sp.gradN.pe}\\
(e\rfloor \nabla_x N)_{\pa} 
& =[\conx_x,e\rfloor N]_{\pa}
\nonumber\\& 
= \frac{1}{\sqrt{\chi}} ( 2\A(\u_{1},\u_{1}) + 2\A(\u_{2},\u_{2}) 
+ \ha \A(\tbu_{1},\tbu_{1}) +\ha \A(\tbu_{2},\tbu_{2}) )
\in\mk{m}_{\pa}.
\label{sp.gradN.pa}
\end{align}
To proceed, it is useful to introduce the linear map 
defined for all $Z\in T_\map M$ by 
\beq
\ad(Z)^2 = -R(\cdot,Z)Z . 
\eeq
Now consider
\beq
e\rfloor \ad(X)^{-2}N =\ad(e_x)^{-2}e\rfloor N= -\ad(e_x)\inv \conx_x
=\sqrt{\chi} \big( (\ha\u_{1},\ha\u_{2}), (-\tbu_{1},-\tbu_{2}) \big) 
\in\mk{m}_{\pe}. 
\eeq
Hence we have 
\begin{align}
& e\rfloor(\ad(\ad(X)^{-2} N)^2 X)_\pe
= (\ad(\ad(e_x)\inv \conx_x)^2 e_x)_\pe = [\conx_x,\ad(e_x)\inv \conx_x]_\pe
\nonumber\\&\qquad
= -\frac{3\sqrt{\chi}}{2}\big( (0,0), (\u_{1}\tbu_{1}-\u_{2}\btbu_{2},\u_{1}\tbu_{2}+\u_{2}\btbu_{1}) \big) 
\in\mk{m}_{\pe}
\label{sp.adsq.N.pe}
\end{align}
and
\begin{align}
& e\rfloor(\ad(\ad(e_x)\inv \conx_x)^2 e_x)_\pa 
= (\ad(e\rfloor\ad(X)^{-2} N)^2 e_x)_\pa = [\conx_x,\ad(e_x)\inv \conx_x]_\pa
\nonumber\\&\qquad
= -\frac{\sqrt{\chi}}{2} \big( \A(\u_{1},\u_{1}) +\A(\u_{2},\u_{2}) +\A(\tbu_{1},\tbu_{1})+\A(\tbu_{2},\tbu_{2}) \big) 
\in\mk{m}_{\pa}
\label{sp.adsq.N.pa}
\end{align}
by means of the Lie brackets \eqref{inclu.12.sp.1} and \eqref{inclu.12.sp.2}.
In addition, we have 
\begin{align}
& g(N,\ad(X)^{-2} N)= g(X,\ad(\ad(X)^{-2} N)^2 X)
= -\<e_x,[\conx_x,\ad(e_x)\inv \conx_x]\>
\nonumber\\&\qquad
= -\frac{\chi}{2} ( \A(\u_{1},\u_{1}) +\A(\u_{2},\u_{2}) +\A(\tbu_{1},\tbu_{1})+\A(\tbu_{2},\tbu_{2}) ) 
\label{sp.N.adsq.N}
\end{align}
since the inner product is $\ad$-invariant. 

Comparing equations \eqref{sp.et.pe}--\eqref{sp.et.pa} 
with equations \eqref{sp.gradN.pe}--\eqref{sp.gradN.pa} and \eqref{sp.adsq.N.pe}--\eqref{sp.N.adsq.N},
we see that 
\begin{align}
& (e_t)_{\pe} +4(e_t)_{\pa} 
= e\rfloor \ad(X)^{-2}\nabla_x N -2 e\rfloor \ad(\ad(X)^{-2} N)^2 X , 
\\
& 2(e_t)_{\pa} = -g(N,\ad(X)^{-2} N) e\rfloor X . 
\end{align}
This yields
\beq
e\rfloor\map_t = (e_t)_{\pe} + (e_t)_{\pa} 
= e\rfloor \ad(X)^{-2}\nabla_x N  -2 e\rfloor \ad(\ad(X)^{-2} N)^2 X 
+\frac{3}{2}g(N,\ad(X)^{-2}N) e\rfloor X .
\eeq
Hence the flow vector $\map_t=Y_{(1)}$ satisfies 
\beq\label{g=sp.mkdvmap}
\map_t = \ad(\map_x)^{-2}\nabla_x^2\map_x 
-2 \ad(\ad(\map_x)^{-2}\nabla_x\map_x )^2 \map_x 
+\frac{3}{2}g(\nabla_x\map_x,\ad(\map_x)^{-2}\nabla_x\map_x ) \map_x,
\quad
|\map_x|=1, 
\eeq
which is a $Sp(n+1)$-invariant curve flow equation
called the {\it non-stretching mKdV map} on $M=Sp(n+1)/Sp(1)\times Sp(n)$.
The nonlinearities in this equation are more complicated than 
in the mKdV map \eqref{g=su.mkdvmap} on $M=SU(2n)/Sp(n)$ 
because of the algebraic property that here the vector spaces 
$\mk{m}_{\pe}\simeq\mk{h}_{\pe}$ each split into two orthogonal eigenspaces 
under the linear map $\ad(e_x)^2$.

\section{Concluding Remarks}
\label{remarks}

The Riemannian symmetric spaces $M=SU(2n)/Sp(n),Sp(n+1)/Sp(1)\times Sp(n)$ 
describe curved generalizations of Euclidean geometries 
in which the Euclidean isometry group is replaced by the Lie group 
$G=SU(2n),Sp(n+1)$ 
and the Euclidean frame rotation gauge group is replaced by
a symplectic subgroup $H=Sp(n),Sp(1)\times Sp(n)$ in $G$.
For arclength-parameterized curves in these geometries, 
the components of the Cartan connection in a suitably defined parallel frame
\cite{AncoJGP} 
along the curve represent differential covariants of the curve,
which can be related to standard differential invariants
by a generalized Hasimoto transformation.
In both geometries these covariants are determined uniquely from 
the curve up to the action of a rigid equivalence group 
$H_\pa = Sp(1)\times Sp(n-1)$ in $H$. 
In particular, when $H=Sp(n)$, the covariants transform as 
a pair of complex vectors having a total of $4n-4$ real components, 
whereas when $H=Sp(1)\times Sp(n)$, the covariants transform as 
an imaginary scalar and a complex scalar, plus a pair of complex vectors,
comprising $4n-1$ real components in total. 

For curves undergoing geometric flows described by 
the non-stretching mKdV map equation \cite{AncoJGP} 
and the non-stretching wave map equation,
the covariants of the curve respectively satisfy bi-Hamiltonian 
mKdV and SG equations that exhibit invariance under the symplectic group 
$Sp(1)\times Sp(n-1)$. 
The simplest cases of these equations occur for the low-dimensional 
Riemannian symmetric spaces $M=SU(4)/Sp(2),Sp(2)/Sp(1)\times Sp(1)$. 

In the case of $M=SU(4)/Sp(2)$, 
the covariants reduce to a pair of complex scalars $u_1,u_2 \in \Cnum$
that transform as a representation of the symplectic group $Sp(1)\times Sp(1)$. 
The resulting multi-component $Sp(1)\times Sp(1)$-invariant 
mKdV and SG equations (cf \eqref{g=su.mkdv} and \eqref{su.sgeq}) 
for this pair of variables 
are equivalent to well-known $SO(4)$-invariant equations 
$\tbu_{t}=\tbu_{xxx}+|\tbu|^2 \tbu_{x}$ 
and $\tbu_{tx}=\pm\sqrt{1-|\tbu_{t}|^2} \tbu$ 
for the 4-component vector variable 
$\tbu=({\rm Re}u_1,{\rm Im}u_1,{\rm Re}u_2,{\rm Im}u_2)$. 
This equivalence is a consequence of the local isomorphisms 
$SU(4)\simeq SO(6)$ and $Sp(2)\simeq SO(5)$ which imply that 
$M=SU(4)/Sp(2)$ is locally isometric to $SO(6)/SO(5)$. 

In the case of $M=Sp(2)/Sp(1)\times Sp(1)$, 
the covariants reduce to an imaginary scalar $u_1\in \i\Rnum$
plus a complex scalar $u_2\in \Cnum$,
transforming as a representation of the symplectic group $Sp(1)$. 
As a consequence of the local isomorphisms 
$Sp(2)\simeq SO(5)$ and $Sp(1)\times Sp(1) \simeq SO(4)$, 
which imply $M=Sp(2)/Sp(1)\times Sp(1) \simeq SO(5)/SO(4)$, 
the resulting multi-component $Sp(1)$-invariant 
mKdV and SG equations (cf \eqref{g=sp.scal.mkdv}--\eqref{g=sp.vec.mkdv} 
and \eqref{g=sp.scal.sg}--\eqref{g=sp.vec.sg}) for these variables 
are equivalent to $SO(3)$-invariant equations 
$\tbu_{t}=\tbu_{xxx}+|\tbu|^2 \tbu_{x}$ 
and $\tbu_{tx}=\pm\sqrt{1-|\tbu_{t}|^2} \tbu$ 
for the 3-component vector variable 
$\tbu=({\rm Im}u_1,{\rm Re}u_2,{\rm Im}u_2)$. 
This is a reduction of the 4-component vector equations in the previous case. 

For all other cases, 
the multi-component $Sp(1)\times Sp(n-1)$-invariant 
mKdV and SG equations that arise from the geometries 
$M=SU(2n)/Sp(n)$ when $n>2$ and $M=Sp(n+1)/Sp(1)\times Sp(n)$ when $n>1$ 
are new and different from each other. 

As will be explained by general results presented elsewhere, 
no NLS equations arise from these geometries 
$M=SU(2n)/Sp(n)$ and $M=Sp(n+1)/Sp(1)\times Sp(n)$
since neither of them has a hermitian structure. 
The same statement applies to the geometries $M=SO(n+1)/SO(n)$ and $M=SU(n)/SO(n)$
considered in earlier work \cite{AncoSIGMA,AncoJPA}. 
Nevertheless, symplectically-invariant NLS equations can be derived 
from the corresponding Lie groups $G=SU(2n)$ and $G=Sp(n+1)$, 
in analogy with the derivation of unitarily-invariant NLS equations 
from $G=SO(n+1)$ and $G=SU(n)$ carried out in \Ref{AncoIMA} by means of 
a suitable parallel frame formulation for non-stretching geometric curve flows
in these Lie groups.

\section*{Acknowledgments}
SCA is supported by an NSERC grant. 
EA is grateful for support from the Mathematics Department at Brock University
for a research visit which initiated this work.

\end{document}